\documentclass[epj,final,numbook,nopacs]{svjour}

\usepackage{graphicx}%
\usepackage{multirow}%
\usepackage{amsmath,amssymb,amsfonts}%
\usepackage[title]{appendix}%
\usepackage{braket}
\usepackage{hyperref}
\hypersetup{colorlinks=true,
            linkcolor=blue,          
            citecolor=blue,        
            filecolor=blue,         
            urlcolor=blue}        

\usepackage{fancyhdr}
\usepackage{cite}

\begin{document}

\title{Stochastic Mean-Field Theory and Applications to Multinucleon Transfer and Kinetic Energy Dissipation Processes in Heavy-Ion Collisions }
\headnote{Review article}
\author{S. Ayik${ }^{1}$, M. Arik${ }^{2}$, O. Yilmaz${ }^{3}$, A. S. Umar${ }^{2}$}

\institute{Physics Department, Tennessee Technological University, Cookeville, TN 38505, USA \and Department of Physics and Astronomy, Vanderbilt University, Nashville, TN 37235, USA \and Physics Department, Middle East Technical University, 06800, Ankara, Turkey }

\date{\today}
\abstract{In this Review article, a brief description of the stochastic mean-field theory (SMF) for describing reaction dynamics in low-energy heavy-ion collisions at bombarding energies in the vicinity of the Coulomb barrier is presented. In these collisions, as a result of strong Pauli blocking, binary nucleon collisions do not have a significant effect on the dissipation and fluctuations. At low energies, the mean-field fluctuations, due to initial correlations, have a dominant effect on fluctuations of macroscopic variables. The SMF theory proposes the determination of an ensemble of single-particle density matrices by specifying random initial fluctuations according to a distribution law. Employing an ensemble of single-particle density matrices, not only the mean values but also the distribution functions of the one-body observables can be determined. If the di-nuclear structure is maintained in heavy-ion collisions, such as deep inelastic collisions and fast quasi-fission reactions, a much simpler description of the reaction mechanism can be derived in terms of several macroscopic variables such as mass and charge asymmetry, and relative linear and relative angular momentum. In this case, by geometric projection of the SMF equations, it is possible to derive the quantal Langevin equations for macroscopic variables. As an application of quantal transport description, an analysis of multinucleon transfers and kinetic energy dissipation and fluctuations is presented for selected quasi-fission reactions.}


\maketitle

\section{Introduction}\label{sec1}
\renewcommand{\thefigure}{1.\arabic{figure}} 
\setcounter{figure}{0}

In this Review, we present a microscopic approach to study dissipative heavy-ion collisions at energies near the Coulomb barrier. This approach is referred to as stochastic mean-field theory (SMF). In quasi-fission reactions,
at Coulomb barrier energies, the identities of projectile and target nuclei are preserved to a large extent, but many nucleons transfer between the projectile-like and target-like nuclei during the contact time. This provides a pathway to the production of unknown neutron-rich isotopes, which are essential for developing our understanding of the physics of atomic nuclei, nuclear structure, and astrophysics. Fragmentation, fission, and fusion processes have been successful in extending the nuclear map of known isotopes~\cite{thoennessen2024}. However, there are regions where these methods have difficulty producing unstable nuclei, typically located in the northeastern part of the nuclear landscape~\cite{zagrebaev2008c}. Investigations of nuclear shape and shell evolution, in the regions of neutron-rich nuclei and the island of stability in the superheavy region, will improve our understanding of the structure of atomic nuclei~\cite{otsuka2020,hofmann2000,oganessian2017}. In particular, the properties of neutron-rich nuclei along the neutron magic number $\mathrm{N}=126$ are crucial to understanding the nucleosynthesis pathways through the r-process~\cite{kajino2019}. Similarly, multinucleon transfer reactions might be a possible alternative to produce neutron-rich superheavy nuclei in the yet unreached island of stability. For experimental projects of new isotope production with multinucleon transfer reactions, we refer the reader to recent reviews~\cite{adamian2020,heinz2022}.

Several phenomenological models have been developed for studying such reactions, including the GRAZING model~\cite{yanez2015}, and the di-nuclear system model (DNS)~\cite{zhang2019,chen2020,wen2017b,diaz-torres2001}. A Langevin type dynamical model has also been successful in describing multinucleon transfer, quasi-fission, and fusion, in a unified 
way~\cite{abe1996,frobrich1998,karpov2002,zagrebaev2007b,karpov2017,saiko2019,saiko2022,saiko2024}. 
There are also microscopic models that explicitly treat nucleonic degrees of freedom, such as the improved quantum molecular dynamics model (ImQMD)~\cite{li2019b}. Although the above-mentioned approaches have been extensively developed and successfully applied, they are based on phenomenology to some extent. On the other hand, microscopic time-dependent self-consistent mean-field theories, such as time-dependent Hartree-Fock theory (TDHF) and its extensions, contain no adjustable parameters and no artificial restrictions on reaction dynamics~\cite{sekizawa2019}.

The TDHF approach~\cite{simenel2012,nakatsukasa2016,simenel2018,simenel2025} can properly describe the most probable dynamical path in low-energy heavy-ion collisions, providing a good description of the total kinetic energy loss (TKEL), the scattering angle, as well as the average neutron and proton numbers of the primary reaction products. 
The TDHF approach has been used to compute multinucleon transfer probabilities \cite{koonin1977,simenel2010,simenel2011,sekizawa2013,scamps2013a,sonika2015,sekizawa2016,scamps2017a,scamps2017b,sekizawa2017,sekizawa2017a,regnier2018}, and to investigate particle number distributions in fission fragments~\cite{scamps2015a,tanimura2017,williams2018}.
Applications of time-dependent microscopic methods based on TDHF to quasi-fission has also attracted lots of efforts in the past few years
\cite{golabek2009,kedziora2010,wakhle2014,oberacker2014,umar2015a,hammerton2015,umar2016,sekizawa2016,yu2017,morjean2017,wakhle2018}.
Employing the particle number projection method, it is possible to extract probability distribution of the reaction products~\cite{simenel2012,simenel2010,sekizawa2014,sekizawa2016}. However, the TDHF approach cannot describe fluctuations in collective variables and severely underestimate fragment mass and charge dispersions. For a reliable description of the collision dynamics, one must improve the approach beyond the standard TDHF description. Employing an extended variational principle, Balian and V\'en\'eroni developed the time-depended random-phase approximation (TDRPA), which significantly improves the mean-field description~\cite{balian1981,balian1984b,balian1985,balian1992}. The TDRPA approach takes into account one-body fluctuations and correlations about the average TDHF trajectory, and has been extensively utilized in nuclear physics
\cite{troudet1985,marston1985,bonche1985,zielinska1988,broomfield2008,broomfield2009,simenel2011,scamps2015a,williams2018,godbey2020b,gao2025}. Recently, using this approach, the mass and charge dispersions of fragments in several systems, including ${}^{60}\mathrm{Ni}+{}^{60}\mathrm{Ni}$~\cite{williams2018} and ${}^{176}\mathrm{Yb}+{}^{176} \mathrm{Yb}$~\cite{godbey2020}, have been calculated and compared with the available data. However, note that the TDRPA method in its current form cannot be applied to asymmetric collisions.

In this work, we present an alternative approach, referred to as the stochastic mean-field theory (SMF), proposed by Ayik in 2008~\cite{ayik2008}, which incorporates mean-field fluctuations and correlations into the description of these reactions. Subsequently, it was shown that the SMF theory includes more than just one-body fluctuations and correlations, through a simplified Bogoliubov-Born-Greenwood-Yvon (BBGKY) hierarchy~\cite{lacroix2016,lacroix2014}. In Section~\ref{sec2a}, we discuss the main concept for the development of the SMF theory. In Section~\ref{sec2b}, we present the derivation of the Langevin description for macroscopic variables by projecting SMF equations on the macroscopic variables such as neutron and proton numbers of projectile-like and target-like fragments and the relative linear momentum-angular momenta of the colliding ions. In Section~\ref{sec3a}, we introduce the Langevin description for macroscopic variables. In Section~\ref{sec3b}, we briefly discuss the numerical aspects of the present calculations in connection with the TDHF formalism. In Section~\ref{sec4}, we present an analysis of the multinucleon transfer mechanism in selected quasi-fission reactions. While in Section~\ref{sec5}, we extend the description to obtain the kinetic energy dissipation and fluctuations in quasi-fission reactions. In Section~\ref{sec6}, we provide our summary and conclusions. We also mention that the SMF theory has also been applied in other contexts such as spinodal instabilities of nuclear matter~\cite{ayik2008a,ayik2009a,ayik2011,yilmaz2011a,yilmaz2013,yilmaz2015,acar2015}, symmetry breaking~\cite{lacroix2012}, Fermionic Hubbard clusters~\cite{lacroix2014b}, as well as nuclear fission~\cite{tanimura2017}. These applications are not discussed in this review. For other mean-field approaches with stochastic extension, see Ref.~\cite{lacroix2014}.

\section{Stochastic Mean-Field Theory}\label{sec2}
\renewcommand{\thefigure}{2.\arabic{figure}} 
\setcounter{figure}{0}

\subsection{Stochastic description of initial correlations}\label{sec2a}
The TDHF approach provides a reduced description of the many-body problem in terms of a single-particle density matrix. In the mean-field approximation, the time-dependent single-particle density matrix is expressed in terms of time-dependent single-particle wave functions. These wave functions are determined by the time-dependent Schr\"odinger equation with a self-consistent mean-field Hamiltonian. In the mean-field approximation, this reduced single-particle approach describes the most probable path of collision dynamics for collision energies on the order of binding energy per nucleon (i.e. $8-10 \mathrm{MeV} /$ nucleon) and
provides a good approximation for the one-body dissipation mechanism. However, the fluctuations of collective dynamics are severely underestimated in the mean-field approximation. Using the well-known projection mechanism of Nakajima and Zwanzig~\cite{nakajima1958,zwanzig1966}, it is possible to carry out a more general reduction of many-body dynamics to the single-particle level. Such a general reduction gives rise to a quantal Langevin description for the reduced single-particle density matrix (see Eq. 2.15 in Ref.~\cite{ayik1980}). These Langevin equations contain, in addition to the self-consistent mean-field Hamiltonian, a binary collision term due to residual interactions and the initial correlation term which acts like a random force. For a detailed description of such a reduction, we refer the reader to Refs.~\cite{ayik1980}. In the semi-classical limit, this equation is referred to as the  Boltzmann-Langevin equation~\cite{ayik1988,ayik1990}. Such a general projection of the many-body density matrix does not lead to a single trajectory, but an ensemble of trajectories for single-particle density matrices. The Langevin equation, in the quantal or semi-classical limit, involves one body dissipation due to the mean-field and collisional dissipation due to the collision term. In a similar manner, fluctuations developed in time by propagation of initial correlation due to residual interactions is referred to as the collisional fluctuations. similarly, initial correlations are also propagated by the self-consistent mean-field, which leads to the mean-field fluctuations. At bombarding energies in the range of the Fermi energy per nucleon, the collisional dissipation and collisional fluctuations make the dominant contributions to the reaction dynamics. On the other hand, for low-energy heavy-ion collisions, at bombarding energies per nucleon in the range of nucleonic binding energy, due to strong Pauli blocking, collisional dissipation and fluctuations are not important. Therefore, at low energies one body dissipation and mean-field fluctuations make the dominant effects in collision dynamics.

In SMF theory~\cite{ayik2008}, the binary collision term and the propagation of initial correlations through residual interactions are omitted. Only propagation of initial correlations by means of self-consistent mean-field are considered. In SMF, initial correlations are incorporated by considering a distribution of Slater determinants. It is well known that initial correlations can be simulated in a stochastic description~\cite{mori1965}. Therefore, it is possible to generate an ensemble of initial single-particle density matrices in terms of a complete set of initial single-particle wave functions and the stochastic elements of the initial density matrix. A member of the initial ensemble, indicated by the label $\lambda$,is expressed as,
\begin{align}
\rho^{\lambda}\left(\boldsymbol{r}, \boldsymbol{r}^{\prime}, t_{0}\right)=\sum_{i, j} \phi_{i}^{*}\left(\boldsymbol{r}, t_{0}\right) \rho_{i j}^{\lambda} \phi_{j}\left(\boldsymbol{r}^{\prime}, t_{0}\right) . \label{eq1}
\end{align}
Here, $\rho_{i j}^{\lambda}$ represents stochastic elements of the initial density matrix, which should be determined according to a suitable distribution function. Initial fluctuations are introduced in the following way: In each generated event $\lambda$, the expectation value of a one-body observable $\hat{Q}$ is calculated as $\langle Q\rangle^{\lambda}=\operatorname{Tr}\left[\rho^{\lambda} \hat{Q}\right]$. In the SMF theory, the original quantum mechanical framework is replaced with a statistical treatment. At the initial state, the expectation values and the variances of one-body observables are evaluated as~\cite{ulgen2019},
\begin{align}
\overline{\langle Q\rangle_{0}^{\lambda}}=\operatorname{Tr}\left[\overline{\rho_{0}^{\lambda}} \hat{Q}\right]=\sum_{i j} \overline{\rho_{i j}^{\lambda}} Q_{j i}^{0} ,\label{eq2} 
\end{align}
and
\begin{align}
\begin{split}
\overline{\left(\langle Q\rangle_{0}^{\lambda}\right)^{2}}-\left(\overline{\langle Q\rangle_{0}^{\lambda}}\right)^{2}
& =\overline{\left(\operatorname{Tr}\left[\delta \rho_{0}^{\lambda} \hat{Q}\right]\right)^{2}} \\
& =\sum_{i j k l} \overline{\delta \rho_{i j}^{\lambda} \delta \rho_{k l}^{\lambda}} Q_{j i}^{0} Q_{l k}^{0} , \label{eq3}
\end{split}
\end{align}
where $\delta \rho_{i j}^{\lambda}=\rho_{i j}^{\lambda}-\overline{\rho_{i j}^{\lambda}}$ is the fluctuating part of the element's initial density matrix, and the quantity $Q_{i j}^{0}=\bra{\phi_{i}\left(t_{0}\right)}\hat{Q}\ket{ \phi_{j}\left(t_{0}\right)}$ represents the initial matrix element of the one-body operator. Here and below, the bar over the quantities indicates the ensemble average over the stochastically generated events. On the other hand, for the natural basis $\bra{\phi_{i}\left(t_{0}\right)}\rho\ket{ \phi_{j}\left(t_{0}\right)}=n_{i} \delta_{i j}$, at the initial time, so that the initial time quantum mechanical expression of the expectation value and the variance of a one-body observable are given by~\cite{ulgen2019}
\begin{align}
\langle Q\rangle_{0}=\sum_{i} n_{i} Q_{i i}^{0} , \label{eq4}
\end{align}
and
\begin{align}
\begin{split}  
\left\langle Q^{2}\right\rangle_{0}-\langle Q\rangle_{0}^{2}=\frac{1}{2} \sum_{i j} & \left[n_{i}\left(1-n_{j}\right)  \right.\\
   & \left. +n_{j}\left(1-n_{i}\right)\right] Q_{j i}^{0} Q_{i j}^{0} . \label{eq5}
\end{split}
\end{align}

The essential point of the SMF theory is that it is designed in such a way that the expectation value and the variance with the statistical treatment coincide with the quantum expressions at the initial time. To satisfy this requirement, initial matrix elements are specified according to uncorrelated Gaussian distributions, each with a mean value and a variance given by~\cite{ayik2008},
\begin{align}
    \overline{\rho_{i j}^{\lambda}}=n_{i} \delta_{i j} , \label{eq6}
\end{align}
and
\begin{align}
\begin{split}
\overline{\delta \rho_{i j}^{\lambda} \delta \rho_{k l}^{\lambda}}=\frac{1}{2}\left[n_{i}\left(1-n_{j}\right)+n_{j}\left(1-n_{i}\right)\right] \delta_{k j} \delta_{l i}  .  \label{eq7}
\end{split}    
\end{align}

Since the fluctuating components of the density matrix have a zero mean by construction, the ensemble average of those events reproduces the ordinary mean-field (TDHF) results. We note that when the initial state has zero temperature, such as the ground state of the projectile and target nuclei before collision, the average occupation numbers $n_{i}$ are zero or one. If an observable is diagonal at the initial state, $Q_{i j}^{0}=Q_{i}^{0} \delta_{i j}$, such as the particle number operators for projectile and target, as seen from Eq.~\eqref{eq5}, the variance of such observables are strictly zero, and therefore they do not exhibit fluctuations at the initial state. If the initial state has finite temperature, in the case of induced fission of a compound nucleus for instance, the average values of the occupation numbers are determined by the Fermi-Dirac distribution.

As mentioned in the Introduction, it is worth nothing here that although the SMF theory was originally proposed to take into account one-body (mean-field) fluctuations at the initial time, it has been shown that it grasps part of many-body correlations through the BBGKY hierarchy~\cite{lacroix2016}. In addition, in the original formulation of the SMF theory~\cite{ayik2008}, the stochastic matrix elements $\delta \rho_{i j}^{\lambda}$ are assumed to be uncorrelated Gaussian numbers with zero mean. It has been shown that the description can be further improved by relaxing the Gaussian assumption. In this review, however, we adopt the Gaussian assumption for the stochastic matrix elements, which allows us to formulate a quantal diffusion description for multinucleon exchanges, as described in Sec.~\ref{sec3}.

Note that in the SMF theory the stochasticity is introduced only at the initial time, the evolution of the mean-field itself is not a stochastic process. In each event, the initial fluctuations are propagated by the meal-field, and the set of time-dependent single-particle wave functions are determined according to
\begin{align}
i \hbar \frac{\partial}{\partial t} \Phi_{j}(\boldsymbol{r}, t ; \lambda)=h\left[\rho^{\lambda}\right] \Phi_{j}(\boldsymbol{r}, t ; \lambda) , \label{eq8}
\end{align}
where, $h\left(\rho^{\lambda}\right)$ denotes the self-consistent mean-field Hamiltonian in event $\lambda$. As a result of the propagation of the initial correlations, even the occupied single-particle wave functions are modified in a nontrivial manner as compared to the standard TDHF wave functions. In a manner like TDHF, we can express the equation of motion of the SMF in terms of density matrix as,
\begin{align}
i \hbar \frac{\partial}{\partial t} \rho^{\lambda}(t)=\left[h\left(\rho^{\lambda}\right), \rho^{\lambda}(t)\right] , \label{eq9}
\end{align}
with density matrix in an event $\lambda$ is given by
\begin{align}
\rho^{\lambda}\left(\boldsymbol{r}, \boldsymbol{r}^{\prime}, t\right)=\sum_{i, j} \phi_{i}^{*}(\boldsymbol{r}, t ; \lambda) \rho_{i j}^{\lambda} \phi_{j}\left(\boldsymbol{r}^{\prime}, t ; \lambda\right) . \label{eq10}
\end{align}
The density matrix fluctuates from event to event due to fluctuations of the element of the initial density matrix and fluctuations of the single-particle wave functions.

\subsection{Fluctuations of one-body observables}\label{sec2b}
By numerical computation of Eq.~\eqref{eq8}, we can generate an ensemble of single-particle density matrices $\left\{\rho^{\lambda}\left(\boldsymbol{r}, \boldsymbol{r}^{\prime}, t\right)\right\}$. Therefor, in the SMF theory, it is possible to calculate not only the mean value, but the probability distribution of an observable represented by a one-body operator $\hat{Q}$. The expectation value of a one-body observable $\hat{Q}$ in an event is determined as,
\begin{align}
Q^{\lambda}(t)=\sum_{i j}\bra{\Phi_{i}(t ; \lambda)}\hat{Q} \ket{\Phi_{j}(t ; \lambda)}\rho_{i j}^{\lambda} . \label{eq11}
\end{align}
We note that influence of the initial fluctuations appears in two different ways: (i) from stochastic elements of the initial density matrix and (ii) fluctuating matrix elements of the observable. Even if the magnitude of initial fluctuations is small, the stochastic mean-field evolution can enhance the fluctuations, and hence events can substantially deviate from one another. The mean value of the observable is determined by the average over the ensemble generated by numerical computation, $Q(t)=\overline{Q^{\lambda}}(t)$. In a similar manner, the variance of the observable is calculated by,
\begin{align}
\sigma_{Q}^{2}(t)=\overline{\left[Q^{\lambda}(t)-Q(t)\right]^{2}} .\label{eq12}
\end{align}

Here we consider a special case of small amplitude fluctuations around the mean value, $\rho^{\lambda}(t)=\rho(t)+\delta \rho^{\lambda}(t)$, where $\delta \rho^{\lambda}(t)$ represents the small amplitude fluctuations. In this case, in Eq.~\eqref{eq11}, the correlation between the matrix elements of the observable and the initial density matrix can be neglected and the mean value of the observable is calculated in the usual manner, and given by the standard TDHF result,
\begin{align}
Q(t)=\sum_{j}\bra{\Phi_{j}(t)} \hat{Q} \ket{\Phi_{j}(t)} n_{j} . \label{eq13}
\end{align}
Here, the single-particle wave functions are determined according to the standard TDHF equations with the standard mean-field Hamiltonian. To calculate the variance of a one-body observable, we need to determine the fluctuations of the density matrix $\delta \rho^{\lambda}(t)$. Small amplitude fluctuations of the density matrix are determined by the time-dependent RPA (TDRPA) equation, which is obtained by linearizing the SMF Eq.~\eqref{eq9} around the average evolution,
\begin{align}
i \hbar \frac{\partial}{\partial t} \delta \rho^{\lambda}(t)=\left[\delta h^{\lambda}(t), \rho(t)\right]+\left[h(\rho), \delta \rho^{\lambda}(t)\right] . \label{eq14}
\end{align}
Here, $\delta h^{\lambda}(t)=(\partial h / \partial \rho) \cdot \delta \rho^{\lambda}(t)$ denotes the fluctuating part of the mean-field Hamiltonian. Employing a complete set of stationary single-particle states, we can express the matrix elements of TDRPA equations as,
\begin{align}
i \hbar \frac{\partial}{\partial t} \delta \rho_{i j}^{\lambda}(t)=\sum_{k l} R_{i j, k l}(t) \delta \rho_{k l}^{\lambda}(t) . \label{eq15}
\end{align}
Here, $\delta \rho_{i j}^{\lambda}(t)=\bra{\Phi_{i}}\delta \rho^{\lambda}(t)\ket{\Phi_{j}}$ denotes the elements of density matrix, and the matrix and the elements of $R(t)$ are given by,
\begin{align}
\begin{split}
R_{i j, k l}(t)= & \bra{\Phi_{i}}h(\rho) \ket{\Phi_{l}} \delta_{k j}- \bra{\Phi_{k}}h(\rho) \ket{\Phi_{j}}\delta_{i l} \\
& + \sum_{n}\left[ \bra{\Phi_{i} \Phi_{k}} \partial h / \partial \rho \ket{\Phi_{n} \Phi_{l}} \rho_{n j}(t)\right. \\
& \left. - \bra{\Phi_{n} \Phi_{k}}\partial h / \partial \rho \ket{\Phi_{j} \Phi_{l}} \rho_{i n}(t)\right] . \label{eq16}
\end{split}
\end{align}
Using super-space notation, Eq.~\eqref{eq15} can be written in a compact form as,
\begin{align}
i \hbar \frac{\partial}{\partial t}\bra{i j }\ket{\delta \rho^{\lambda}(t)}=\bra{i j}R(t) \ket{\delta \rho^{\lambda}(t)} , \label{eq17}
\end{align}
or
\begin{align}
i \hbar \frac{\partial}{\partial t}\ket{\delta \rho^{\lambda}(t)}=R(t) \ket{\delta \rho^{\lambda}(t)} , \label{eq18}
\end{align}
where $\ket{\delta \rho^{\lambda}(t)}$ acts like a vector with double indices and $R(t)$ is a matrix in the super-space. Because of its linear form, formal solution of this equation is given by,
\begin{align}
\ket{\delta \rho^{\lambda}(t)}=\exp \left[-\frac{i}{\hbar} \int_{t_{0}}^{t} R(s) d s\right] \ket{\delta \rho^{\lambda}\left(t_{0}\right)} . \label{eq19}
\end{align}
The fluctuating part of a one-body observable is calculated according to
\begin{align}
\delta Q^{\lambda}\left(t_{1}\right)=\bra{Q}\exp \left[-\frac{i}{\hbar} \int_{t_{0}}^{t_{1}} R(s) d s\right]\ket{\delta \rho^{\lambda}\left(t_{0}\right)} , \label{eq20}
\end{align}
where $t_{1}$ represents the final time at which the observation is made and $t_{0}$ is the initial time. It may be more convenient to introduce a time-dependent one-body operator $B(t)$ according to
\begin{align}
\bra{B(t)}= \bra{Q\exp \left[-\frac{i}{\hbar} \int_{t}^{t_{1}} R(s) d s\right]}  .\label{eq21}
\end{align}
It is easy to show that time evolution of the Heisenberg operator $B(t)$ is determined by the dual of the TDRPA equations according to
\begin{align}
i \hbar \frac{\partial}{\partial t} B(t)=[h(\rho), B(t)]+\operatorname{Tr}\left(\frac{\partial h}{\partial \rho}\right) \cdot[B(t), \rho] .\label{eq22}
\end{align}
The solution is determined by a backward evolution of this equation with the boundary condition $B\left(t_{1}\right)=Q$. In an event $\lambda$, the expectation value of the observable can be expressed as,
\begin{align}
\begin{split}  
\delta Q^{\lambda}\left(t_{1}\right)& = \bra{B\left(t_{0}\right) } \ket{\delta \rho^{\lambda}\left(t_{0}\right)} \\
& =\sum_{i j}\bra{\Phi_{i}\left(t_{0}\right)}B\left(t_{0}\right) \ket{\Phi_{j}\left(t_{0}\right)}\delta \rho_{j i}^{\lambda}\left(t_{0}\right) .\label{eq23}
\end{split}
\end{align}
Then the variance of the observable becomes,
\begin{align}
\begin{split}
\sigma_{Q}^{2}\left(t_{1}\right) & = \sum\left[\bra{\Phi_{i}\left(t_{0}\right)}B\left(t_{0}\right)\ket{\Phi_{j}\left(t_{0}\right)} \right. \\
& \times \left. \bra{\Phi_{j^{\prime}}\left(t_{0}\right)} B\left(t_{0}\right) \ket{\Phi_{i^{\prime}}\left(t_{0}\right)}\overline{\delta \rho_{j i}^{\lambda}\left(t_{0}\right) \delta \rho_{i^{\prime} j^{\prime}}^{\lambda}\left(t_{0}\right)}\right] \\
& = \sum\left|\bra{\Phi_{i}\left(t_{0}\right)} B\left(t_{0}\right)\ket{\Phi_{j}\left(t_{0}\right)}\right|^{2} n_{i}\left(1-n_{j}\right),\label{eq24}    
\end{split}
\end{align}
where the last equality is obtained using  Eq.~\eqref{eq7}. This result is identical with formula derived in a previous work employing a variational approach by R. Balian and M. V\'en\'eroni~\cite{balian1981}.

\section{Quantum Langevin Equations for Macroscopic Variables}\label{sec3}
\renewcommand{\thefigure}{3.\arabic{figure}} 
\setcounter{figure}{0}
\subsection{Langevin equation for neutron and proton transfer}\label{sec3a}

\begin{figure}[!htb]
\includegraphics*[width=0.48\textwidth]{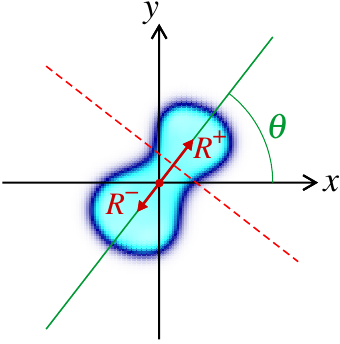}
\caption{Geometry of di-nuclear complex for a non-central collision with beam direction along the x-axis. Symmetry axis and window plane indicated by green solid line, and window plane by red dashed line in reaction plane.}
\label{fig:3.1}
\end{figure}

During the near-barrier energy collisions of heavy-ions, such as deep-inelastic collisions and quasifission reactions, projectile and target form a di-nuclear structure and interact mainly via multinucleon transfer. When a di--nuclear structure is maintained during the collision dynamics, we do not need to generate an ensemble of stochastic mean-field events. In this case, we can describe the collision dynamics in terms of much easier transport description for the relevant macroscopic variables, such as the charge and mass asymmetry, and relative linear and angular momentum of colliding ions. It is possible to deduce the quantal Langevin equations for macroscopic variables by geometric projection of the SMF equations with the help of the window dynamics~\cite{ayik2017,ayik2018,yilmaz2018,ayik2019,sekizawa2020,ayik2020b,yilmaz2020}. For describing nucleon diffusion mechanism, we consider neutron number and proton number of the projectile-like or target-like fragments as relevant macroscopic variables. Here, we take the neutron $N_{1}^{\lambda}$ and proton $Z_{1}^{\lambda}$ numbers of the target-like fragments as the macroscopic event variables. In each event $\lambda$, the neutron and proton numbers are determined by integrating the nucleon density over the projectile side of the window between the colliding nuclei,
\begin{align}
\binom{N_{1}^{\lambda}(t)}{Z_{1}^{\lambda}(t)}=\int d^{3} r \Theta\left[x^{\prime}(t)\right]\binom{\rho_{n}^{\lambda}(\boldsymbol{r}, t)}{\rho_{p}^{\lambda}(\boldsymbol{r}, t)}, \label{eq25}
\end{align}
where $x^{\prime}(t)=\left[y-y_{0}(t)\right] \sin \theta+\left[x-x_{0}(t)\right] \cos \theta$. The $(x, y)$-plane represents the reaction plane with $x$-axis being the beam direction in the center of mass frame (COM) of the colliding ions. The window plane is perpendicular to the symmetry axis and its orientation is specified by the condition $x^{\prime}(t)=\left[y-y_{0}(t)\right] \sin \theta+\left[x-x_{0}(t)\right] \cos \theta$. In this expression, $x_{0}(t)$ and $y_{0}(t)$ denote the coordinates of the window center relative to the origin of the COM frame, $\boldsymbol{\theta}(\boldsymbol{t})$ is the smaller angle between the orientation of the symmetry axis and the beam direction. We neglect fluctuations in the orientation of the window and determine the mean evolution of the window dynamics by diagonalizing the mass quadrupole moment tensor of the system for each impact parameter $b$ or the initial orbital angular momentum $\ell$, and time-step, as described in Appendix~\ref{sec:appA}, [also see Appendix A of~\cite{ayik2018}]. In terms of TDHF description, it is possible to determine the time evolution of the rotation angle $\boldsymbol{\theta}(t)$ of the symmetry axis. The coordinates $x_{0}(t)$ and $y_{0}(t)$ of the center point of the window are located at the center of the minimum density slice on the neck between the colliding ions. For deformed nuclei, the outcome of the collisions depends on the relative orientation of the projectile and target. Usually, we consider following collision geometries: (i) side-side geometry in which the deformation axes of both the projectile and the target are perpendicular to the beam direction, (ii) tip-tip geometry in which the deformation axes of both the projectile and the target are parallel to the beam direction, (iii) tip-side geometry in which the deformation axis of the target is perpendicular while the projectile axis is parallel to the beam direction, and (iv) side-tip geometry in which directions of target and projectile are interchanged. Fig.~\ref{fig:3.1} shows the density profile of a colliding system in the reaction plane. The window plane and the symmetry axis of the di-nuclear complex are indicated by thick and dash lines.

In the following, all quantities are calculated for a given initial orbital angular momentum $\ell$, but for the purpose of clarity of expressions, we do not attach the angular momentum label to these quantities. The quantity in Eq.~\eqref{eq25},
\begin{align}
\rho_{\alpha}^{\lambda}(\boldsymbol{r}, t)=\sum_{i j \in \alpha} \Phi_{j}^{*}(\boldsymbol{r}, t ; \lambda) \rho_{j i}^{\lambda} \Phi_{i}^{\alpha}(\boldsymbol{r}, t ; \lambda) , \label{eq26}
\end{align}
denotes neutron and proton densities in the event $\lambda$ of the ensemble. Here, and in the rest of the article, we use the notation $\alpha=n, p$ as the neutron and proton labels.

Here, we briefly discuss the derivation of the Langevin equation for neutron and proton numbers of the projectile-like fragments, and for details we refer to Refs.~\cite{ayik2018,yilmaz2018}. The rate of change of the neutron and proton numbers for the target-like fragments are given by,
\begin{align}
\frac{d}{d t}\binom{N_{1}^{\lambda}(t)}{Z_{1}^{\lambda}(t)}=-\int d^{3} r \Theta\left[x^{\prime}(t)\right]\binom{\boldsymbol{\nabla} \cdot \boldsymbol{j}_{n}^{\lambda}(t)}{\boldsymbol{\nabla} \cdot \boldsymbol{j}_{p}^{\lambda}(t)} . \label{eq27}
\end{align}
In obtaining this expression we neglect the term arising from the rate of change of the position and the rotation of the window plane, and employ the continuity equation with the fluctuating current densities of the neutrons and protons given by,
\begin{align}
\boldsymbol{j}_{\alpha}^{\lambda}(\boldsymbol{r}, t)=\frac{\hbar}{m} \sum_{i j \in \alpha} \operatorname{Im}\left(\Phi_{j}^{* \alpha}(\boldsymbol{r}, t ; \lambda) \boldsymbol{\nabla} \Phi_{i}^{\alpha}(\boldsymbol{r}, t ; \lambda) \rho_{j i}^{\lambda}\right) . \label{eq28}
\end{align}
By carrying out a partial integration, one obtains a coupled set of Langevin equations for the macroscopic variables $N_{1}^{\lambda}(t)$ and $Z_{1}^{\lambda}(t)$,
\begin{align}
\begin{split}
\frac{d}{d t}\binom{N_{1}^{\lambda}(t)}{Z_{1}^{\lambda}(t)} &=\int d^{3} r g\left(x^{\prime}\right)\binom{\boldsymbol{\hat{e}} \cdot \boldsymbol{j}_{n}^{\lambda}(\boldsymbol{r}, t)}{\boldsymbol{\hat{e}} \cdot \boldsymbol{j}_{p}^{\lambda}(\boldsymbol{r}, t)} \\
& =\binom{v_{n}^{\lambda}(t)}{v_{p}^{\lambda}(t)} , \label{eq29}
\end{split}
\end{align}
where $\boldsymbol{\hat{e}}$ is the unit vector along the symmetry axis with components $\boldsymbol{\hat{e}}_{x}=\cos \theta$ and $\boldsymbol{\hat{e}}_{y}=\sin \theta$. In the integrand, we replace the delta function by a smoothing function $\delta\left(x^{\prime}\right) \rightarrow g\left(x^{\prime}\right)$ in terms of a Gaussian $g(x)=(1 / \kappa \sqrt{2 \pi}) \exp \left(-x^{2} / 2 \kappa^{2}\right)$ with dispersion $\boldsymbol{\kappa}$. The Gaussian behaves almost like a delta function for sufficiently small $\boldsymbol{\kappa}$ values. In the numerical calculations dispersion of the Gaussian is taken in the order of the lattice spacing $\boldsymbol{\kappa}=1.0 \mathrm{fm}$. The right side of Eq.~\eqref{eq29} defines the fluctuating drift coefficients $v_{\alpha}^{\lambda}(t)$ for neutrons and protons. As seen from the current density, there are two different sources for fluctuations of the drift coefficients: (i) Fluctuations due to different set of wave functions in each event $\lambda$. This part of the fluctuations can be approximately described in terms of the fluctuating macroscopic variables as $v_{\alpha}^{\lambda}(t) \rightarrow v_{\alpha}\left(N_{1}^{\lambda}(t), \mathrm{Z}_{1}^{\lambda}(t)\right)$, and (ii) fluctuations introduced by the stochastic part $\delta \rho_{j i}^{\lambda}=\rho_{j i}^{\lambda}-\delta_{j i} n_{j}$ of the density matrix at the initial state. In this work, we consider small amplitude fluctuations and linearize the Langevin Eq.~\eqref{eq29} around the mean values of the macroscopic variables $\delta N_{1}^{\lambda}=N_{1}^{\lambda}-\bar{N}_{1}$ and $\delta Z_{1}^{\lambda}=\mathrm{Z}_{1}^{\lambda}-\bar{Z}_{1}$. The mean values $\bar{N}_{1}=\overline{N_{1}^{\lambda}}$ and $\bar{Z}_{1}=\overline{Z_{1}^{\lambda}}$ are determined by the mean-field description of the TDHF approach. The fluctuations evolve according to the linearized coupled Langevin equations,
\begin{align}
\begin{split}
\frac{d}{d t}
\begin{pmatrix}
    \delta Z_{1}^{\lambda} \\ \delta N_{1}^{\lambda}
\end{pmatrix}
& = \begin{pmatrix}
    \frac{\partial v_{p}}{\partial Z_{1}}\left(Z_{1}^{\lambda}-\overline{Z}_{1}\right)+\frac{\partial v_{p}}{\partial N_{1}}\left(N_{1}^{\lambda}-\overline{N}_{1}\right) \\
    \frac{\partial v_{n}}{\partial Z_{1}}\left(Z_{1}^{\lambda}-\overline{Z}_{1}\right)+\frac{\partial v_{n}}{\partial N_{1}}\left(N_{1}^{\lambda}-\overline{N}_{1}\right)
\end{pmatrix} \\
 &\quad \quad +  \begin{pmatrix}
\delta v_{p}^{\lambda}(t) \\
\delta v_{p}^{\lambda}(t)
\end{pmatrix} ,
\end{split}
\label{eq30}
\end{align}
where the derivatives of drift coefficients are evaluated at the mean values $\bar{N}_{1}$ and $\bar{Z}_{1}$. The linear limit provides a good approximation for small amplitude fluctuations, and it becomes even better if the driving potential energy has nearly harmonic behavior around the mean values. There are two different contributions to the stochastic part of the drift coefficients: (i) density fluctuations in the vicinity of rotating window plane which involves collective velocity of the window and (ii) current density fluctuations across the rotating window. Since the nucleon flow velocity through the window is much larger than the collective velocity of the window, the current density fluctuations make the dominant contributions. Therefore, in the analysis, we retain only the current density fluctuations in the stochastic part of the drift coefficients,
\begin{align}
\begin{split}
\delta \nu_{\alpha}^{\lambda}(t)& =\frac{\hbar}{m} \sum_{i j \in \alpha} \int d^{3} r g\left(x^{\prime}\right) \\
& \quad \times \operatorname{Im}\left(\Phi_{j}^{* \alpha}(\boldsymbol{r}, t) \boldsymbol{\hat{e}} \cdot \boldsymbol{\nabla} \Phi_{i}^{\alpha}(\boldsymbol{r}, t) \delta \rho_{j i}^{\lambda}\right) . \label{eq31}
\end{split}
\end{align}

According to the basic postulate of the SMF approach the stochastic elements of the initial density matrix $\delta \rho_{j i}^{\lambda}$ are specified in terms of uncorrelated Gaussian distributions, it then follows that the stochastic part of the neutron and proton drift coefficients $\delta v_{\alpha}^{\lambda}(t)$ are determined by uncorrelated Gaussian distributions with variances to be discussed in the following section. There are two competing effects in the Langevin Eqs.~\eqref{eq30}. The first terms drive the system toward local equilibrium, and it is determined by the gradient of the potential energy surface in ( $\mathrm{N}-\mathrm{Z}$ ) plane. The stochastic part of the current density provides the source for the growing fluctuations of charge and mass-asymmetry of the di-nuclear system.

In principle, one can determine the distribution function for the neutron and proton numbers of the primary fragments by numerically generating an ensemble of solutions of the coupled Langevin equations~\eqref{eq30}. It is well known that the Langevin equations for macroscopic variables is equivalent to the Fokker Planck equation for the distribution function of this variable~\cite{risken1996}. In certain cases, it is much easier to determine the distribution function from the solution of the Fokker-Plank equation. In the coupled Langevin equations~\eqref{eq30}, drift coefficients are linearly dependent on the macroscopic variables and diffusion coefficients, as we will see in the next section, they are independent of macroscopic variables. In this case, the solution of coupled Langevin equations for a given initial angular momentum is given by a nearly analytical form of a correlated Gaussian distribution. In most applications, the linear form of the coupled Langevin equation appears to provide good approximation compared to the data. However, it is possible to improve the description by including non-linear terms to the coupled Langevin equations, which could provide a better description for the large-amplitude fluctuations. If the system under investigation is in the vicinity of a saddle point, potential energy could have a negative curvature. In this case, both the driving force and the random force tend to increase the variance of charge and mass distributions. If the contact time is sufficiently short, the linear approximation may also provide a useful approach in this case.

\subsection{Quantal diffusion coefficients of neutron and proton}\label{sec3b}

Stochastic part of the drift coefficients, $\delta v_{p}^{\lambda}(t)$ and $\delta v_{n}^{\lambda}(t)$, are specified by uncorrelated Gaussian distributions. Stochastic drift coefficients have zero mean values $\overline{\delta v_{p}^{\lambda}}(t)=0$, $\overline{\delta v_{n}^{\lambda}}(t)=0$ and the associated correlation functions~\cite{gardiner1991,weiss1999}
\begin{align}
\int_{0}^{t} d t^{\prime} \overline{\delta v_{\alpha}^{\lambda}(t) \delta v_{\alpha}^{\lambda}\left(t^{\prime}\right)}=D_{\alpha \alpha}(t), \label{eq32}
\end{align}
determine the diffusion coefficients, $D_{\alpha \alpha}(t)$, for proton and neutron transfers. In the stochastic part of the drift coefficients $\delta v_{\alpha}^{\lambda}(t)$ in Eq.~\eqref{eq31}, we impose a physical constraint on the summations of single-particle sates. Self-transitions among the single-particle states of the projectile or target nuclei among themselves do not contribute to the nucleon exchange mechanism. Therefore, in Eq.~\eqref{eq31}, we restrict summations as follows: when the summation $i \in T$ runs over the states originating from the target nucleus, the summation $j \in P$ runs over the states originating from the projectile, and vice versa. Using the main postulate of the SMF approach given by Eq.~\eqref{eq31}, we can calculate the correlation functions of the stochastic part of the drift coefficients. At zero temperature, since the average occupation factors are zero or one, we find the correlation functions are expressed as,
\begin{align}
\begin{split}
\overline{\delta v_{\alpha}^{\lambda}(t) \delta v_{\alpha}^{\lambda}\left(t^{\prime}\right)}  = &  \operatorname{Re}\left( \sum_{p \in P,  h \in T} A_{p h}^{\alpha}(t) A_{p h}^{* \alpha}\left(t^{\prime}\right) \right. \\
  & \qquad \left. + \sum_{p \in T,  h \in P} A_{p h}^{\alpha}(t) A_{p h}^{* \alpha}\left(t^{\prime}\right) \right) . \label{eq33}
\end{split}
\end{align}
Here, the individual matrix elements are given by
\begin{align}
\begin{split}
A_{p h}^{\alpha}(t)=\frac{\hbar}{2 m} \int d^{3} r g\left(x^{\prime}\right) & \left(\Phi_{p}^{* \alpha}(\boldsymbol{r}, t) \boldsymbol{\hat{e}} \cdot \nabla \Phi_{h}^{\alpha}(\boldsymbol{r}, t)\right. \\
& - \left. \Phi_{h}^{\alpha}(\boldsymbol{r}, t) \boldsymbol{\hat{e}} \cdot \nabla \Phi_{p}^{* \alpha}(\boldsymbol{r}, t)\right) . \label{eq34}  
\end{split}
\end{align}
We note that employing an integration by parts, we can put this expression in to the following form,
\begin{align}
\begin{split}
A_{p h}^{\alpha}(t)=\frac{\hbar}{m} \int & d^{3} r  g\left(x^{\prime}\right) \Phi_{p}^{* \alpha}(\boldsymbol{r}, t) \\
 & \times \left(\boldsymbol{\hat{e}} \cdot \nabla \Phi_{h}^{\alpha}(\boldsymbol{r}, t)-\frac{x^{\prime}}{2 \kappa^{2}} \Phi_{h}^{\alpha}(\boldsymbol{r}, t)\right) .\label{eq35}
\end{split}
\end{align}
To evaluate the correlation functions in Eq.~\eqref{eq33}, of the stochastic drift coefficient, we introduce the following approximate treatment. In the first term of the right-hand side of Eq.~\eqref{eq33}, we add and subtract the hole contributions to give,
\begin{align}
\begin{split}
 \sum_{p \in P,  h \in T} A_{p h}^{\alpha}(t) A_{p h}^{* \alpha}\left(t^{\prime}\right)= & \sum_{a \in P,  h \in T} A_{a h}^{\alpha}(t) A_{a h}^{* \alpha}\left(t^{\prime}\right) \\ 
 & - \sum_{h^{\prime} \in P, h \in T} A_{h^{\prime} h}^{\alpha}(t) A_{h^{\prime} h}^{* \alpha}\left(t^{\prime}\right) . \label{eq36} 
\end{split}
\end{align}
Here, the summation $a$ is over the complete set (particle and hole) of states originating from the projectile. In the first term, we cannot use the closure relation to eliminate the complete set of single-particle states because the wave functions are evaluated at different times. However, we note that the time-dependent single-particle wave functions during short time intervals exhibit nearly diabatic behavior~\cite{norenberg1981}. In other words, during short time intervals the nodal structure of time-dependent wave functions does not change appreciably. Most dramatic diabatic behavior of the time-dependent wave-functions is apparent in the fission dynamics. The Hartree-Fock solutions force the system to follow the diabatic path, which prevents the system from breaking into fragments. As a result of these observations, we introduce, for short time $\tau=t-t^{\prime}$ evolutions on the order of the correlation time, a diabatic approximation into the time-dependent wave-functions by shifting the time backward (or forward) according to,
\begin{align}
\Phi_{a}\left(\boldsymbol{r}, t^{\prime}\right) \approx \Phi_{a}(\boldsymbol{r}-\boldsymbol{u} \tau, t) , \label{eq37}
\end{align}
where $\boldsymbol{u}$ denotes a suitable flow velocity of nucleons through the window. Now, we can employ the closure relation to obtain,
\begin{align}
\sum_{a} \Phi_{a}^{*}\left(\boldsymbol{r}_{1}, t\right) \Phi_{a}\left(\boldsymbol{r}_{2}-\boldsymbol{u} \tau, t\right)=\delta\left(\boldsymbol{r}_{1}-\boldsymbol{r}_{2}+\boldsymbol{u} \tau\right), \label{eq38}
\end{align}
where, the summation $a$ runs over the complete set of states originating from target or projectile, and the closure relation is valid for each set of the spin-isospin degrees of freedom. The flow velocity $\boldsymbol{u}(\boldsymbol{r}, T)$ may depend on the mean position $\boldsymbol{r}=\left(\boldsymbol{r}_{1}+\boldsymbol{r}_{2}\right) / 2$ and the mean-time $T=\left(t+t^{\prime}\right) / 2$. Employing this closure relation in the first term on the right-hand side of Eq.~\eqref{eq36}, we find
\begin{align}
\begin{split}
\sum_{a, h} & A_{a h}^{\alpha}(t) A_{a h}^{* \alpha}\left(t^{\prime}\right)=\sum_{h} \int d^{3} r_{1} d^{3} r_{2} \\
&\times \delta\left(\boldsymbol{r}_{1}-\boldsymbol{r}_{2}+\boldsymbol{u}_{h} \tau\right) W_{h}^{\alpha}\left(\boldsymbol{r}_{1}, t\right) W_{h}^{* \alpha} \left(\boldsymbol{r}_{2}, t^{\prime}\right) . \label{eq39}
\end{split}
\end{align}
The closure relation in Eq.~\eqref{eq38} is valid for any choice of the flow velocity. The most suitable choice is the flow velocity of the hole state $\boldsymbol{u}_{h}(\boldsymbol{r}, T)$ in each term in the summation, which is used in this expression. In this manner the complete set of single-particle states is eliminated, and the calculations of the quantal diffusion coefficients are greatly simplified. In fact, to calculate this expression, we only need the hole states originating from target and projectile, which are provided by the TDHF description. The local flow velocity of each wave-function is specified by the usual expression of the current density divided by the particle density as given in Eq.~\eqref{eqb8} in Appendix~\ref{sec:apPb}. The quantities $W_{h}^{\alpha}\left(\boldsymbol{r}_{1}, t\right)$ are,
\begin{align}
\begin{split}
W_{h}^{\alpha}\left(\boldsymbol{r}_{1}, t\right)=\frac{\hbar}{m} g\left(x_{1}^{\prime}\right)& \left(\boldsymbol{\hat{e}} \cdot \nabla_{1} \Phi_{h}^{\alpha}\left(\boldsymbol{r}_{1}, t\right) \right. \\
&- \left. \frac{x_{1}^{\prime}}{2 \kappa^{2}} \Phi_{h}^{\alpha}\left(\boldsymbol{r}_{1}, t\right)\right) , \label{eq40}  
\end{split}
\end{align}
and $W_{h}^{* \alpha}\left(\boldsymbol{r}_{2}, t^{\prime}\right)$ is given by a similar expression. A detailed analysis of Eq.~\eqref{eq39} is presented in Appendix~\ref{sec:apPb}. The result of this analysis as given by Eq.~\eqref{eqb19} is,
\begin{align}
\begin{split}
\sum_{a \in P, h \in T} A_{a h}^{\alpha}(t) A_{a h}^{* \alpha}\left(t^{\prime}\right)&= \int d^{3} r \tilde{g}\left(x^{\prime}\right) \\
& \times G_{T}(\tau) J_{\perp, \alpha}^{T}(\boldsymbol{r}, t-\tau / 2). \label{eq41}
\end{split}
\end{align}
Here, $J_{\perp, \alpha}^{T}(\boldsymbol{r}, t-\tau / 2)$ represents the sum of the magnitude of current densities perpendicular to the window due to each wave functions originating from the target,
\begin{align}
\begin{split}
 J_{\perp, \alpha}^{T}(\boldsymbol{r}, t-\tau / 2)= \frac{\hbar}{m} \sum_{h \in T}& \left|\operatorname{Im} \Phi_{h}^{* \alpha}(\boldsymbol{r}, t-\tau / 2) \right. \\
& \left. \times \left(\boldsymbol{\hat{e}} \cdot \nabla \Phi_{h}^{\alpha}(\boldsymbol{r}, t-\tau / 2)\right)\right| . \label{eq42}   
\end{split}
\end{align}
The quantity $G_{T}(\tau)$ is given by Eq.~\eqref{eqb20}, and it is the average value of the memory kernels $G_{T}^{h}(\tau)$ of Eq.~\eqref{eqb14}. It is possible to carry out a similar analysis in the second term in the right side of Eq.~\eqref{eq33} to give,
\begin{align}
\begin{split}
\sum_{a \in T, h \in P} A_{a h}^{\alpha}(t) & A_{a h}^{* \alpha}\left(t^{\prime}\right)=\int  d^{3} r \tilde{g}\left(x^{\prime}\right) \\
& \times G_{P}(\tau) J_{\perp, \alpha}^{P}(\boldsymbol{r}, t-\tau / 2) . \label{eq43}
\end{split}
\end{align}
In a similar manner, $J_{\perp, \alpha}^{P}(\boldsymbol{r}, t-\tau / 2)$ is determined by the sum of the magnitude of the current densities due wave functions originating from projectile. As a result, the quantal expressions of the proton and neutron diffusion coefficients takes the form,
\begin{align}
\begin{split}
D_{\alpha \alpha}(t) & =  \int_{0}^{t} d \tau \int d^{3} r \tilde{g}\left(x^{\prime}\right)  \left(G_{T}(\tau) J_{\perp, \alpha}^{T}(\boldsymbol{r}, t-\tau / 2) \right. \\ 
& \qquad \qquad \left. + G_{P}(\tau) J_{\perp, \alpha}^{P}(\boldsymbol{r}, t-\tau / 2)\right) \\ 
& - \int_{0}^{t} d \tau  \operatorname{Re}\left(\sum_{h^{\prime} \in P, h \in T} A_{h^{\prime} h}^{\alpha}(t) A_{h^{\prime} h}^{* \alpha}(t-\tau) \right. \\
& \left. \qquad \qquad \qquad + \sum_{h^{\prime} \in T, h \in P} A_{h^{\prime} h}^{\alpha}(t) A_{h^{\prime} h}^{* \alpha}(t-\tau)\right) . \label{eq44}    
\end{split}
\end{align}

In general, we observe that there is a close analogy between this quantal expression and the classical diffusion coefficient in the random walk problem~\cite{gardiner1991,weiss1999,randrup1979}. The first line in the quantal expression gives the sum of the nucleon currents across the window from the target-like fragment to the projectile-like fragment and from the projectile-like fragment to the target-like fragment, which is integrated over the memory. This is analogous to the random walk problem, in which the diffusion coefficient is given by the sum of the rate for the forward and backward steps. The second line in the quantal diffusion expression represents the Pauli blocking effects in the nucleon transfer mechanism, which do not have a classical counterpart. It is important to note that the quantal diffusion coefficients are entirely determined in terms of the occupied single-particle wave functions obtained from the TDHF solutions. The quantities in the Pauli blocking factors are determined by
\begin{align}
\begin{split}
A_{h^{\prime} h}^{\alpha}(t)=\frac{\hbar}{2 m} \int d^{3} r & g\left(x^{\prime}\right) \left[\Phi_{h^{\prime}}^{* \alpha}(\boldsymbol{r} t) \boldsymbol{\hat{e}} \cdot \boldsymbol{\nabla} \Phi_{h}^{\alpha}(\boldsymbol{r}, t)\right.  \\ 
& \left. -\Phi_{h}^{\alpha}(\boldsymbol{r}, t) \boldsymbol{\hat{e}} \cdot \boldsymbol{\nabla} \Phi_{h^{\prime}}^{* \alpha}(\boldsymbol{r}, t)\right] , \label{eq45}
\end{split}
\end{align}
and the memory kernels $G_{T}(\tau)$ in Eq.~\eqref{eq44} are given by
\begin{align}
G_{T}(\tau)=\frac{1}{\sqrt{4 \pi}} \frac{1}{\tau_{T}} \exp \left[-\left(\tau / 2 \tau_{T}\right)^{2}\right] , \label{eq46}
\end{align}
with the memory time determined by the average flow velocity $u_{T}$ of the target nucleons across the window according to $\tau_{T}=\kappa /\left|u_{T}(t)\right|$, and $G_{P}(\tau)$ is given by a similar expression. In a previous work~\cite{ayik2016,ayik2017}, we have estimated the memory time to be about $\tau_{T}=\tau_{P} \approx 25\; \mathrm{fm} / \mathrm{c}$, which is much shorter than the contact time of about $600\; \mathrm{fm} / \mathrm{c}$. As a result, the memory effect is not as important for diffusion coefficients. The fact that diffusion coefficients are being determined by the mean-field properties is consistent with the fluctuation dissipation theorem of non-equilibrium statistical mechanics, and it greatly simplifies calculations of quantal diffusion coefficients. Diffusion coefficients include the quantal effects due to shell structure, Pauli blocking, and full effects of the collision geometry without any adjustable parameters.

It is possible to deduce coupled differential equations for neutron and proton variances, $\sigma_{N N}^{2}(\ell)=$ $\overline{\delta N^{\lambda} \delta N^{\lambda}},\\ \sigma_{Z Z}^{2}(\ell)=\overline{\delta Z^{\lambda} \delta Z^{\lambda}}$, and co-variances $\sigma_{N Z}^{2}(\ell)=\overline{\delta N^{\lambda} \delta Z^{\lambda}}$ by multiplying the Langevin Eq.~\eqref{eq30} with $\delta N^{\lambda}$ and $\delta Z^{\lambda}$, and taking an average over the ensemble, which is generated from the solution of the Langevin equation. These coupled equations are presented in Refs.~\cite{ayik2018,yilmaz2018,ayik2019,sekizawa2020}. For completeness here, we provide these differential equations,
\begin{align}
\frac{\partial}{\partial t} \sigma_{N N}^{2}=2 \frac{\partial v_{n}}{\partial N_{1}} \sigma_{N N}^{2}+2 \frac{\partial v_{n}}{\partial z_{1}} \sigma_{N Z}^{2}+2 D_{N N} , \label{eq47}
\end{align}
\begin{align}
\frac{\partial}{\partial t} \sigma_{Z Z}^{2}=2 \frac{\partial v_{p}}{\partial Z_{1}} \sigma_{Z Z}^{2}+2 \frac{\partial v_{p}}{\partial N_{1}} \sigma_{N Z}^{2}+2 D_{Z Z} ,\label{eq48}
\end{align}
and
\begin{align}
\begin{split}
\frac{\partial}{\partial t} \sigma_{N Z}^{2}  =\frac{\partial v_{p}}{\partial N_{1}} & \sigma_{N N}^{2} +\frac{\partial v_{n}}{\partial Z_{1}} \sigma_{Z Z}^{2} \\
&+\sigma_{N Z}^{2}\left(\frac{\partial v_{p}}{\partial Z_{1}}+\frac{\partial v_{n}}{\partial N_{1}}\right). \label{eq49}
\end{split}
\end{align}
The above set of coupled equations are also familiar from the phenomenological nucleon
exchange model, and they were derived from the Fokker-Planck equation for the
neutron and proton distribution of fragments in deep-inelastic heavy-ion
collisions~\cite{schroder1981,merchant1982}. Nucleon drift mechanism, as well as
dispersions of fragment distributions, are determined in terms of two competing
effects: (i) Nucleon diffusion tends to increase dispersion of distribution
functions and (ii) Potential energy of the di-nuclear system $U\left(N_{1},
Z_{1}\right)$ on the neutron and proton plane that controls the mean nucleon drift
and determines saturation values of dispersions.

Neutron and proton drift coefficients are determined by the rate of change of
potential energy of di-nuclear system in (N, Z)-plane. Asymptotic values of
variances and co-variances are determined by the curvature of the potential
energy surface around the local equilibrium value. Curvature parameters are the
second derivatives of the potential energy surface, or the first derivative of the
drift coefficients, in the vicinity of the local equilibrium. In addition to the
diffusion coefficients, we need curvature parameters to determine the neutron,
proton, and mixed dispersions from Eqs.~(\ref{eq47}-\ref{eq49}). Potential
energy of the di-nuclear system consists mainly of surface energy, electrostatic
energy, symmetry energy, and centrifugal potential energy. TDHF theory includes
different energy contributions on a microscopic level. Furthermore, TDHF
calculations illustrate that the potential energy depends on the geometry of the di-nuclear
system. It is possible to extract the curvatures of the potential energy with the help
of Einstein relation in the over damped
limit~\cite{ayik2018,yilmaz2018,ayik2019,sekizawa2020}. In the over damped
limit, drift coefficients are related to the potential energy surface in $(N,
Z)$-plane as

\begin{eqnarray}
v_{n}(t)&=&-\frac{D_{N N}}{T^{*}} \frac{\partial}{\partial N_{1}} U\left(N_{1}, N_{1}\right) , \label{eq50} \\
v_{p}(t)&=&-\frac{D_{\mathrm{ZZ}}}{T^{*}} \frac{\partial}{\partial Z_{1}} U\left(N_{1}, Z_{1}\right) , \label{eq51} 
\end{eqnarray}  
where $T^{*}$ represent the effective temperature of the system, and ( $N_{1}, Z_{1}$ ) indicate neutron-proton numbers of one of the fragments in the di-nuclear system. The details of Einstein's relations in the overdamped limit are given in Appendix~\ref{sec:appE}

\subsection{Potential energy of the di-nuclear system}\label{sec3c}
\begin{figure}[!htb]
\includegraphics*[width=0.48\textwidth]{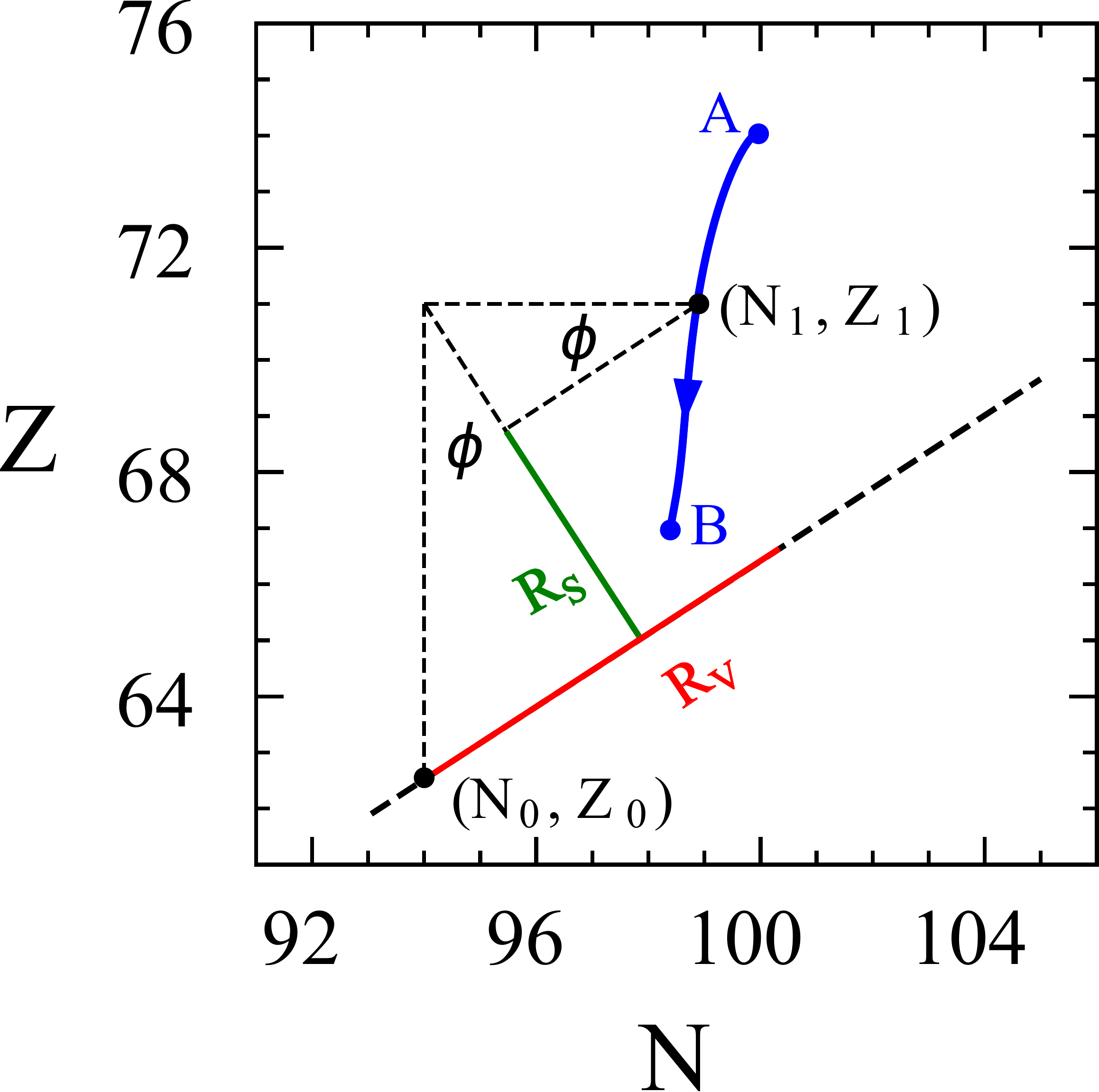}
\caption{Typical drift path in the $(N-Z)$ plane. Points $\left(N_{0}, Z_{0}\right)$ and $\left(N_{1}, Z_{1}\right)$ represent the local equilibrium state and neutron and proton numbers of the target-like or the projectile-like fragments. $R_S$ and $R_V$ denote the iso-vector and iso-scalar distances, respectively and their expressions are given in Eq.~\eqref{eq53} and Eq.~\eqref{eq54}. Please see text for further details.}
\label{fig:3.2}
\end{figure}

When the projectile and the target have different charge asymmetry values, colliding system rapidly evolves toward equilibration. Subsequently, the nucleon transfer continues and the system drifts towards the nearly constant charge asymmetry line, with a value around $(N-Z) /(N+Z)=0.20-0.23$, depending on the system. We refer to this line as the iso-scalar drift path, which runs nearly parallel to the bottom of the stability valley until the system reaches a local equilibrium state. In Fig.~\ref{fig:3.2}, iso-scalar drift path is represented schematically by a thick dashed straight line.  In this figure ($N_{0}, Z_{0}$) represents the local equilibrium state, and $\left(N_{1}, Z_{1}\right)$ is an arbitrary fragment on the drift path. 
The line between points $A$ and $B$ of Fig.~\ref{fig:3.2} depicts the actual path that may be taken by the target
like fragment.
Local equilibrium state is specified by shell closure or symmetric division of the colliding system. In the vicinity of the local equilibrium, we approximate the shape of the potential energy, in $(N-Z)$ plane, in terms of two parabolic forms,
\begin{align}
U\left(N_{1}, Z_{1}\right)=\frac{1}{2} a R_{S}^{2}\left(N_{1}, Z_{1}\right)+\frac{1}{2} b R_{V}^{2}\left(N_{1}, Z_{1}\right) . \label{eq52}
\end{align}
Here, $R_S$ represents perpendicular distances of fragment with neutron and proton numbers ($N_{1}, Z_{1}$) from iso--scalar path, and it is referred to as iso-vector path. $R_{V}$ indicates distance from the local equilibrium state $\left(N_{0}, Z_{0}\right)$ along the iso-scalar path. Because of the sharp increase of the asymmetry energy, the iso-vector curvature parameter "$a$" is usually much larger than the iso-scalar curvature parameter "$b$". It is possible to express iso-vector and iso-scalar distances in terms of neutron and proton numbers of the fragment and neutron and proton numbers of the equilibrium states as,
\begin{align}
R_{V}(t)=\left(N_{1}(t)-N_{0}\right) \cos \phi+\left(Z_{1}(t)-Z_{0}\right) \sin \phi ,\label{eq53}
\end{align}
and
\begin{align}
R_{S}(t)=\left(Z_{1}(t)-Z_{0}\right) \cos \phi-\left(N_{1}(t)-N_{0}\right) \sin \phi .    \label{eq54}
\end{align}

The angle $\phi$ is the angle between the iso-scalar path and $N$-axis. Depending on the collision's geometry, it is possible to have sign changes in the expressions for the iso-scalar and the iso-vector paths. Since the potential energy has an analytical form, we can immediately determine the derivatives of the drift coefficients by employing the Einstein's relations, Eqs.~(\ref{eq50}--\ref{eq51}), as
\begin{align}
\frac{\partial v_{n}}{\partial N_{2}} & =-D_{N N}\left(\alpha \sin ^{2} \phi+\beta \cos ^{2} \phi\right)  \label{eq55} , \\ 
\frac{\partial v_{p}}{\partial Z_{2}} & =-D_{Z Z}\left(\alpha \cos ^{2} \phi+\beta \sin ^{2} \phi\right)   \label{eq56} , \\
\frac{\partial v_{n}}{\partial Z_{2}} & =-D_{N N}(\beta-\alpha) \sin \phi \cos \phi \label{eq57} , \\
\frac{\partial v_{p}}{\partial N_{2}} & =-D_{Z Z}(\beta-\alpha) \sin \phi \cos \phi .    \label{eq58}
\end{align}
Here, $\alpha=a / T^{*}$ and $\beta=b / T^{*}$ indicate the reduced curvature coefficients. Only ratios of the curvature parameters and the effective temperature appear. As a result, the effective temperature is not a parameter in the description. These derivatives are valid for different collision geometries, for standard and inverse quasi fission reactions. Derivatives are also valid for symmetric collisions, in which there is no net drift. In this case, it is possible to estimate the reduced curvature parameters by employing suitable set of neighboring reactions.

Einstein relations in Eqs.~(\ref{eq50}-\ref{eq51}) are not exact relations between drift, diffusion, and the derivative of the potential energy, but they are valid in the overdamped limit of the nucleon diffusion mechanism (See Appendix~\ref{sec:appE} for detailed discussion on the overdamped limit). The overdamped limit of diffusion mechanism is approximately satisfied during the strong overlap time interval between the projectile-like and target-like fragments. By inverting the Einstein relations, we obtain expressions for curvature parameters in terms of transport coefficients,
\begin{align}
\beta(t)=-\left(\frac{v_{n} \cos \phi}{D_{N N}(t)}+\frac{v_{p} \sin \phi}{D_{Z Z}(t)}\right) \frac{1}{R_{V}(t)} , \label{eq59}
\end{align}
\begin{align}
\alpha(t)=+\left(\frac{v_{n} \sin \phi}{D_{N N}(t)}-\frac{v_{p} \cos \phi}{D_{Z Z}(t)}\right) \frac{1}{R_{S}(t)} . \label{eq60}
\end{align}

In most cases local equilibrium is a stable equilibrium of the potential energy. For the drift path shown in Fig.~\ref{fig:3.2}, neutron drift coefficient has a negative sign, and proton drift has a larger magnitude then the neutron drift. Note that there a is positive sign in front of the parentheses in Eq.~\eqref{eq59}, and negative sign in Eq.~\eqref{eq60}. Therefore, curvature parameters should be positive. Since it is possible to calculate curvature parameters in terms of transport coefficient, we can extract information about the curvature parameters of the macroscopic potential energy from TDHF calculations. As a result of shell effects in the microscopic description of the collision dynamics in the TDHF approach, it appears that the reduced curvature parameters depend on time. cross-sections for production of primary fragments are determined by the asymptotic values of the dispersions. In the asymptotic limit, the rate of change of dispersion is nearly zero, $\frac{\partial}{\partial t} \sigma_{\mathrm{NN}}^{2}(t) \rightarrow 0, \frac{\partial}{\partial t} \sigma_{\mathrm{ZZ}}^{2}(t) \rightarrow$ $0, \frac{\partial}{\partial t} \sigma_{\mathrm{NZ}}^{2}(t) \rightarrow 0$. Therefore, it is possible that the asymptotic values of the dispersions can be determined from quasi-static solutions of Eqs.~(\ref{eq55}-\ref{eq58}) by setting them equal to zero and employing the nearly asymptotic values of the curvature parameters. According to the Einstein relations, Eqs.~(\ref{eq50}-\ref{eq51}), the expressions for the reduced curvature parameters Eq.~\ref{eq59} and Eq.~\ref{eq60} are valid in the overdamped limit, which occurs during the strong overlap interval of the projectile and target. Therefore, we cannot estimate the asymptotic values of the reduced curvature parameters directly from these expressions. We employ the following approximation: We calculate the average values of the reduced iso-scalar and reduced iso-vector curvature parameters over the strong overlap intervals and use these average values over the entire contact time in Eqs.~(\ref{eq55}-\ref{eq58}). The average values of the curvature parameters are usually determined as,
\begin{align}
\bar{\beta}=\frac{1}{t_{B}-t_{A}} \int_{t_{A}}^{t_{\mathrm{B}}} \beta(t) d t ,\label{eq61}
\end{align}
and
\begin{align}
\bar{\alpha}=\frac{1}{t_{B}-t_{A}} \int_{t_{A}}^{t_{B}} \alpha(t) d t . \label{eq62}
\end{align}

We specify the strong overlap time interval with the help of diffusion coefficients, which are calculated for a typical initial angular momentum. In evaluating variances from solutions of Eqs.~(\ref{eq47}-\ref{eq49}), we use the average values of the curvature parameters over the complete contact time. Also, we note that the approximate Einstein relations may lead to singular behavior of curvature parameters in Eqs.~(\ref{eq61}--\ref{eq62}) when distance $R_{S}$ and $R_{V}$ become very small. At the instant when the drift path crosses the iso-scalar path, the iso-vector distances become zero and the curvature parameter becomes infinite. Therefore, in calculating average values of curvature parameters, the averaging interval should safely avoid the singular behavior in Eqs.~(\ref{eq61}-\ref{eq62}). In some cases, to avoid singular behavior, we approximate the average values in Eqs. ~(\ref{eq61}-\ref{eq62}) by moving average values of the iso-scalar and the iso-vector distances outside the integral. For a heavy di-nuclear systems, rotational energy does not have a sizable effect on the curvature parameters. Therefore, we often use head-on collisions for estimating the curvature parameters. When drift occurs toward asymmetry, the iso-scalar distance given by Eq.~\eqref{eq53} changes sign, and consequently, there is sign change in Eq.~\eqref{eq59}. In collisions of symmetric systems, and in certain reactions due to shell closure effects, collisions do not exhibit sizable drift, and hence it is not possible to determine the curvature parameters. In those situations, it is possible to determine the curvature parameters by employing collisions of a suitable neighboring system.

\subsection{Cross-sections of primary fragments}\label{sec3d}
In general, the joint probability distribution function, $P_{\ell}(N, Z)$, for producing a binary fragment with neutron number $N$ and proton number $Z$, is determined by generating a large number of solutions of the Langevin Eq.~\eqref{eq30}. It is well known that the Langevin equation is equivalent to the Fokker-Planck equation for the distribution function of the macroscopic variables~\cite{risken1996}. In the case when the drift coefficients are linear functions of macroscopic variables, as we have in Eq.~\eqref{eq30}, the proton and neutron distribution function for initial angular momentum $l$ is given as a correlated Gaussian function described by the mean values, the neutron, proton, and mixed dispersions as,
\begin{align}
P_{\ell}(N, Z)=\frac{1}{2 \pi \sigma_{N N}(\ell) \sigma_{Z Z}(\ell) \sqrt{1-\rho_{\ell}^{2}}} \exp \left(-C_{\ell}\right)  .  \label{eq63}
\end{align}
Here the exponent $C_{\ell}$ for each impact parameter is given by,
\begin{align}
\begin{split}
C_{\ell}=\frac{1}{2\left(1-\rho_{\ell}^{2}\right)} & \left[\left(\frac{Z-Z_{\ell}}{\sigma_{Z Z}(\ell)}\right)^{2} +\left(\frac{Z-Z_{\ell}}{\sigma_{Z Z}(\ell)}\right)^{2} \right. \\ 
& \left. -2 \rho\left(\frac{Z-Z_{\ell}}{\sigma_{Z Z}(\ell)}\right)\left(\frac{N-N_{\ell}}{\sigma_{N N}(\ell)}\right)\right] , \label{eq64}
\end{split}
\end{align}
with the correlation coefficient defined as $\rho_{\ell}={\sigma_{N Z}^{2}(\ell)}/{ \sigma_{Z Z}(\ell) \sigma_{N N}(\ell)}$. The quantities $N_{\ell}=\bar{N}_{\ell}^{\lambda}, Z_{\ell}=$ $\bar{Z}_{\ell}^{\lambda}$ denote the mean neutron and proton numbers of the target-like or project-like fragments. These mean values are determined from the TDHF calculations.

Variances and co-variances are determined from the solutions of the coupled differential Eqs.~(\ref{eq47}-\ref{eq49}) with the initial conditions $\sigma_{N N}^{2}(t=0)=0, \;\sigma_{N N}^{2}(t=0)=0$, and $\sigma_{N N}^{2}(t=0)=0$ for each angular momentum. Production cross-sections of primary isotopes are calculated using the standard expression,
\begin{align}
\sigma^{s}(N, Z)=\frac{\pi \hbar^{2}}{2 \mu E_{c . m.}} \sum_{\ell_{\text {min }}}^{\ell_{\text {max }}}(2 l+1) P_{\ell}^{S}(N, Z) , \label{eq65}
\end{align}
with,
\begin{align}
P_{\ell}^{s}(N, Z)=\frac{1}{2}\left[P_{\ell, p r o}^{s}(N, Z)+P_{\ell, t a r}^{s}(N, Z)\right] .\label{eq66}
\end{align}
In these expressions, label "$s$" indicates different geometries of collision. Quantities $P_{\ell, p r o}^{s}(N, Z)$ and $P_{\ell, t a r}^{s}(N, Z)$ denote normalized probabilities of producing projectile-like and target-like fragments. These probabilities are given by Eq.~\eqref{eq63} with mean values being projectile-like and target-like fragments, respectively. The factor of $1 / 2$ is introduced to make the total primary fragment distribution normalized to unity. In summation over $\ell$, the range of initial angular momentum also depends on the detector geometry in the laboratory frame. In calculations, we perform the summation in the range of $\ell_{\min }$ to $\ell_{\max }$, the range of $\ell$ values specified by the angular acceptance of the detector. If only mass numbers of fragments are identified, double probability in the cross-section expression is replaced by the mass number distribution,
\begin{align}
P_{\ell}^{s}(A)=\frac{1}{2}\left[P_{\ell, p r o}^{s}(A)+P_{\ell, \text { tar }}^{s}(A)\right] .
\label{eq67}
\end{align}
Probability distribution of mass number of produced fragments is determined by summing over $N$ or $Z$ and keeping the total mass number, $\mathrm{A}=N+Z$, constant
\begin{align}
\begin{split}
P_{\ell, p r o}^{S}(A)= & \frac{1}{\sqrt{2 \pi}} \frac{1}{\sigma_{A A}^{S}(\ell)} \\
& \times\exp \left[-\frac{1}{2}\left(\frac{A-A_{\ell}, p r o}{\sigma_{A A}^{S}(\ell)}\right)^{2}\right] , \label{eq68}   \end{split}
\end{align}
where mass variance is given by $\sigma_{A A}^{2}(\ell)=\sigma_{N N}^{2}(\ell)+\sigma_{Z Z}^{2}(\ell)+2 \sigma_{N Z}^{2}(\ell)$. If the projectile and the target are spherical, there is no averaging needed over different collision symmetries. If the projectile and target are deformed, usually we calculate tip-tip, side-side, side-tip and tip-side geometries. The cross-sections of primary fragment production are calculated by weighted averaging over different collision geometries. Primary fragments are excited and cool down by particle emissions and secondary fission. It is possible to follow de-excitation process and calculate cross-sections of secondary isotope production with the help of statistical GEMINI++ code~\cite{charity2008}. We discuss calculations of primary and secondary cross-sections for some selected systems in Section~\ref{sec4}.

\section{Applications of SMF theory on Multinucleon Transfer Reactions in Heavy-ion Collisions}\label{sec4}
\renewcommand{\thefigure}{4.\arabic{figure}} 
\setcounter{figure}{0}
Reaction outcomes and the dominant reaction mechanisms depend heavily on the chosen projectile-beam combination in the entrance reaction channel. For this reason, we have divided this section into two different subsections. In section~\ref{sec4a} the systems in which quasi-fission and fusion-fission mechanisms compete are given. In section~\ref{sec4b} we focus our discussion of isotope production via MNT reactions in neutron-rich and super-heavy regions, as well as the effect of relative orientation of the reaction partners on the reaction outcome. All of the numerical computations in this section and are carried out by extending the three-dimensional TDHF program of Umar \textit{et al.}~\cite{umar1991a,umar2006c} to incorporate the quantal transport calculations, and using the SLy4d Skyrme energy density functional~\cite{chabanat1998a,kim1997}.

\subsection{Competition between QF and FF}\label{sec4a}

In this section, we focus on systems in which both QF and FF mechanisms come into play. In QF reactions, usually a large number of nucleon transfer takes place from the heavier target towards to the lighter projectile, driving the system toward more mass symmetric exit channel. If the charge product $Z_pZ_t$ of the composite system is small enough to allow the system to fuse for near central collisions, the mass symmetric fission fragments are also produced via the fusion-fission mechanism. Since sub-barrier fusion via tunneling is not included in the TDHF calculations, it is not feasible to determine a precise initial angular momentum range corresponding to fusion. Nevertheless, one can get the fusion-fission primary product distribution via GEMINI++ code and combine it with SMF results to cover both fusion-fission and quasi-fission processes. A comprehensive description of both mechanisms can be achieved by combining the stochastic mean–field (SMF) transport coefficients with the post–scission fragment distributions generated by the \textsc{GEMINI++} statistical–decay code~\cite{charity2008}.  

First, we present the results for the $^{48}$Ca+$^{238}$U system  at $E_{\mathrm{c.m.}}=193$~MeV since it provides a textbook example of multinucleon transfer in collisions and application of SMF. The ground‐state Hartree–Fock calculations yield a spherical $^{48}$Ca projectile and a strongly prolate $^{238}$U target. In what follows we focus on the MNT channel obtained in the tip orientation of the prolate deformed $^{238}$U nucleus; TDHF trajectories show that only this alignment populates scattering angles that fall inside the angular acceptance spectrometer (25$^{\circ}$–65$^{\circ}$ in the laboratory frame) in Ref.~\cite{kozulin2012}, therefore, the side orientation of the ${}^{238}\mathrm{U}$ was not included in this work. TDHF and SMF calculations were carried out for a range of initial angular momenta $ 38 \hbar\leq  \ell \leq 56\hbar$, for the tip orientation of the $^{238}$U target. 
\begin{figure}[!htb]
\centering
\includegraphics*[width=0.48\textwidth]{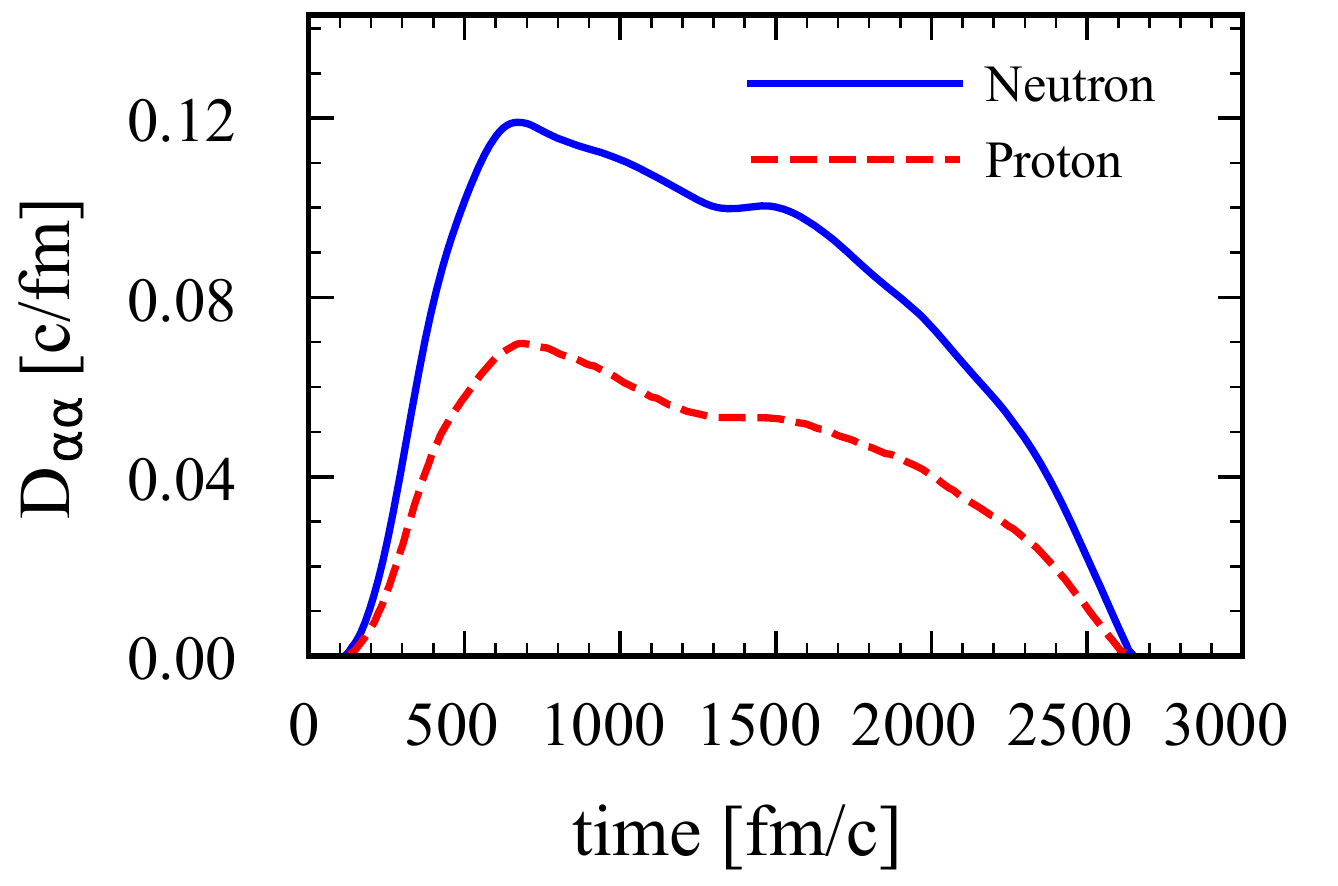}
\caption{Neutron and proton diffusion coefficients as a function of time for the collision 
$^{48}$Ca+$^{238}$U at $E_{c.m.}$ = 193~MeV for the tip geometry of the uranium and for initial orbital angular momentum $\ell = 40\hbar$.}
\label{fig:4.1}
\end{figure}

\begin{figure}[!htb]
\centering
\includegraphics*[width=0.48\textwidth]{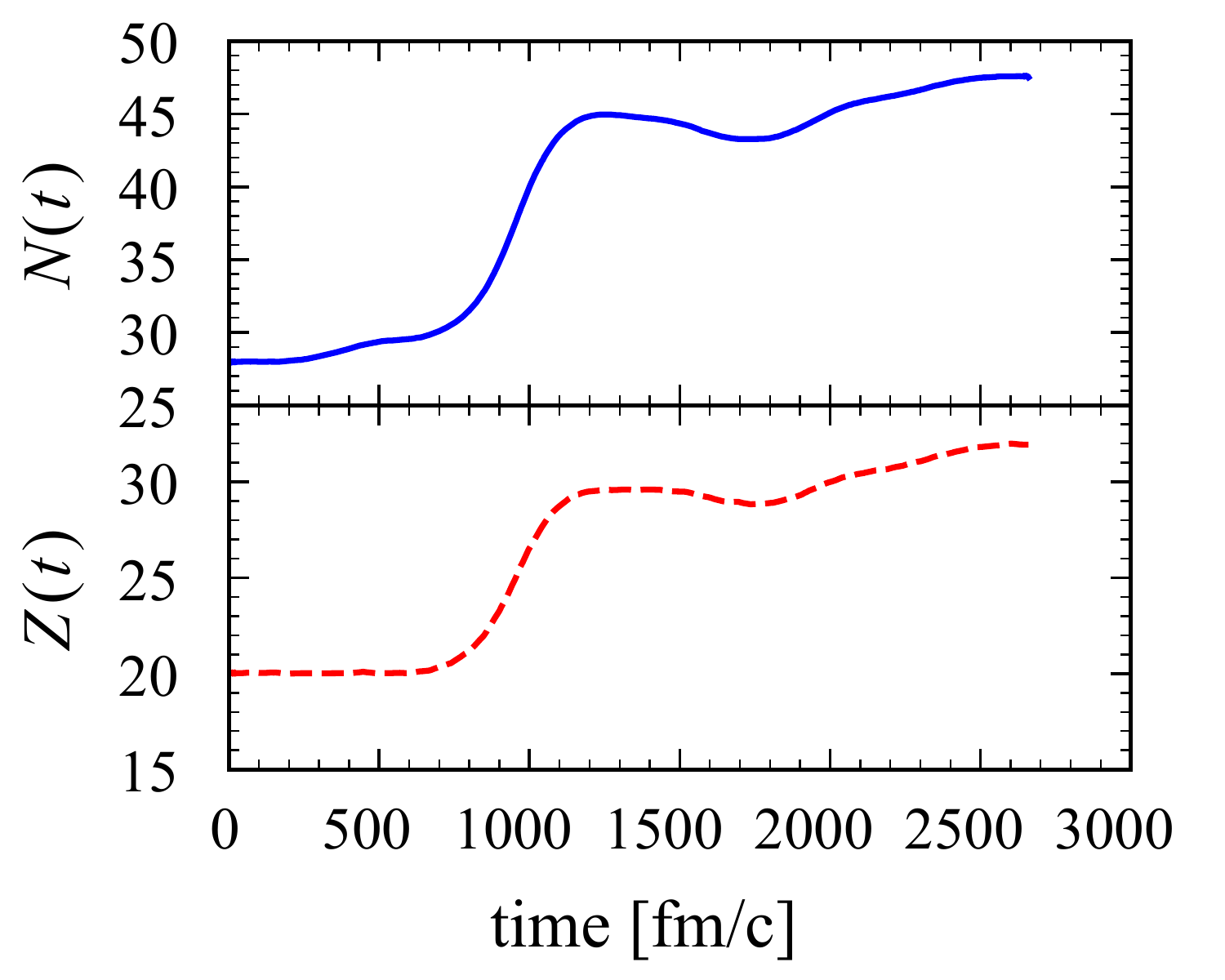}
\caption{Mean values of neutron and proton numbers of projectile-like fragments at initial angular momentum $\ell$ = $40\hbar$ for the $^{48}$Ca+$^{238}$U system at $E_{c.m.}$ = 193~MeV are shown as a function of time. Solid blue lines denote the neutron numbers and dashed red lines denote the proton numbers of target-like fragments. The labels $t_A$, $t_B$ , and $t_C$ indicate the projection of the time intervals used to determine the curvature parameters. }
\label{fig:4.2}
\end{figure}

\begin{figure}[!htb]
\centering
\includegraphics*[width=0.48\textwidth]{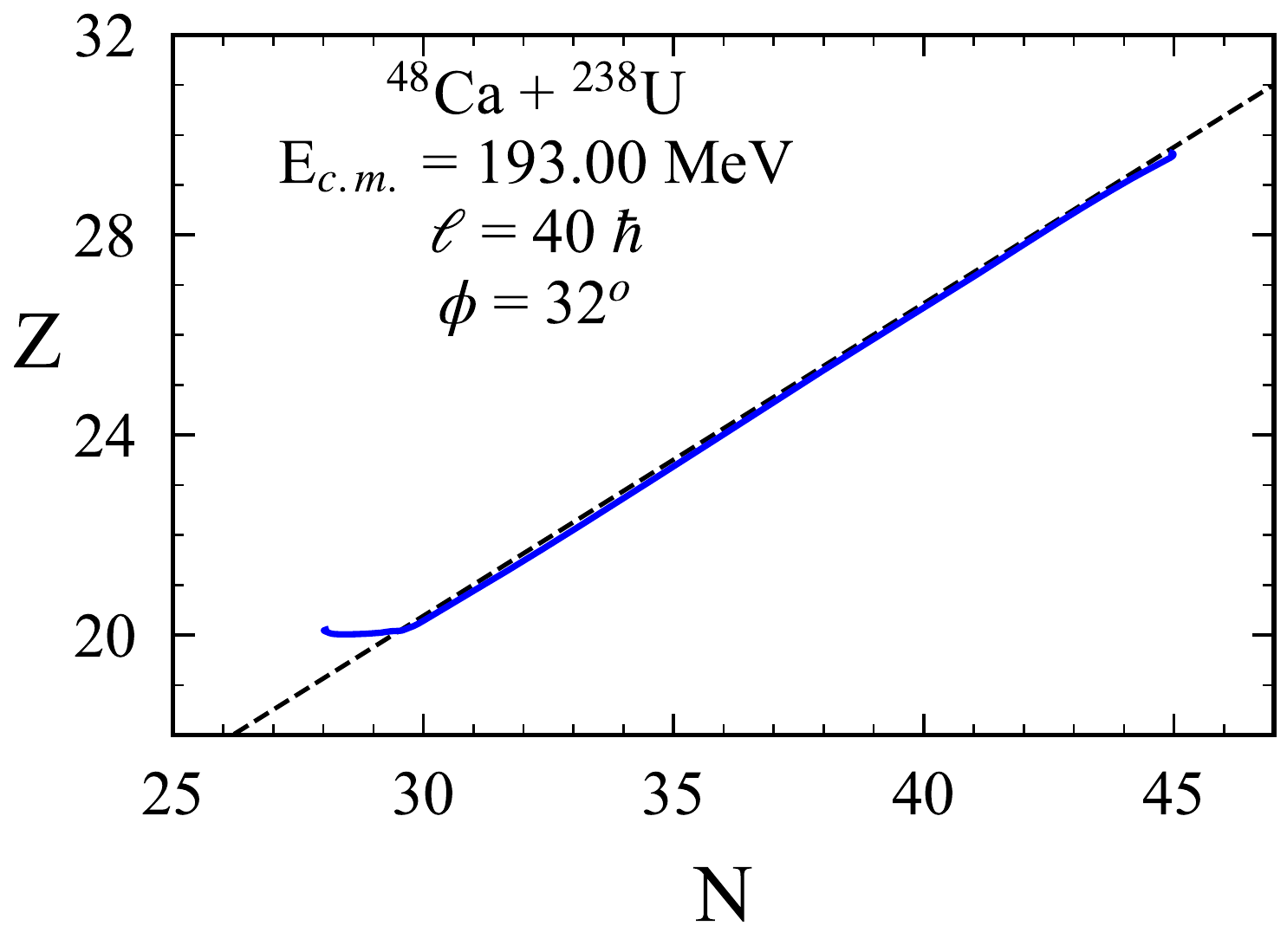}
\caption{The mean-drift path in the (N, Z)-plane for the $^{48}$Ca+$^{238}$U
collision with tip geometry of the uranium at bombarding energy
$E_{c.m.}$ = 193 MeV and initial angular momentum $\ell$ = $40\hbar$.}
\label{fig:4.3}
\end{figure}
\begin{figure*}[!htb]
\centering
\includegraphics*[width=1.0\textwidth]{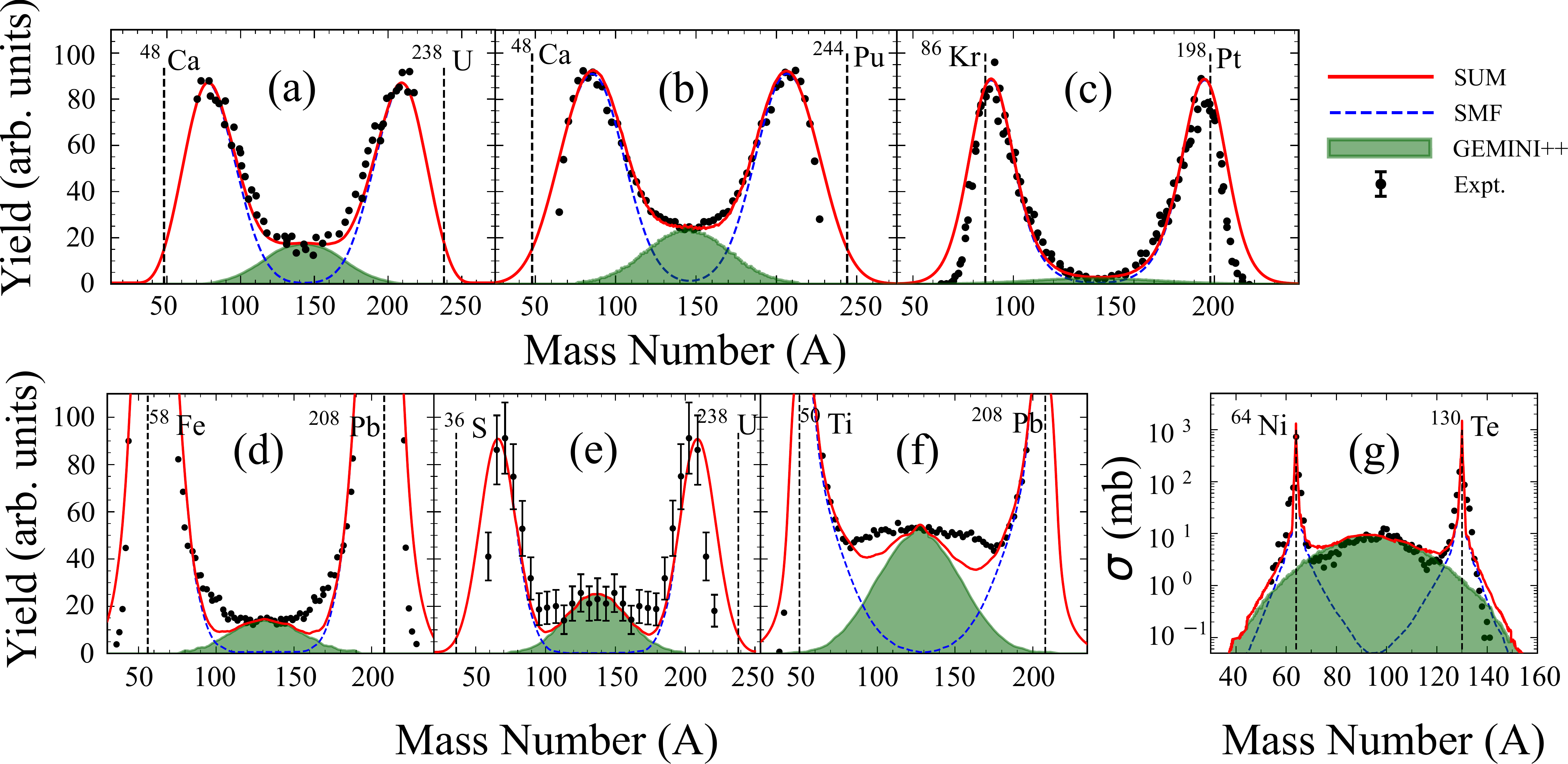}
\caption{Solid red line shows the combined primary yield of multinucleon transfer (dashed blue line in center region) and binary fission
(green shaded region) as function of fragment mass $A$ for $^{48}$Ca+$^{238}$U (a), $^{48}$Ca+$^{244}$Pu (b), $^{86}$Kr+$^{198}$Pt (c), $^{58}$Fe+$^{208}$Pb (d), $^{36}$S+$^{238}$U (e), $^{50}$Ti+$^{208}$Pb (f), and $^{64}$Ni+$^{130}$Te (g) systems~\cite{ayik2021,arik2023,kayaalp2024,kayaalp2024b}. Black filled circles show the available experimental data~\cite{kozulin2010,kozulin2019,sen2022,itkis2022,itkis2011,appannababu2016,krolas2010}.}
\label{fig:4.4}	
\end{figure*}
In Fig.~\ref{fig:4.1}, we show the neutron and proton diffusion coefficients, calculated using Eq.~\eqref{eq44}, as a function of time for the initial angular momentum $\ell=40 \hbar$. Diffusion coefficients play an important role since they appear: (i) directly in the coupled differential equations for dispersion values in Eqs.(\ref{eq47}--\ref{eq49}), (ii) in calculating the reduced curvature parameters in Eqs.(\ref{eq59}--\ref{eq60}). To calculate the charge dispersions, we need to determine the average values of curvature parameters for this binary system, which explained in detail in Sec.~\ref{sec3c}. For this purpose, we have used the collision  at $\ell=40 \hbar$. As discussed in Appendix~\ref{sec:appA}, the window separating the projectile-like and target-like fragments is determined by diagonalizing the mass quadrupole moment tensor and integrating the charge densities on either side of the window at each time step. In Fig.~\ref{fig:4.2} the mean values of neutron and proton numbers of projectile-like fragments are given at initial angular momentum $\ell = 40 \hbar $ as a function of time. During the collision, roughly~12 protons (accompanied by~20 neutrons) are transferred from the heavier target-like fragment to the lighter projectile-like fragment. To gain more insight, we can easily eliminate the time dependence and plot the charge numbers in $N-Z$ plane. Fig.~\ref{fig:4.3} presents the drift path of the projectile-like fragment in the $N-Z$ plane for this system. If we denote the charge asymmetry of the reaction fragments with $\delta$ =$\frac{N-Z}{A}$, the charge asymmetry values of reaction partners are $\delta(\mathrm{Ca}) \simeq 0.167$ and $\delta(\mathrm{U}) \simeq 0.227$. When the reaction partners have different charge asymmetries, such as in this system, the binary system undergoes a rapid approach towards the stability line (which is denoted by the dashed black line in Fig.~\ref{fig:4.3}, where fragments have equal charge asymmetry values of $\delta = 0.217$. After reaching charge asymmetry equilibrium, the system drifts along the iso-scalar valley with a slope $\phi\approx32^{\circ}$. This rapid charge transfer towards the charge asymmetry equilibrium, followed by slower transfer along the stability line (shown by the solid line) is observed in TDHF solutions to various other MNT reactions~\cite{ayik2018,ayik2023,ayik2023b,arik2023,ocal2025,sekizawa2020,kayaalp2024,kayaalp2024b}. As a next step, the reduced curvatures $\alpha$ and $\beta$ are deduced from the Einstein relations Eqs.~(\ref{eq50}-\ref{eq51}) by averaging Eqs.~(\ref{eq61}--\ref{eq62}) over the strong‐overlap window where neutron and proton drift is most vigorous. We obtain $\bar{\alpha}=0.142$ and $\bar{\beta}=0.027$; their ratio confirms that the restoring force perpendicular to the iso‐scalar path dominates the slow collective motion towards the same charge asymmetry value. 
Using the diffusion coefficients and the reduced curvature parameters, the task is to solve Eqs.~(\ref{eq47}--\ref{eq49}) with initial conditions $\sigma_{NN}^{2}=\sigma_{ZZ}^{2}=\sigma_{NZ}^{2}=0$ and use their asymptotic values to calculate the primary product probabilities, given by Eq~\ref{eq63}. The TDHF+SMF results, for the final orbital angular momentum $\ell_f$, final average total kinetic energy $TKE$, scattering angles in center of mass and lab frames $\theta_{c.m.},\theta_1^{lab}$, $\theta_2^{lab}$, final mass and charge values of the fragments, and neutron, proton, mixed variances are calculated for each initial angular momentum and given in detail in Ref.~\cite{ayik2021}. 

The primary fragment's charge and mass cross-sections are obtained by summing over the relevant angular momenta range given in Eq.~\eqref{eq65} (with a similar expression for yield). To investigate the distribution resulting from fusion-fission contribution, we employ the GEMINI++ code~\cite{charity2008}. Total yield of the primary fragment mass
distribution is then given by,
\begin{align}
    Y(A) = \eta \left( P^{MNT}(A) + P^{FF}(A)\right),
\end{align}
where $P^{MNT}(A)$ is given by Eq.~\eqref{eq67} and $P^{FF}(A)$ is the fragment probability distribution calculated by GEMINI++. Here, the excitation energy of the
compound nuclei is estimated by, $E_{CN} = E_{c.m.} + Q_{gg}$ where $ Q_{gg}$ stands for released disintegration energy in fusion
reactions. The number of simulations is set to 100000, which is sufficient to get a statistical distribution in this region. 

The methodology explained thus far is employed to investigate different projectile-target combinations for MNT reactions using the SMF framework~\cite{arik2023,ayik2023,ayik2023b,kayaalp2024,kayaalp2024b,ocal2025}, and the procedure is more or less similar to one for the $^{48}$Ca+$^{238}$U system~\cite{ayik2021}. For details of the calculations and calculated values, we suggest the reader consult the original papers. In works ~\cite{arik2023,kayaalp2024,kayaalp2024b}, a very similar approach is applied to $^{48}$Ca+$^{244}$Pu, $^{86}$Kr+$^{198}$Pt, $^{58}$Fe+$^{208}$Pb, $^{50}$Ti+$^{208}$Pb, $^{36}$S+$^{238}$U, and $^{64}$Ni+$^{130}$Te systems, where QF and FF mechanisms come into play and the computational results are given in detail in the original papers~\cite{arik2023,ayik2021,kayaalp2024,kayaalp2024b}. In Fig.~\ref{fig:4.4} we show the primary product mass distribution for these systems, including the $^{48}$Ca+$^{238}$U system.

\begin{figure*}[!htb]
	\includegraphics*[width=1.0\textwidth]{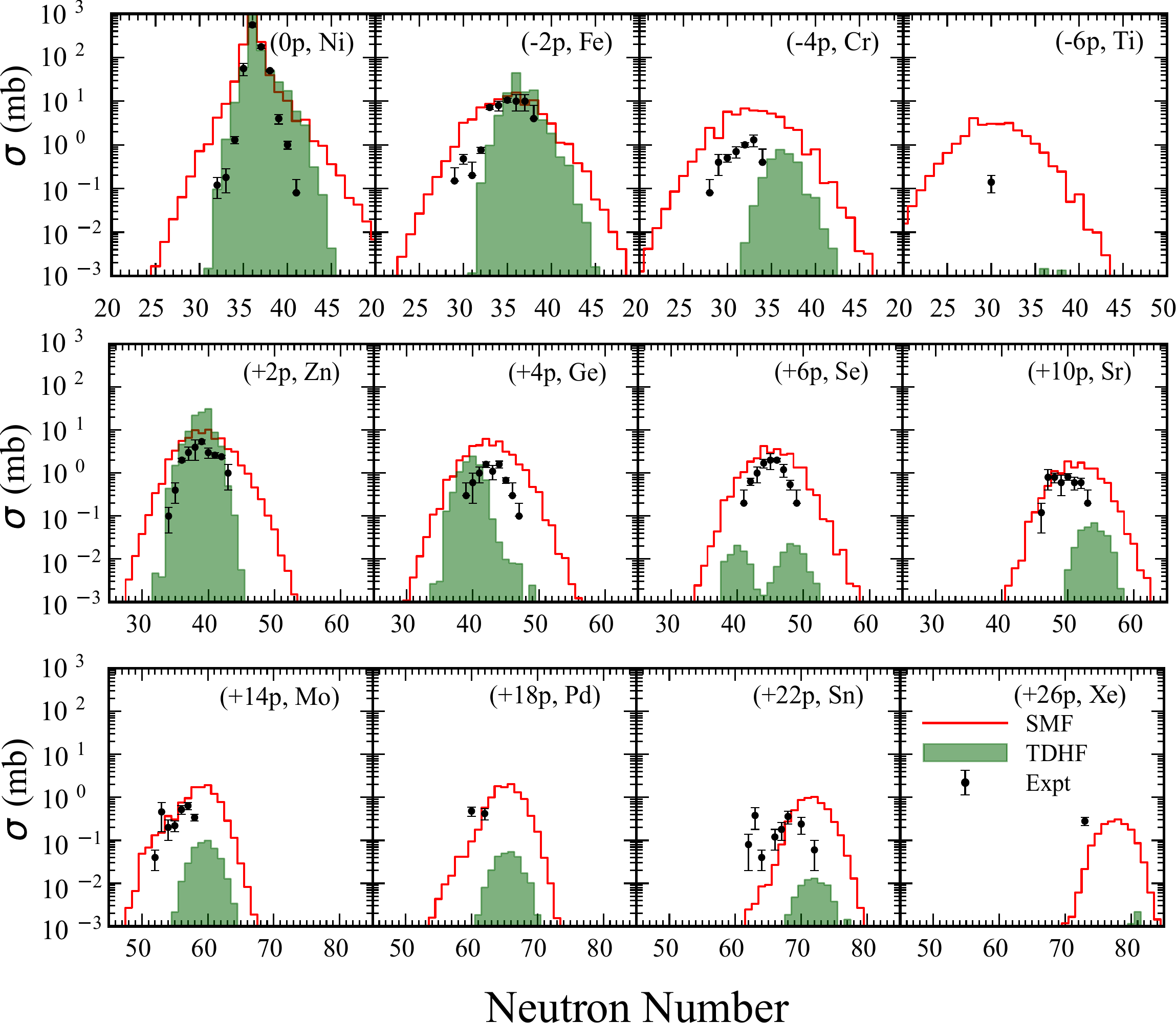}
	\caption{Selection of secondary production cross-sections for lighter (projectile-like) fragments for the ${}^{64}\text{Ni}+{}^{208}\text{Pb}$ reaction at $E_\text{c.m.}=268$~MeV. In each panel, production cross-sections for different isotopes [as indicated by (±xp; X ), where x indicates the number of transferred protons and X
		stands for the corresponding element] are shown as a function of the neutron number. Blue thick histograms show the results of SMF+GEMINI++
		calculations, while red thin histograms show those of TDHF+GEMINI++. Red solid circles show the experimental data with error bars, taken from
		Ref.~\cite{krolas2003}. The down arrows indicate upper bound of the measured cross-sections~\cite{krolas2003}.}
	\label{fig:4.5}
\end{figure*}

In Fig.~\ref{fig:4.4}, dashed blue lines indicate the primary product mass distributions calculated within the
SMF framework, whereas filled green areas represent the fusion–fission fragment distribution calculated using the GEMINI++ code. The summation of the yield distributions calculated by Eq.~\eqref{eq67} is indicated by solid red lines. The available experimental data for each system is denoted by black filled circles. In Fig.~\ref{fig:4.4}, one can see that the mass symmetry region $(A_{CN}/2 \pm 20 )$ is heavily populated and it mainly consist of contributions from symmetric fission fragments and overlapping tails of the MNT reaction products. For $^{48}$Ca+$^{244}$Pu, $^{86}$Kr+$^{198}$Pt, $^{48}$Ca+$^{238}$U and $^{36}$S+$^{238}$U systems, one can calculate the ratio integrated yield between the interval $A_{CN}/2 \pm 20$ to the total integrated yield under the experimental values. In $^{48}$Ca+$^{244}$Pu, 11\%, $^{86}$Kr+$^{198}$Pt \%3 in $^{36}$S+$^{238}$U 14.9\%,  $^{48}$Ca+$^{238}$U 11\%. If we compare the charge product $Z_pZ_t$ with the calculated ratio values, we see that (i) $^{48}$Ca+$^{244}$Pu and $^{48}$Ca+$^{238}$U systems have roughly equal charge product and ratio, and (ii) as the charge product increases the ratio of calculated FF yield to total yield decreases. This clearly shows how the charge product influences the competition between QF and FF.

\subsection{Heavy isotope production}
\label{sec4b}

The multinucleon transfer reactions might be a possible alternative to produce
neutron-rich heavy nuclei and possibly superheavy nuclei in the yet-unreached island of
stability. For this purpose, MNT reactions involving heavy projectile and target combinations have been extensively studied experimentally at near-barrier energies
over the last two decades~\cite{kozulin2012,kozulin2014,kratz2013,watanabe2015,devaraja2015,desai2019,birkenbach2015,vogt2015,galtarossa2018,kalandarov2020}. To investigate the primary and secondary isotope distributions in MNT reactions, the SMF theory has been applied to various systems at near-barrier energies over the last decade. Here, we share some of these results, for a detailed description of the systems and calculations we suggest the reader to refer to the original papers.
In Fig.~\ref{fig:4.5} we plot the secondary product isotope distributions of lighter (projectile-like) fragments for the ${}^{64}\text{Ni}+{}^{208}\text{Pb}$ system at $E_\text{c.m.}=268$~MeV. For this system the production yields were extracted for a wide range of isotopes, encompassing both projectile-like fragments (PLFs) and target-like fragments (TLFs), over the full angular distribution, and originating from multiple reaction mechanisms, including not only deep-inelastic collisions but also the transfer-induced fission processes. In this work, the secondary product distributions calculated by the SMF theory are compared with the available experimental data~\cite{krolas2003} and the TDHF+PNP method in which the transfer probabilities are extracted from TDHF framework by particle number projection (PNP) method of Simenel~\cite{simenel2010}.  The comparison between the experimental data and the theoretical predictions provides a critical insight into the validity and limitations of the existing models. First, it is evident that both the TDHF+PNP method and the SMF theory are adequate in describing channels involving the transfer of a few nucleons. However, as the number of transferred nucleons increases, the TDHF+PNP method systematically underestimates the cross-sections. This behavior of the TDHF+PNP approach has also been observed in other studies. In contrast, the SMF theory has been shown to yield results of the correct order of magnitude even for channels involving the transfer of as many as 26 protons (along with a substantial number of neutrons), demonstrating consistency with experimental data.

Notably, the SMF approach succeeds in reproducing nucleon transfers in both projectile-to-target and target-to-projectile directions, a result attributed to the underlying quantal diffusion mechanism. The agreement between SMF predictions and experimental data highlights the importance of accurately modeling the formation and subsequent de-excitation of heavy nuclei in order to account for the observed yields of isotopes ranging from technetium (Tc) to xenon (Xe).

\begin{figure}[!htb]
\includegraphics*[width=0.48\textwidth]{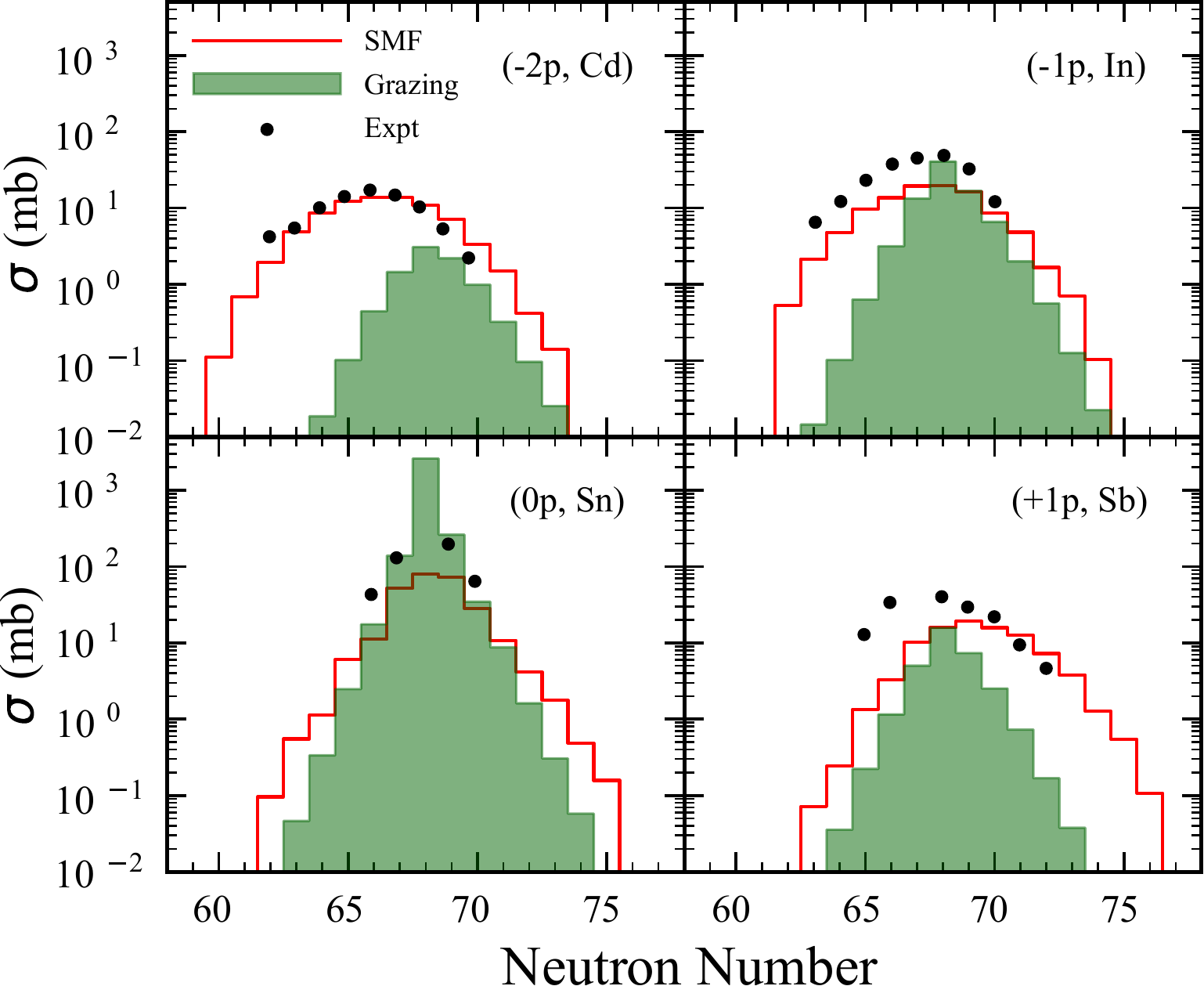}
\caption{Observed and calculated cross-sections for Sn-like fragments with $Z=$48, 49, 50, 51 for the ${}^{206}\text{Pb}+{}^{118}\text{Sn}$ reaction at $E_\text{c.m.}=436.8$~MeV~\cite{kayaalp2024b}, as a function of neutron
number. Experimental data were taken from Ref.~\cite{diklic2023}. The NRV site is used with default parameters~\cite{karpov2017b} for GRAZING calculations.}
\label{fig:4.6}
\end{figure}

In a separate study~\cite{kayaalp2024b}, we investigated the secondary product isotope cross-sections in the ${}^{206}\text{Pb}+{}^{118}\text{Sn}$ system at the center-of-mass energy of $E_{\text{c.m.}} = 436.8$~MeV, and compared the results with both the available experimental data and the open-source semi-classical GRAZING model. The results are presented in Fig.~\ref{fig:4.6}, where it can be observed that while the GRAZING model provides a reasonable description of the no-proton-transfer channel, it significantly underestimates both the width and the magnitude of the distributions in channels involving proton transfer. In contrast, the SMF theory successfully reproduces the overall structure and yields across all transfer channels, demonstrating its robustness in describing multinucleon transfer dynamics beyond the elastic and quasi-elastic regime.

\begin{figure}[!htb]
\includegraphics*[width=0.48\textwidth]{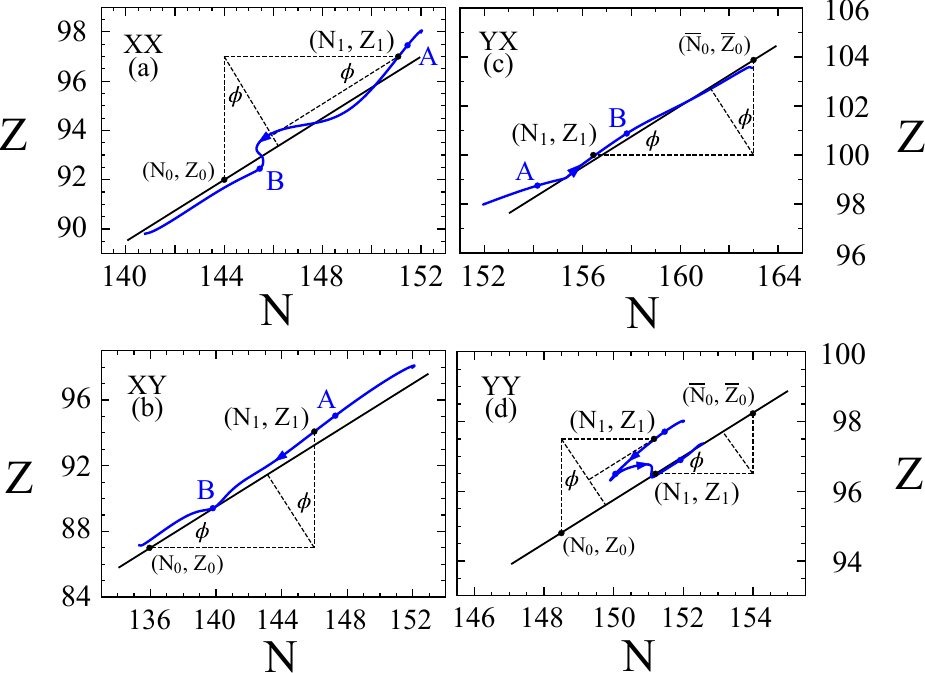}
\caption{Blue curves show the drift path of Cf-like fragments for the head-on
collision of ${}^{250}\text{Cf}+{}^{232}\text{Th}$ system at $E_\text{c.m.}=950$~MeV 
in tip-tip (XX), tip-side (XY), side-tip (YX) and side-side (YY) geometries.}
\label{fig:4.7}
\end{figure}

In MNT reactions, the target and the projectile might have large deformations in their ground states. Consequently, the reaction dynamics and the transfer of nucleons depend on the relative alignment of the target and the projectile. We refer to this as the dependence on collision geometry. We denote the initial orientation of either
the projectile or the target principle deformation axis to be in
the beam direction as X, and we denote the case when their
principle axis is perpendicular to the beam direction as Y. As
a result we are usually faced with four distinct orientation possibilities
for the target and the projectile, labeled as YY , XX , XY , and
YX , corresponding to side-side, tip-tip, tip-side, and side-tip
collision geometries, respectively. Here, the first letter stands for the orientation
of the lighter collision partner.  To serve as an example, we show the drift path of Cf-like fragments in the head-on
collision of ${}^{250}\text{Cf}+{}^{232}\text{Th}$ system at $E_\text{c.m.}=950$~MeV 
in tip-tip (XX), tip-side (XY), side-tip (YX) and side-side (YY) geometries in Fig.~\ref{fig:4.7}.

\begin{figure*}[!htb]
\includegraphics*[width=1.0\textwidth]{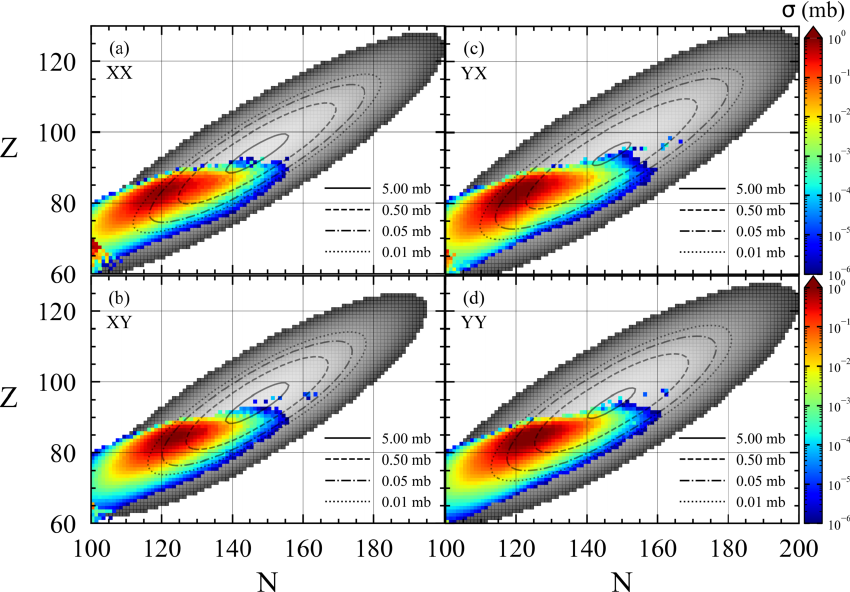}
\caption{Enlarged view of secondary production cross-sections in the $N-Z$ plane for 
${}^{250}\text{Cf}+{}^{232}\text{Th}$ system at $E_\text{c.m.}=950$~MeV  in tip-tip
(XX), tip-side (XY), side-tip (YX) and side-side (YY) geometries. The shaded background highlights the primary product cross-section, see the text for details.}
\label{fig:4.8}
\end{figure*}

In these figures, the thick black lines indicate the equilibrium charge asymmetry
with $(N+Z)/(N-Z)= 0.22$.  We observe that in all
geometries, Cf-like fragments drift nearly along the isoscalar
direction with charge asymmetry approximately equal to 0.22.
Figure~\ref{fig:4.7}(a) shows the drift path for the tip-tip collision. As
usually observed in quasifission reactions, Cf-like heavy fragments lose nucleons and the system drifts toward symmetry.
In the side-tip collision, shown in Fig.~\ref{fig:4.6}(c), the nucleon
drift mechanism is very different from the tip-tip geometry.
Here, the heavy fragment gains neutrons and protons and
the system drifts along the isoscalar path toward asymmetry. This kind of drift path is not very common, and is referred to as the inverse quasifission reaction. In Fig.~\ref{fig:4.7}(b), in the tip-side collision, the nucleon drift
mechanism is very different from that of other geometries.
The di-nuclear system drifts along the isoscalar path with the
same charge asymmetry toward symmetry. However, Cf-like
heavy fragments continue to lose neutrons and protons until
they nearly reach thorium at the exit channel. In Ref.~\cite{kedziora2010},
this type of drift was named as the swap inverse quasifission
reaction. In this type of reaction, the lighter (heavier) fragment in the entrance channel becomes the heavier (lighter) one in the exit channel. Fig.~\ref{fig:4.7}(d) illustrates the drift path in a side-side collision. Unlike other collision geometries, the side–side case shows only a small transfer of nucleons, with the neutron and proton numbers of the final fragments remaining nearly unchanged from their initial values. This suggests that the shell effects play a distinct role in this geometry. Consequently, the di-nuclear system tends to remain near a local equilibrium point on the potential energy surface in the $N-Z$ plane, without significant drift toward symmetry or asymmetry.

Different relative orientations lead to different TDHF charge values on the exit channel and transport coefficients calculated by SMF, resulting in different primary and secondary isotope distribution. In Ref.~\cite{ayik2023}, we have studied the secondary isotope production in ${}^{250}\text{Cf}+{}^{232}\text{Th}$ system at $E_\text{c.m.}=950$~MeV. Fig.~\ref{fig:4.8} shows the enlarged view of secondary production cross-sections in the $N-Z$ plane for 
${}^{250}\text{Cf}+{}^{232}\text{Th}$ system at $E_\text{c.m.}=950$~MeV in tip-tip
(XX), tip-side (XY), side-tip (YX), and side-side (YY) geometries. The shaded background in Fig.~\ref{fig:4.6} shows the primary product cross-sections for each orientation in the range $10^3$mb to $10^{-6}$mb. As can be seen from the figure, the primary isotopes near the superheavy region do not survive the de-excitation process, resulting in an accumulation near the doubly-magic ${}^{208}\mathrm{Pb}$. Still, the primary product cross-sections in side-tip and side-side geometries are nearly an order of magnitude larger for heavy-fragment yields calculated in tip-tip and tip-side, reflecting the enhanced inverse quasi-fission in these geometries. Below $N=100$, and $Z=60$, decay products consist of secondary fission of excited heavy fragments. Above this region the cross-sections are populated by light particle emission including neutron, proton, and alpha particles. Calculations predict production of broad range of neutron-rich isotopes for nuclei with proton numbers in the range of $Z=70-90$, with cross-sections on the order of several hundred  micro barns. Our calculations indicate a number of neutron-rich heavy isotopes with sizable cross-sections, including $^{246}\text{Cm}$ with a cross-section of 159~nanobarn for the side-side collision, $^{248}\text{Cm}$ with a cross-section of 80 nanobarn for the tip-side collision, and $^{253}\text{Cm}$ with a cross-section of 252 nanobarn for the side-tip collision. In reality, all collision orientations contribute to the cross-sections and one should weigh each one with an appropriate geometric factor. As a result, the values provided here should be seen as an order of magnitude as well as upper limits. Since there is no experimental work performed for this system yet, we were unable to test our results. 

\begin{figure}[!htb]
\includegraphics*[width=0.48\textwidth]{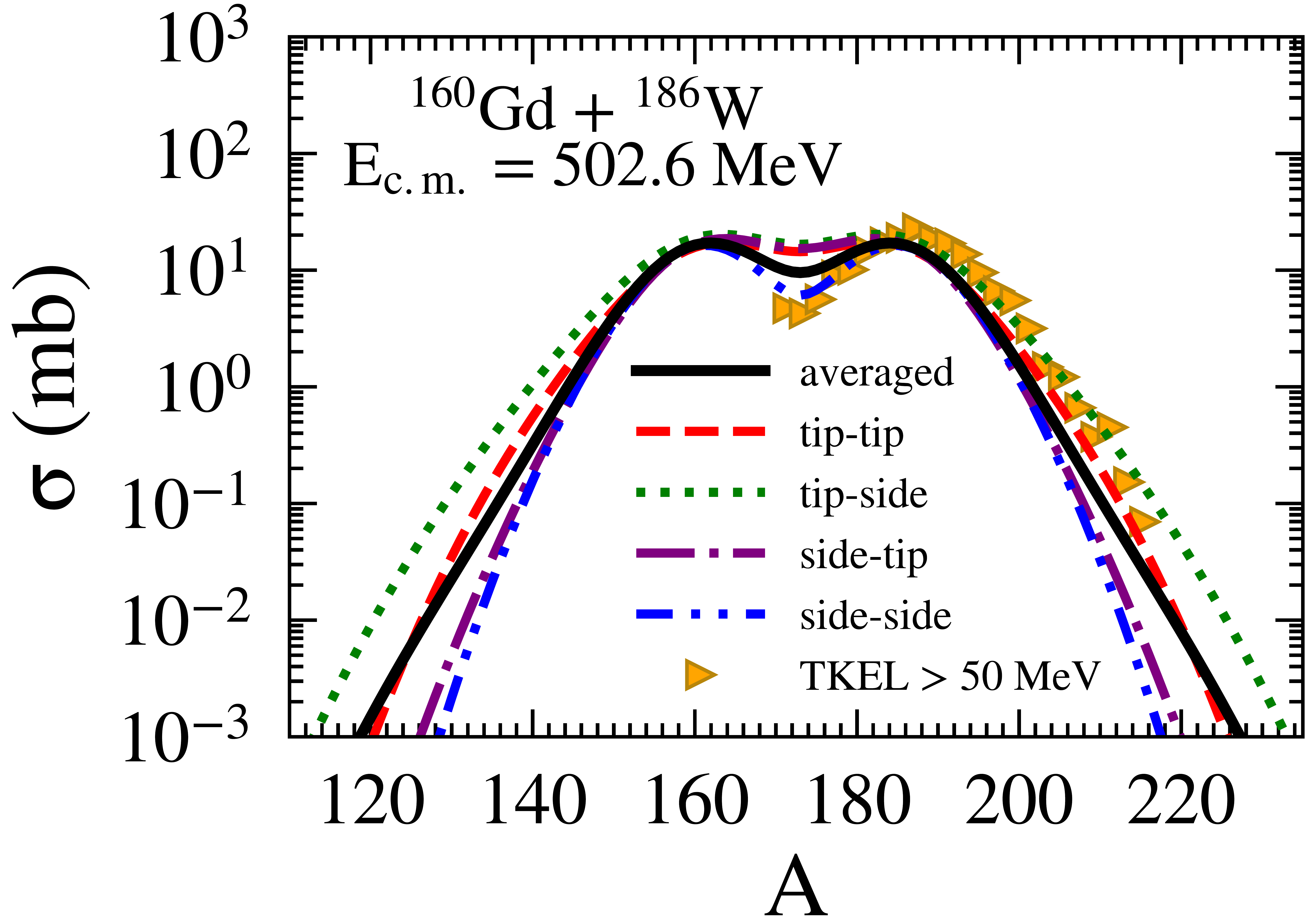}
\caption{Total cross-section in collision of ${}^{160} \text{Gd}+{}^{186} \text{W}$ system at $E_\text{c.m.}=502.6$~MeV
as a function of mass number A~\cite{ayik2023b}. Different geometries are indicated by dashed (red), dotted (green),
dashed-dotted (brown) and dashed-dot-dotted (blue) lines. Average cross-section and experimental data taken from
Ref.~\cite{kozulin2017} are indicated by solid black line and yellow triangles, respectively.}
\label{fig:4.9}
\end{figure}

In another work, we have investigated the ${}^{160}\text{Gd}+{}^{186}\text{W}$ system at $E_\text{c.m.}=502.6$~MeV and $E_\text{c.m.}=461.9$~MeV energies for four different relative orientations and a broad range of angular momenta~\cite{ayik2023b}. In this work, primary mass distributions arising from multinucleon transfer reactions are investigated for each relative orientation and compared with the available experimental data~\cite{kozulin2017}. The primary mass distributions resulting from different collision geometries are averaged by following a similar approach to Ref.~\cite{hinde1996}. The results are given in Fig.~\ref{fig:4.9}.
We note that the results are in good agreement with the experimental data. In ${}^{250}\text{Cf}+{}^{232}\text{Th}$ reaction, inverse quasi-fission mechanism is observed when lighter (heavier) fragment is oriented in side (tip) orientation. We should mention that some orientations and initial angular momentum are more favorable to produce neutron-rich isotopes. Analysis of these systems underline the importance of quasi-fission and inverse quasi-fission mechanisms in enhancing the neutron-rich heavy isotope production.

\begin{figure}[!htb]
\includegraphics*[width=0.48\textwidth]{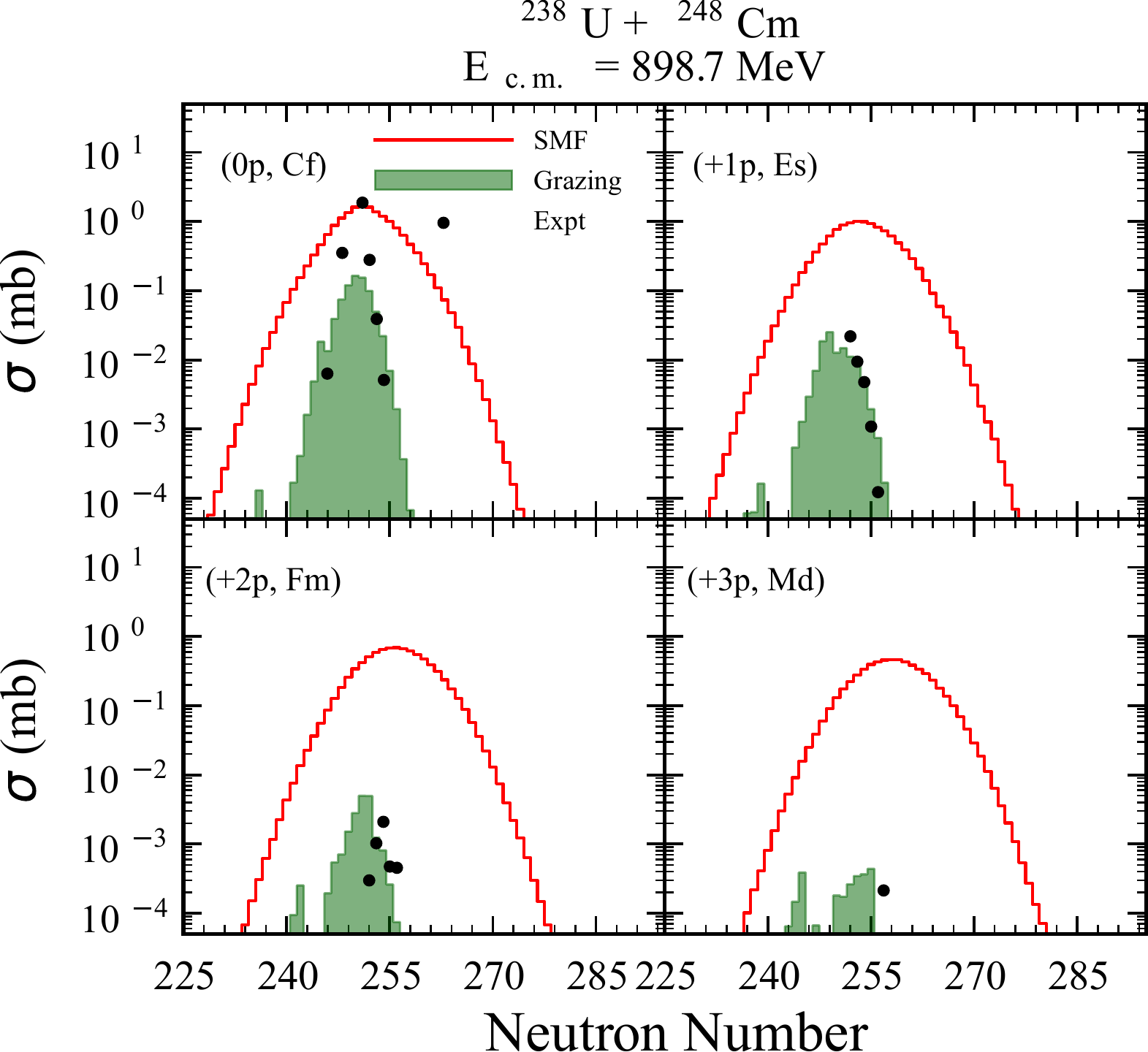}
\caption{ Production cross-sections of transuranium isotopes for the ${ }^{238} \mathrm{U}+{ }^{248}\mathrm{Cm}$ system at E$_{\mathrm{c.m.}}$ = 898.7~MeV }
\label{fig:4.10}
\end{figure}

Recently, we have also investigated the secondary isotope production in ${}^{238}\text{U}+{}^{248}\text{Cm}$ system at $E_\text{c.m.}=898.7$~MeV energy and compared the results with the available experimental data~\cite{ocal2025}. The results are given in Fig~\ref{fig:4.9}. Similar to ${}^{250}\text{Cf}+{}^{232}\text{Th}$ system, the reaction products in the exist channel de-excite mostly via fission. We can see that there is a 2-3 orders of magnitude difference between the primary and secondary isotopes in the superheavy region. This highlights the dominant effect of  de-excitation process in collisions involving superheavies.
Overall, these results clearly demonstrate that, depending on the deformation and
relative orientation of the reaction partners, the outcome of the reaction differs considerably. This effect supports the idea that MNT
reactions can serve as a useful tool to produce neutron-rich
heavy nuclei near the lead region, which may not be available
via fusion-fission and fragmentation. It has been shown that secondary de-excitation of primary fragments decrease the superheavy and heavy neutron rich isotope production substantially. We should also point out that the SMF theory is fully microscopic, including quantal effects such as the shell structure and Pauli blocking, without introducing any adjustable parameters.

\section{Kinetic Energy Dissipation and Fluctuations in SMF}\label{sec5}
\renewcommand{\thefigure}{5.\arabic{figure}} 
\setcounter{figure}{0}

Nuclear dissipation plays a crucial role in both heavy-ion collisions and nuclear fission. To unravel the dissipation mechanisms, extensive experimental and theoretical studies have been conducted over many years~\cite{ayik1976,agassi1977,randrup1978b,randrup1979}. In low-energy heavy-ion reactions, the dominant contribution arises from one-body dissipation and fluctuations driven by nucleon exchange. The TDHF framework provides a microscopic description of the dissipative dynamics in this regime, with the mean-field approach accounting for one-body dissipation and predicting the most likely trajectory of the collision. However, as noted in section~\ref{sec1}, mean-field dynamics significantly underestimates fluctuations. The SMF theory offers a way to incorporate both dissipation and fluctuations of the relative kinetic energy in heavy-ion collisions~\cite{ayik2020b}. Furthermore, in section~\ref{sec5a}, we introduce a quantal diffusion approach to describe dissipation and fluctuations of relative kinetic energy and angular momentum transfer. In Section~\ref{sec5b}, we investigate transport coefficients for linear momentum, derive the kinetic energy distribution function, and in section~\ref{sec5d} apply this framework to study kinetic energy dissipation in the ${ }^{136}\mathrm{Xe}+{ }^{208}\mathrm{Pb}$ system at $E_{c.m.}=526$~MeV.

\subsection{Quantal Langevin equation for relative motion}\label{sec5a}
In Fig.~\ref{fig:3.1}, the vectors $\boldsymbol{R}^{+}$ and $\boldsymbol{R}^{-}$ denote the positions of the mean centers-of-mass of the projectile-like and target-like fragments, respectively, measured relative to the total center of mass of the system. Using the local density and current density, one can express the fragment masses, their center-of-mass positions, and their total linear momenta as follows:
\begin{align}
M_{\lambda}^{\mp}=\int d^{3} x \Theta\left(\mp x^{\prime}\right) \rho^{\lambda}(\boldsymbol{r}, t) , \label{eq70}
\end{align} 
\begin{align}
\boldsymbol{R}_{\lambda}^{\mp}=\frac{1}{M_{\lambda}^{\mp}} \int d^{3} x \Theta\left(\mp x^{\prime}\right) \boldsymbol{r} \rho^{\lambda}(\boldsymbol{r}, t) , \label{eq71}
\end{align}
\begin{align}
 \boldsymbol{P}_{\lambda}^{\mp}=\int d x^{3} \Theta\left(\mp x^{\prime}\right) m \boldsymbol{j}^{\lambda}(\boldsymbol{r}, t). \label{eq72}
\end{align}
Here, the local density and the current density indicate the total values that are summed over neutrons and protons $\rho^{\lambda}(\boldsymbol{r}, t)=\rho_{n}^{\lambda}(\boldsymbol{r}, t)+\rho_{p}^{\lambda}(\boldsymbol{r}, t)$ and $\boldsymbol{j}^{\lambda}(\boldsymbol{r}, t)=\boldsymbol{j}_{n}^{\lambda}(\boldsymbol{r}, t)+\boldsymbol{j}^{\lambda}_{p}(\boldsymbol{r}, t)$.

The local densities and current densities of neutrons and protons, for an event $\lambda$, are defined in Eqs.~~(\ref{eq27}--\ref{eq28}). In this formulation, we neglect fluctuations of the window geometry around its mean position. As noted in Section~\ref{sec3a} and Appendix~\ref{sec:appA}, the mean window position is described by $\Theta(\mp x^{\prime})$, where $x^{\prime}(t)=\left[x-x\_{0}(t)\right]\cos \theta(t)+\left[y-y\_{0}(t)\right]\sin \theta(t)=0$ specifies the time-dependent orientation of the window in the reaction plane. With these definitions, we can introduce the relative coordinate $\boldsymbol{R}_{\lambda}=\boldsymbol{R}_{\lambda}^{+}-\boldsymbol{R}_{\lambda}^{-}$, the reduced mass $\mu_{\lambda}=M_{\lambda}^{-}\cdot M_{\lambda}^{+}/M$, and the relative linear momentum,

\begin{equation}
\begin{split}
\boldsymbol{P}_{\lambda}=\frac{M_{\lambda}^{-} \boldsymbol{P}_{\lambda}^{+}-M_{\lambda}^{+} \boldsymbol{P}_{\lambda}^{-}}{M} & =\mu_{\lambda} \boldsymbol{{\dot{R}}}_{\lambda} \\ 
& =\mu_{\lambda}\left[\dot{R}_{\lambda} \boldsymbol{\hat{e}}_{R}+R_{\lambda} \dot{\theta}_{\lambda} \boldsymbol{\hat{e}}_{\theta}\right]  .
\label{eq73}    
\end{split}
\end{equation}
Here, $M = M_{\lambda}^{-}+M_{\lambda}^{+}$ is the total mass, $\dot{\boldsymbol{R}}_{\lambda}=\dot{\boldsymbol{R}}_{\lambda}^{+}-\dot{\boldsymbol{R}}_{\lambda}^{-}$ denotes the relative velocity vector and the velocities of the projectile- and target-like fragments can be defined by $\boldsymbol{\dot{R}}_{\lambda}^{\mp}=\boldsymbol{P}_{\lambda}^{\mp} / M_{\lambda}^{\mp}$. 
In the last term of Eq.~\eqref{eq72} the relative velocity is decomposed into radial and tangential components with unit vectors in the radial and tangential directions,
\begin{align}
\boldsymbol{\hat{e}}_{R}=\cos \theta \boldsymbol{\hat{i}}+\sin \theta \boldsymbol{\hat{j}} \quad \text { and } \quad \boldsymbol{\hat{e}}_{\theta}=-\sin \theta \boldsymbol{\hat{i}}+\cos \theta \boldsymbol{\hat{j}} . \label{eq74}
\end{align}
By neglecting the time dependence of the reduced mass, the rate of change of the relative momentum can be written as,

\begin{align}
\begin{split}
\frac{d}{d t} \boldsymbol{P}_{\lambda}&=\mu_{\lambda}\left[\left(\ddot{R}_{\lambda}-R_{\lambda} \dot{\theta}_{\lambda}^{2}\right) \boldsymbol{\hat{e}}_{R}+\left(R_{\lambda} \ddot{\theta}_{\lambda}+2 \dot{R}_{\lambda} \dot{\theta}_{\lambda}\right) \boldsymbol{\hat{e}}_{\theta}\right] \\
& =\left(\frac{d K_{\lambda}}{d t}-\frac{L_{\lambda}^{2}}{\mu_{\lambda} R_{\lambda}^{3}}\right) \boldsymbol{\hat{e}}_{R}+\left(\frac{1}{R_{\lambda}} \frac{d L_{\lambda}}{d t}\right) \boldsymbol{\hat{e}}_{\theta} ,\label{eq75}
\end{split}
\end{align}
where we have introduced the radial component of the relative linear momentum $K_{\lambda}=\boldsymbol{\hat{e}}_{R} \cdot \boldsymbol{P}_{\lambda}$, and relative orbital angular momentum defined as $L_{\lambda}=\mu_{\lambda} R_{\lambda}^{2} \dot{\theta}$. The first and second in Eq.~\eqref{eq75} denote the rate of change of the radial and the tangential components, respectively.

The rate of change of the linear momentum of the projectile-like and target-like fragments can be expressed in terms of the current density and its derivative, using the definition in Eq.~\eqref{eq71}, to obtain

\begin{align}
\begin{split}
\frac{d}{d t} \boldsymbol{P}_{\lambda}^{\mp}= & \mp \int d^{3} x \delta\left(x^{\prime}\right) \dot{x}^{\prime} m_{j} \boldsymbol{j}_{\lambda}(\boldsymbol{r}, \boldsymbol{p}, t) \\
& +\int d^{3} x \Theta\left(\mp x^{\prime}\right) \frac{\partial}{\partial t} \boldsymbol{j}_{\lambda}(\boldsymbol{r}, t) . \label{eq76}
\end{split}
\end{align}
Using the TDHF Eqs.~\eqref{eq8} for the single-particle wave functions for the event $\lambda$, it is possible to write down the rate of change of the radial and tangent and components of linear momentum of the fragments in the following form
\begin{align}
\begin{split}
& \boldsymbol{\hat{e}}_{\alpha} \cdot \frac{\partial}{\partial t} \boldsymbol{P}_{\lambda}^{\mp}=\mp \int d^{3} x \delta\left(x^{\prime}\right) \dot{x}^{\prime} m \boldsymbol{\hat{e}}_{\alpha} \cdot \boldsymbol{j}_{\lambda}(\boldsymbol{r}, \boldsymbol{p} t)+\text {pot. terms}\\
& -\int d^{3} x \Theta\left(\mp x^{\prime}\right) \boldsymbol{\nabla} \cdot \sum_{j i}\left(\boldsymbol{V}_{j i}^{\alpha}(\boldsymbol{r}, t ; \lambda)-\boldsymbol{W}_{j i}^{\alpha}(\boldsymbol{r}, t ; \lambda)\right) \rho_{j i}^{\lambda} ,\label{eq77}
\end{split}
\end{align}
with
\begin{align}
\begin{split}
\boldsymbol{V}_{j i}^{\alpha}(\boldsymbol{r}, t ; \lambda)  =& \frac{\hbar^{2}}{4 m} \left(\phi_{i}(\boldsymbol{r}, t ; \lambda) \boldsymbol{\nabla}\left(\boldsymbol{\hat{e}}_{\alpha} \cdot \boldsymbol{\nabla}\right) \phi_{j}^{*}(\boldsymbol{r}, t ; \lambda)\right.  \\ 
& \left.  +\phi_{j}^{*}(\boldsymbol{r}, t ; \lambda) \boldsymbol{\nabla}\left(\boldsymbol{\hat{e}}_{\alpha} \cdot \boldsymbol{\nabla}\right) \phi_{i}(\boldsymbol{r}, t ; \lambda) \right) ,\label{eq78}
\end{split}
\end{align}

\begin{align}
\begin{split}
\boldsymbol{W}_{j i}^{\alpha}(\boldsymbol{r}, t ; \lambda)=& \frac{\hbar^{2}}{4 m}\left(\left(\boldsymbol{\nabla} \phi_{i}(\boldsymbol{r}, t ; \lambda)\right)\left(\boldsymbol{\hat{e}}_{\alpha} \cdot \boldsymbol{\nabla}\right) \phi_{j}^{*}(\boldsymbol{r}, t ; \lambda) \right. \\
& \left. +\left(\boldsymbol{\nabla} \phi_{j}^{*}(\boldsymbol{r}, t ; \lambda)\right)\left(\boldsymbol{\hat{e}}_{\alpha} \cdot \boldsymbol{\nabla}\right) \phi_{i}(\boldsymbol{r}, t ; \lambda)\right) . \label{eq79}
\end{split}
\end{align}
Here, $\boldsymbol{\hat{e}}_{\alpha}$ denotes the unit vectors in the radial $(\alpha=R)$ and the tangential $(\alpha=\theta)$ directions. The "Potential terms" in Eq.~\eqref{eq77} represent terms associated with the mean-field potential, excluding the kinetic terms in the single-particle Hamiltonian for the event $\lambda$.

Combining Eq.~\eqref{eq77} with Eq.~\eqref{eq73}, we can drive a Langevin equation for the rate of change of the relative linear momentum
\begin{align}
\begin{split}
\frac{d}{d t} \boldsymbol{P}_{\lambda}=\int d^{3} x g\left(x^{\prime}\right) \dot{x}^{\prime}  m \boldsymbol{j}_{\lambda}(\boldsymbol{r}, p, t)& +\text{pot. terms }  \\
& +\boldsymbol{f}^{\lambda}(t) ,\label{eq80}  
\end{split}
\end{align}
where the first and second terms represent the forces arising from the motion of the window plane and the potential terms, respectively. The quantity $\boldsymbol{f}^{\lambda}(t)$ is the fluctuating dynamical force due to nucleon exchange between projectile-like and target-like fragments. Its radial and tangent components are given by
\begin{align}
f_{\alpha}^{\lambda}(t)=\sum_{i j} Y_{j i}^{\alpha}(t ; \lambda) \rho_{j i}^{\lambda} ,\label{eq81}
\end{align}
where we introduce a shorthand notation,
\begin{align}
Y_{j i}^{\alpha}(t ; \lambda)=\int d^{3} x g\left(x^{\prime}\right) \boldsymbol{\hat{e}}_{R} \cdot\left(\boldsymbol{V}_{j i}^{\alpha}(\boldsymbol{r}, t ; \lambda)-\boldsymbol{W}_{j i}^{\alpha}(\boldsymbol{r}, t ; \lambda)\right) . \label{eq82}
\end{align}
In obtaining this result, we employed a partial integration in Eq.~\eqref{eq77} and used the following relations,
\begin{align}
\nabla_{x} \Theta\left[\left(x-x_{0}\right) \cos \theta+\left(y-y_{0}\right) \sin \theta\right]=\delta\left(x^{\prime}\right) \cos \theta ,\label{eq83}
\end{align}
and
\begin{align}
\nabla_{y} \Theta\left[\left(x-x_{0}\right) \cos \theta+\left(y-y_{0}\right) \sin \theta\right]=\delta\left(x^{\prime}\right) \sin \theta .\label{eq84}
\end{align}

In a manner similar to Eq.~\eqref{eq29}, in Eq.~\eqref{eq75}, we introduce a smoothing function $\delta\left(x^{\prime}\right) \rightarrow g\left(x^{\prime}\right)$ in terms of a Gaussian $g(x)=(1 / \kappa \sqrt{2 \pi}) \exp \left(-x^{2} / 2 \kappa^{2}\right)$ with a dispersion $\kappa \approx 1.0 \mathrm{fm}$ that is in the order of the lattice spacing in the numerical calculations. By projecting Eq.~\eqref{eq74} along the radial and the tangential directions, together with Eq.~\eqref{eq80}, we obtain two coupled Langevin equations for the radial and relative angular momentum as ~\cite{gardiner1991,weiss1999},
\begin{align}
\begin{split}
\frac{d K_{\lambda}}{d t}-\frac{L_{\lambda}^{2}}{\mu_{\lambda} R_{\lambda}^{3}}= & +\int d^{3} x g\left(x^{\prime}\right) \dot{x}^{\prime} m \boldsymbol{\hat{e}}_{R} \cdot \boldsymbol{j_{\lambda}}(\boldsymbol{r}, t) \\
& + \text { potential terms }+f_{R}^{\lambda}(t) ,
\label{eq85}
\end{split}
\end{align}
and
\begin{align}
\begin{split}
\frac{1}{R_{\lambda}} \frac{d L_{\lambda}}{d t}= & +\int d^{3} x g\left(x^{\prime}\right) \dot{x}^{\prime} m \boldsymbol{\hat{e}}_{\theta} \cdot \boldsymbol{j}_{\lambda}(\boldsymbol{r}, t) \\
& +\text { potential terms }+f_{\theta}^{\lambda}(t) . \label{eq86}
\end{split}
\end{align}
On the right-hand side of these expressions, the first term accounts for the force generated by the motion of the window plane, while the second term corresponds to the conservative forces arising from nuclear and Coulomb potential energies. The fluctuating components, $f_{R}^{\lambda}(t)$ and $f_{\theta}^{\lambda}(t)$, describe the dynamical forces produced by multinucleon exchange between the projectile-like and target-like fragments. These dynamical forces serve as the primary source of dissipation and fluctuations in the relative momentum during damped heavy-ion processes, such as deep-inelastic collisions and quasifission reactions.

For small amplitude fluctuations the ensemble average values of these equations of motion are equivalent to the TDHF description for the radial and angular components of the relative momentum. Consequently, the mean values of the total kinetic energy and the orbital angular momentum are determined by TDHF. We employ the Langevin Eqs.~\eqref{eq85} and Eq.~\eqref{eq86} to describe fluctuations around their mean values. There are two different sources for fluctuations of the dynamical forces, $f_{R}^{\lambda}(t)$, and $f_{\theta}^{\lambda}(t)$, induced by the nucleon exchange: (i) Fluctuations due to different set of wave functions in each event $\lambda$, and (ii) Fluctuations introduced by the stochastic part $\delta \rho_{j i}^{\lambda}$ of the density matrix of the initial state.

Former part of the fluctuations can be approximately described in terms of the fluctuating components of the radial and angular momentum as $f_{R}^{\lambda}(t) \rightarrow f_{R}^{d i s s}\left(K_{\lambda}\right)$, and $f_{\theta}^{\lambda}(t) \rightarrow f_{\theta}^{\text {diss }}\left(L_{\lambda}\right)$. Here, $f_{R}^{\text{diss}}\left(K_{\lambda}\right)$ and $f_{\theta}^{d i s s}\left(L_{\lambda}\right)$ are the mean values of radial and tangential components of the dissipative part of the dynamical force expressed in terms of fluctuating radial and angular momenta, respectively. We assume that the amplitude fluctuations are sufficiently small, so that we can linearize the Langevin equations~\eqref{eq85} and~\eqref{eq86} around their mean values to give,
\begin{align}
\begin{split}
\frac{\partial}{\partial t} \delta K_{\lambda}(t)-\frac{2 L}{\mu R^{3}} \delta L_{\lambda}(t)=& \left(\frac{\partial f_{R}}{\partial K}\right) \delta K_{\lambda}(t) \\ & +\delta f_{R}^{\lambda}(t), \label{eq87}
\end{split}
\end{align}
\begin{align}
\frac{\partial}{\partial t} \delta L_{\lambda}(t)=\left(\frac{\partial f_{\theta}}{\partial L}\right) \delta L_{\lambda}(t)+R(t) \delta f_{\theta}^{\lambda}(t). \label{eq88}
\end{align}

The fluctuating forces arising from the potential energy terms make only a minor contribution to the fluctuations of the relative momentum. In our treatment, we disregard these forces, along with the force associated with the window’s motion represented by the first terms on the right-hand side of Eqs.~\eqref{eq85} and \eqref{eq86}. We also neglect fluctuations in the reduced mass and in the relative separation of the fragment centers. The quantities $\mu(t)$, $R(t)$, and $L(t)$ denote the mean values of the reduced mass, relative distance, and orbital angular momentum of the colliding system, respectively, as determined by the TDHF equations. The derivatives of the dissipative forces on the right-hand side of these equations define the reduced radial and tangential friction coefficients,

\begin{align}
\begin{split}
\left(\frac{\partial f_{R}^{d i s s}}{\partial K}\right)=-\gamma_{R}(t)  \\ \left(\frac{\partial f_{\theta}^{d i s s}}{\partial L}\right)=-\gamma_{\theta}(t). \label{eq89}
\end{split}
\end{align}
Multiplying both sides of Eq.~\eqref{eq87} and Eq.~\eqref{eq88} by $\delta K_{\lambda}(t)$ and $\delta L_{\lambda}(t)$, and taking the ensemble averaging, it is possible to deduce a set of coupled differential equations for the variances ~\cite{schroder1981,merchant1982},
\begin{align}
\begin{split}
& \frac{\partial}{\partial t} \sigma_{K K}^{2}(t)-\frac{4 L}{\mu R^{3}} \sigma_{K L}^{2}(t)=-2 \gamma_{R}(t) \sigma_{K K}^{2}(t)+2 D_{R R}(t) ,\label{eq90}
\end{split}
\end{align}
\begin{align}
\begin{split}
\frac{\partial}{\partial t} \sigma_{L L}^{2}(t)=-2 \gamma_{\theta}(t) \sigma_{L L}^{2}(t)+2 R^{2}(t) D_{L L}(t) ,\label{eq91}
\end{split}
\end{align}
\begin{align}
\begin{split}
\frac{\partial}{\partial t} \sigma_{K L}^{2}(t)-\frac{2 L}{\mu R^{3}} \sigma_{L L}^{2}(t)=& -2\left(\gamma_{R}(t) +\gamma_{\theta}(t)\right) \sigma_{K L}^{2}(t) \\
& + R\left(D_{K L}(t)+D_{L K}(t)\right) .\label{eq92}
\end{split}
\end{align}
Here, variances are defined as $\sigma_{K K}^{2}(t, l)=\overline{\delta K_{\lambda}(t) \delta K_{\lambda}(t)}$, $\sigma_{L L}^{2}(t, l)=\overline{\delta L_{\lambda}(t) \delta L_{\lambda}(t)}$, and $\sigma_{K L}^{2}(t, l)=\overline{\delta K_{\lambda}(t) \delta L_{\lambda}(t)}$ for each initial angular momentum $l$. The bar indicates the average values taken over the ensemble generated in the SMF approach and $D_{\alpha \beta}(t)$ denotes the momentum diffusion coefficients, which we discuss in the next section. We note that we use the same notation $(\alpha, \beta)$ to indicate the radial and the tangential directions $(R, \theta)$ and the radial and angular momenta $(K, L)$.

\subsection{Momentum diffusion coefficients}\label{sec5b}
Momentum diffusion coefficients for the radial and the angular momenta are defined as the time integral over the history of autocorrelation functions of the stochastic forces,
\begin{align}
D_{\alpha \beta}(t)=\int_{0}^{t} d t^{\prime} \overline{\delta f_{\alpha}^{\lambda}(t) \delta f_{\beta}^{\lambda}\left(t^{\prime}\right)}. \label{eq93}
\end{align}
Where $\delta f_{\alpha}^{\lambda}(t)$ is the stochastic part of the dynamical force $f_{\alpha}^{\lambda}(t)$, $\delta f_{\alpha}^{\lambda}(t)=\sum_{i j} Y_{j i}^{\alpha}(t) \delta \rho_{j i}^{\lambda}$. In this expression $Y_{j i}^{\alpha}(\mathrm{t})$ is given in Eq.~\eqref{eq82}, but it is determined in the mean-field approximation. Using the basic postulate of the SMF approach we can evaluate the ensemble averages, and the correlation functions of the random force on the radial and the tangential direction becomes,
\begin{align}
\begin{split}
\overline{\delta f_{\alpha}^{\lambda}(t) \delta f_{\beta}^{\lambda}\left(t^{\prime}\right)}=\operatorname{Re}\left(\sum_{p \in P, h \in T} Y_{h p}^{\alpha}(t) Y_{h p}^{* \beta}\left(t^{\prime}\right)\right. \\
\left.+\sum_{p \in T, h \in P} Y_{h p}^{\alpha}(t) Y_{h p}^{* \beta}\left(t^{\prime}\right)\right)
.\label{eq94}
\end{split}
\end{align}
In this expression, in the first term the summations run
over the particle states originating from projectile $p \in P$,
and the hole states originating from target $h \in T$, and in
the second terms the summations run in the reverse. By
adding and subtracting the hole-hole terms, the first term
in this expression can be written as,
\begin{align}
\begin{split}
\sum_{p \in P, h \in T} Y_{h p}^{\alpha}(t) Y_{h p}^{* \beta}\left(t^{\prime}\right)= & \sum_{h \in T, a \in P} Y_{h a}^{\alpha}(t) Y_{h a}^{* \beta}\left(t^{\prime}\right) \\
& -\sum_{h \in T, h^{\prime} \in P} Y_{h h^{\prime}}^{\alpha}(t) Y_{h h^{\prime}}^{* \beta}\left(t^{\prime}\right) .\label{eq95}
\end{split}
\end{align}
In the first term, the summation $a \in P$ runs over the complete set of states originating from the projectile.
We introduce a similar subtraction in the second term of Eq.~\eqref{eq94}.
As shown in Appendix~\ref{sec:appC},, using the closure relation in a
diabatic approximation of the TDHF wave functions, it is
possible to eliminate the complete set of projectile states
in the first term and the complete set of target states in
the second term. As a result, the radial, the tangential and
the mixed diffusion coefficients are given by the following
compact expression,
\begin{align}
\begin{split}
& D_{\alpha \beta}(t)=\int_{0}^{t} d \tau \int d^{3} r \tilde{g}\left(x^{\prime}\right)\left[G_{T}(\tau) J_{\alpha \beta}^{T}(\boldsymbol{r}, \bar{t})+\right. \\
& \left.G_{P}(\tau) J_{\alpha \beta}^{P}(\boldsymbol{r}, \bar{t})\right]-\int_{0}^{t} d \tau \operatorname{Re}\left(\sum_{h \in T, h^{\prime} \in P} Y_{h h^{\prime}}^{\alpha}(t) Y_{h h^{\prime}}^{* \beta}(t-\tau)\right. \\
& \left.+\sum_{h \in P, h^{\prime} \in T} Y_{h h^{\prime}}^{\alpha}(t) Y_{h h^{\prime}}^{* \beta}(t-\tau)\right) .\label{eq96}
\end{split}
\end{align}
In the first line, the quantity $\mathcal{J}_{\alpha \beta}^{T}(\boldsymbol{r}, t-\tau / 2)$ is given by
\begin{align}
\begin{split}
\mathcal{J}_{\alpha \beta}^{T}(\boldsymbol{r}, \bar{t})=& \frac{\hbar}{m} \sum_{h \in T}\left[m u_{\alpha}^{h}(\boldsymbol{r}, \bar{t})\right]\left[m u_{\beta}^{h}(\stackrel{r}{r}, \bar{t})\right] \\ & \times \left|\operatorname{Im}\left(\phi_{h}^{*}(\boldsymbol{r}, \bar{t}) \boldsymbol{\hat{e}}_{R} \cdot \boldsymbol{\nabla} \phi_{h}(\boldsymbol{r}, \bar{t})\right)\right| ,\label{eq97}
\end{split}
\end{align}
where $\bar{t}=\left(t+t^{\prime}\right) / 2=t-\tau / 2$. This expression represents the magnitude of the nucleon flux that carries the product of the momentum components $m u_{\alpha}^{h}(\bar{t})$ and $m u_{\beta}^{h}(\boldsymbol{r}, \bar{t})$ from the target-like fragment in perpendicular and tangential directions to the window plane. The quantity $\mathcal{J}_{\alpha \beta}^{P}(\boldsymbol{r}, t-\tau / 2)$ is given by a similar expression and represents the magnitude of the nucleon flux that carries the product of the momentum components $m u_{\alpha}^{h}(\boldsymbol{r}, \bar{t})$ and $m u_{\beta}^{h}(\boldsymbol{r}, \bar{t})$ from the projectile-like fragment in the perpendicular and tangent directions to the window plane. The radial and tangent component of the nucleon flow velocities are determined from,
\begin{align}
\begin{split}
u_{\alpha}^{h}(\boldsymbol{r}, \bar{t})=\frac{\hbar}{m} \frac{1}{\left|\phi_{h}(\boldsymbol{r}, \bar{t})\right|^{2}} \operatorname{Im}\left(\phi_{h}^{* \alpha}(\boldsymbol{r}, \bar{t}) \boldsymbol{\hat{e}}_{\alpha} \cdot \boldsymbol{\nabla} \phi_{h}^{\alpha}(\boldsymbol{r}, \bar{t})\right) .\label{eq98}  
\end{split}
\end{align}

We observe that there is a close analogy between the quantal expression and the classical diffusion coefficient in a random walk problem. The first line in the quantal expression gives the sum of the nucleon flux across the window from the target-like fragment to the projectile-like fragment and from the projectile-like fragment to the target-like fragment, which is integrated over the memory. Each nucleon transfer across the window in both directions carries the product of the momentum components, which increases the rate of change of the momentum dispersion. This is analogous to the random walk problem, in which the diffusion coefficient is given by the sum of the rate for the forward and backward steps. The second line in the quantal diffusion expression represents the Pauli blocking effects in nucleon transfer mechanism, which do not have a classical counterpart. The quantities in the Pauli blocking factors are determined by hole-hole elements of the matrix $Y_{h h^{\prime}}^{\alpha}(t)$ and $Y_{h h^{\prime}}^{*}(t)$, which are defined in Eq.~\eqref{eq82}.

\subsection{Total kinetic energy distribution}\label{sec5c}
It is possible to determine the joint probability distribution function $P_{\ell}(K, L)$ of the radial linear momentum $K$ and the orbital angular momentum $L$ using the coupled Langevin equations. It is well known that these coupled Langevin equations are equivalent to the Fokker-Planck description for the joint probability distribution $P_{l}(K, L)$ ~\cite{risken1996}. When the radial and tangential friction forces have linear dependence on the radial and the angular momentum, the solution of the joint probability distribution can be expressed as a correlated Gaussian function,

\begin{align}
\begin{split}
P_{l}(K, L)=\frac{1}{2 \pi \sigma_{K K}(\ell) \sigma_{L L}(\ell) \sqrt{1-\eta_{l}^{2}}} \exp \left[-C_{l}(K, L)\right] .\label{eq99}
\end{split}
\end{align}
Here, the exponent $C_{l}(K, L)$ for each impact parameter is given by
\begin{align}
\begin{split}
C_{l}(K, L)=\frac{1}{2\left(1-\eta_{l}^{2}\right)} & \left[\left(\frac{K-K_{l}}{\sigma_{K K}(\ell)}\right)^{2} +\left(\frac{L-L_{l}}{\sigma_{L L}(\ell)}\right)^{2}  \right. \\ & \left. -2 \eta_{l}\left(\frac{K-K_{l}}{\sigma_{K K}(\ell)}\right)\left(\frac{L-L_{l}}{\sigma_{L L}(\ell)}\right) \right] .\label{eq100}
\end{split}
\end{align}
The quantities $\sigma_{K K}(\ell)$ and $\sigma_{L L}(\ell)$ are the dispersions of the radial momentum and angular momentum distributions for each initial angular momentum and the correlation factor, $\eta_{l}=\sigma_{K L}^{2}(\ell) / \sigma_{K K}(\ell) \sigma_{L L}(\ell)$, is determined by the ratio of the mixed variance $\sigma_{K L}^{2}(\ell)$ to the product of dispersions of the radial momentum and the angular momentum.

The mean values of the radial and angular momentum for each value of initial angular momentum are indicated by $K_{\ell}$ and $L_{\ell}$, which are determined by the TDHF calculations. In terms of radial momentum $K$ and orbital angular momentum $L$, the asymptotic total kinetic energy is given by $E_{k i n}^{\infty}=K^{2} / 2 \mu+L^{2} / 2 \mu R_{f}^{2}+Z_{1} Z_{2} e^{2} / R_{f}$. For a given initial angular momentum $\ell$, we define the TKE distribution $G_{\ell}(E)$ as

\begin{align}
\begin{split}
G_{l}(E)=\int d K d L \delta\left(E-E_{k i n}^{\infty}(K, L)\right) P_{l}(K, L). \label{eq101}    
\end{split}
\end{align}
Note that $K$ and $L$ in this expression correspond to the radial and the angular momentum at the instant $t_{f}$, respectively, and $K$ stands for the asymptotic TKE. It is important to mention that $\mu$ and $Z_{1,2}$ are, in general, $\ell$ dependent quantities, and fluctuations in the mass and charge asymmetries may affect the TKE fluctuations. However, we neglect the effects of mass and charge fluctuations on the TKE distribution and retain the mean values of the mass and charge asymmetry for each angular momentum.

In practice, the mixed diffusion coefficients $D_{K L}(t)$ and $D_{L K}(t)$ are expected to be much smaller than the radial and the angular momentum diffusion coefficients, $D_{K K}(t)$ and $D_{L L}(t)$. Hence, in the present work, we neglect the mixed term $\sigma_{K L}(t)$ in Eq.~\eqref{eq92}. In such a case, the expression can be greatly simplified by taking the asymptotic limit, $R \rightarrow \infty$, leading to 
\begin{align}
G_{\ell}(E)=\int d K^{\infty} \delta\left(E-E_{k i n}^{\infty}\left(K^{\infty}\right)\right) P_{\ell}\left(K^{\infty}\right) ,\label{eq102}    
\end{align}
where $E_{\text {kin}}^{\infty}(K)=K^{2} / 2 \mu$ and $P_{\ell}(K)$ is the probability distribution of the radial momentum. Notice that by taking the limit $R \rightarrow \infty$, the centrifugal part of the kinetic energy and Coulomb energy transformed into the radial TKE and $K^{\infty}$ in Eq.~\eqref{eq102} corresponds to the asymptotic value of the radial momentum for $R \rightarrow \infty$. After taking the integral over the angular momentum variable, the asymptotic radial momentum distribution becomes a simple Gaussian
\begin{align}
P_{\ell}(K)=\frac{1}{\sqrt{2 \pi}} \frac{1}{\sigma_{K K}(\ell)} \exp \left[-\frac{1}{2}\left(\frac{K-K(\ell)}{\sigma_{K K}(\ell)}\right)^{2}\right] ,\label{eq103}
\end{align}
where the mean value of the mean value of the asymptotic radial momentum is related to the mean asymptotic total kinetic energy from TDHF, $E_{\text {kin}}^{\infty}(\ell)$ by, $K_{\ell}^{\infty}=\left(2 \mu E_{k i n}^{\infty}(\ell)\right)^{1 / 2}$. After trivial integration, we obtain the asymptotic TKE distribution,
\begin{align}
G_{\ell}(E)=\frac{1}{\sqrt{8 \pi E}} \frac{1}{2 \tilde{\sigma}_{K K}(\ell)} \exp \left[-\frac{1}{2}\left(\frac{\sqrt{E}-\sqrt{E_{k i n}^{\infty}(\ell)}}{\tilde{\sigma}_{K K}(\ell)}\right)^{2}\right] \label{eq104}
\end{align}
where, $\tilde{\sigma}_{K K}(\ell)=\frac{1}{\sqrt{2 \mu}} \sigma_{K K}(\ell)$. To obtain the radial dispersion $\sigma_{K K}(t)$, we solve the quantal diffusion equation for the radial component;
\begin{align}
\frac{d}{d t} \sigma_{K K}^{2}(t)=-2 \gamma_{R}(t) \sigma_{K K}^{2}(t)+2 D_{K K}(t) .\label{eq105}
\end{align}
We note that the unit of the TKE distribution, $G_{\ell}(E)$, is $1 /$MeV, hence the fraction of events with final TKE in the energy range $\Delta E$ in MeV is given by $G_{\ell}(E)  \Delta E$.

\subsection{Results for ${ }^{136}\mathrm{Xe}+{ }^{208}\mathrm{P b}$ Collisions}\label{sec5d}
In this section, as the first application of the proposed formalism given in the preceding sections, we present calculations of the TKE distribution for the ${ }^{136} \mathrm{Xe}+{ }^{208} \mathrm{Pb}$ reaction system at the $E_\mathrm{c.m.}=526$~MeV, for which extensive experimental data, reported by Kozulin \textit{et al.}~\cite{kozulin2012}, are available. TDHF calculations were performed for a range of initial orbital angular momentum $L$. Table~\ref{tab:table1} shows the results of TDHF calculations for a set of observables in ${ }^{136} \mathrm{Xe}+{ }^{208} \mathrm{Pb}$ reaction the $E_\mathrm{c.m.}=526$~MeV. We mention here that for the ${ }^{136} \mathrm{Xe}+{ }^{208} \mathrm{Pb}$ system, the average number of transferred nucleons are small, reflecting a small charge asymmetry and possible shell effects in the reactants. Nucleons are, however, actively exchanged during the collision, which is the source of dissipation and fluctuations of the observables, such as mass, charge, TKE, and scattering angles, in low-energy heavy-ion reactions. For this reaction, mean TKE reaches around 175~MeV for small angular momenta, while the contact time is rather short ($<2 \mathrm{zs}$). Because of the short contact time the di-nuclear system does not rotate much in the reaction plane. We note that fragments are emitted outside the experimental coverage range (25$^{\circ}-70^{\circ}$ in the laboratory frame) in events below $\ell<100 \hbar$ in TDHF calculations.

To evaluate TKE distributions, three-dimensional TDHF code of K. Sekizawa was further extended to incorporate the SMF approach~\cite{ayik2020b}, and applied to various systems~\cite{williams2018,sekizawa2014,sonika2015,sekizawa2017}. For computational details we refer the readers to article~\cite{ayik2020b}. To obtain the kinetic energy distribution, Eq.~\eqref{eq104}, we need to evaluate the asymptotic value of the radial momentum dispersion, $\sigma_{K K}(t)$, for each value of the initial angular momentum $\ell$ by solving Eq.~\eqref{eq105}. The radial momentum diffusion coefficients, $D_{K K}(t)$, is directly computed from occupied single-particle orbitals within the TDHF approach with the quantal expression given by Eq.~\eqref{eq95}. However, it is not trivial to determine the radial friction coefficient directly from TDHF. Nevertheless, using the analogy to the random walk problem, it is possible to extract from TDHF an approximate expression for the radial friction force and the radial friction coefficient. Details of this analysis are given in the Appendix~\ref{sec:appD}.

\begin{table*}
\begin{center}
\caption{A list of numerical results of the TDHF calculations for the ${ }^{136} \mathrm{Xe}+{ }^{208} \mathrm{Pb}$ reaction at $E_\mathrm{c.m.}=526$~MeV : From left to right columns, the initial angular momentum $\ell$ in $\hbar$, the final average relative angular momentum $\ell_{f}$ in $\hbar$, neutron and proton numbers of projectile-like (target-like) fragments $N_{1}$ and $Z_{1}\left(N_{2}\right.$ and $Z_{2}$ ), mean total kinetic energy loss (TKEL) in MeV, contact time in $t_{\text {contact }}$ in $\mathrm{fm} / \mathrm{c}$, scattering angles in center-of-mass frame, $\theta_\mathrm{c.m.}$, and those in laboratory frame for project-like (target-like) fragments $\theta_{1}^{l a b}\left(\theta_{2}^{l a b}\right)$ in degrees, The contact time is defined as duration time in which the minimum density between two fragments exceed half of the saturation density, $\rho_{\text {sat}} / 2=0.08\, \mathrm{fm}^{-3}$}
\begin{tabular}{c|c|c|c|c|c|c|c|c|c|c}
\hline
$\ell(\hbar)$ & $\ell_{f} (\hbar)$ & $N_{1}$ & $Z_{1}$ & $N_{2}$ & $Z_{2}$ & TKEL (MeV) & $t_{\text {contact }}$  (fm/c) & $\theta_{\text {c.m.}}$ (deg) & $\vartheta_{1}^{\text {lab }}$ (deg) & $\vartheta_{2}^{\text {lab}}$ (deg) \\
\hline
0 & 0 & 82.3 & 55.1 & 125.0 & 80.9 & 173.1 & 661.4 & 180.0 & 180.0 & 0.0 \\
\hline
50 & 39 & 82.4 & 55.1 & 124.9 & 80.8 & 177.0 & 646.0 & 150.1 & 96.4 & 13.5 \\
\hline
100 & 78 & 82.0 & 54.5 & 125.4 & 81.5 & 175.6 & 611.8 & 123.1 & 72.9 & 25.4 \\
\hline
110 & 85 & 81.9 & 54.3 & 125.5 & 81.6 & 175.1 & 593.2 & 118.2 & 69.5 & 27.5 \\
\hline
120 & 95 & 81.8 & 54.3 & 125.6 & 81.7 & 175.0 & 591.4 & 113.3 & 66.2 & 29.6 \\
\hline
130 & 104 & 81.7 & 54.3 & 125.7 & 81.7 & 175.3 & 586.6 & 108.6 & 63.0 & 31.6 \\
\hline
140 & 114 & 81.9 & 54.4 & 125.5 & 81.5 & 175.0 & 571.0 & 104.0 & 60.0 & 33.6 \\
\hline
150 & 124 & 82.4 & 54.7 & 125.0 & 81.2 & 174.3 & 555.2 & 99.7 & 57.1 & 35.6 \\
\hline
160 & 134 & 83.1 & 55.2 & 124.3 & 80.7 & 171.9 & 538.6 & 95.9 & 54.6 & 37.4 \\
\hline
170 & 142 & 83.7 & 55.6 & 123.7 & 80.3 & 169.4 & 529.4 & 92.5 & 52.5 & 39.0 \\
\hline
180 & 149 & 83.9 & 55.8 & 123.5 & 80.2 & 168.5 & 517.6 & 89.7 & 50.7 & 40.3 \\
\hline
190 & 158 & 83.6 & 55.6 & 123.9 & 80.3 & 167.4 & 474.0 & 87.1 & 49.3 & 41.3 \\
\hline
200 & 166 & 83.0 & 55.4 & 124.5 & 80.6 & 166.0 & 462.4 & 85.2 & 48.4 & 42.0 \\
\hline
210 & 173 & 82.7 & 55.2 & 124.9 & 80.7 & 161.0 & 440.2 & 84.1 & 47.9 & 42.6 \\
\hline
220 & 179 & 82.3 & 55.0 & 125.3 & 80.9 & 154.9 & 409.4 & 83.3 & 47.8 & 43.1 \\
\hline
230 & 185 & 81.8 & 54.8 & 125.8 & 81.2 & 149.1 & 378.4 & 82.5 & 47.6 & 43.5 \\
\hline
240 & 194 & 81.5 & 54.6 & 126.2 & 81.3 & 140.7 & 343.6 & 81.7 & 47.4 & 44.1 \\
\hline
250 & 203 & 81.3 & 54.5 & 126.4 & 81.5 & 129.8 & 303.2 & 81.0 & 47.4 & 44.8 \\
\hline
260 & 214 & 81.2 & 54.4 & 126.6 & 81.5 & 117.1 & 257.2 & 80.5 & 47.5 & 45.5 \\
\hline
270 & 225 & 81.2 & 54.4 & 126.6 & 81.5 & 103.5 & 226.0 & 80.0 & 47.5 & 46.2 \\
\hline
280 & 240 & 81.4 & 54.5 & 126.5 & 81.5 & 86.0 & 192.4 & 79.5 & 47.6 & 47.2 \\
\hline
290 & 258 & 81.5 & 54.5 & 126.4 & 81.5 & 66.8 & 146.8 & 79.0 & 47.8 & 48.2 \\
\hline
300 & 277 & 81.5 & 54.4 & 126.4 & 81.6 & 48.9 & 100.8 & 78.5 & 47.9 & 49.0 \\
\hline
310 & 295 & 81.6 & 54.3 & 126.4 & 81.7 & 32.7 & 44.2 & 78.1 & 48.0 & 49.8 \\
\hline
320 & 311 & 81.7 & 54.2 & 126.2 & 81.8 & 16.9 & 0.0 & 77.9 & 48.2 & 50.4 \\
\hline
330 & 325 & 81.9 & 54.1 & 126.1 & 81.9 & 8.0 & 0.0 & 77.4 & 48.1 & 51.0 \\
\hline
340 & 337 & 81.9 & 54.0 & 126.1 & 82.0 & 4.7 & 0.0 & 76.4 & 47.5 & 51.6 \\
\hline
350 & 347 & 81.9 & 54.0 & 126.1 & 82.0 & 3.3 & 0.0 & 75.2 & 46.7 & 52.3 \\
\hline
    \end{tabular}
    \label{tab:table1}
\end{center}
\end{table*}

\begin{figure}[!htb]
\includegraphics*[width=0.48\textwidth]{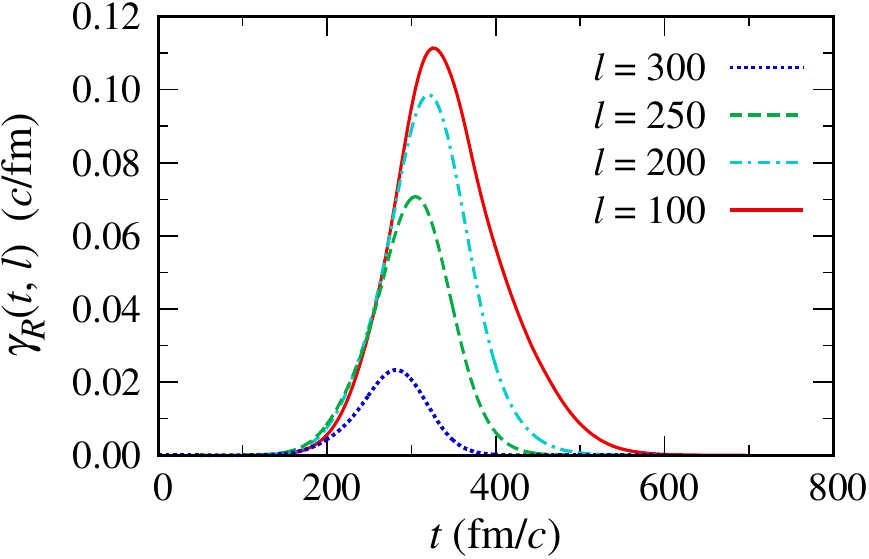}
\caption{Reduced friction coefficients for the ${ }^{136} \mathrm{Xe}+{ }^{208}\mathrm{Pb}$ reaction at $E_\mathrm{c.m.}=526$~MeV with initial orbital angular momenta $\ell=100,200,250$, and 300 (in units of $\hbar$) are shown as functions of time.}
\label{fig:5.1}
\end{figure}

\begin{figure}[!htb]
\includegraphics*[width=0.48\textwidth]{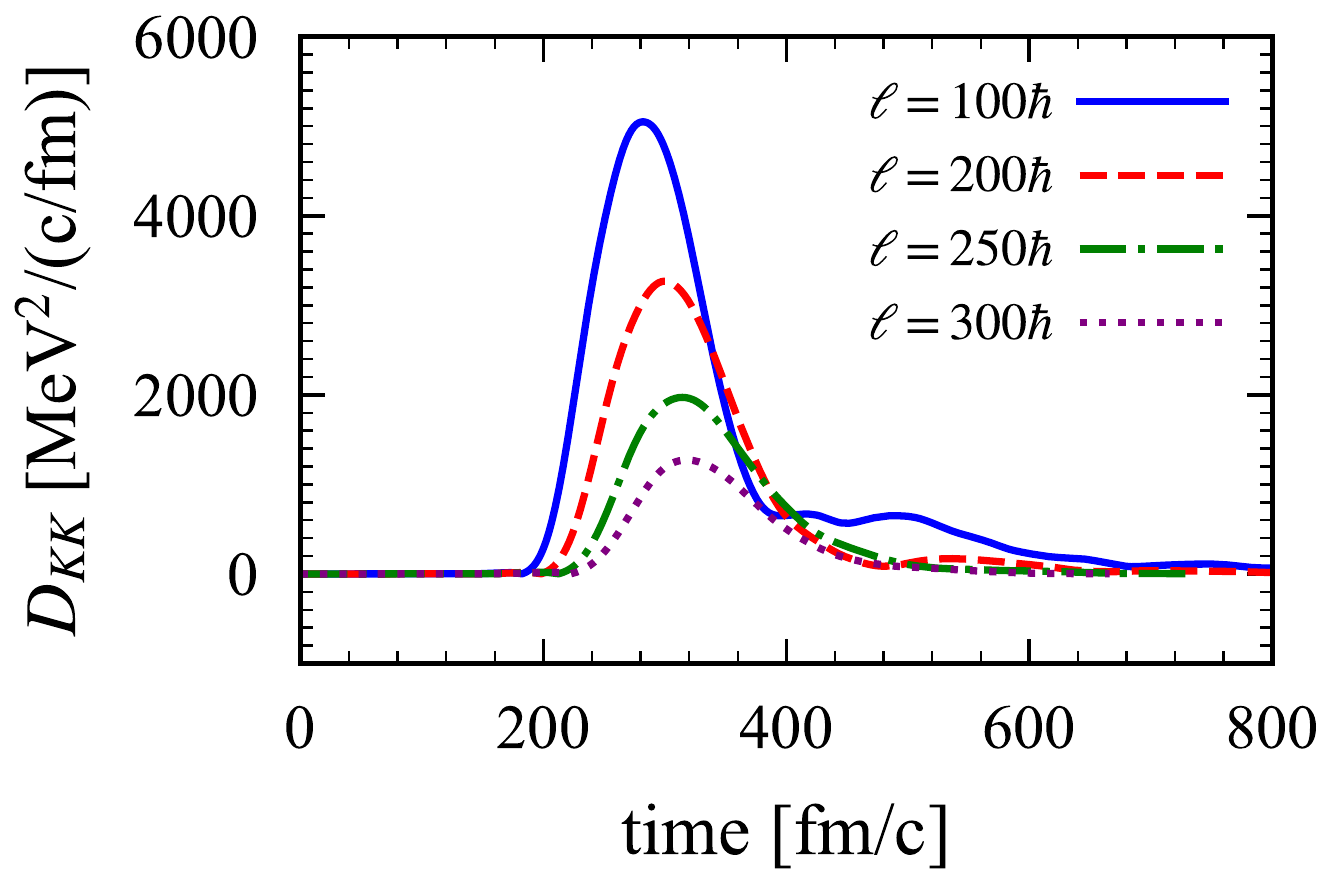}
\caption{Radial momentum diffusion coefficients for the ${ }^{136} \mathrm{Xe}+{ }^{208} \mathrm{Pb}$ reaction at $E_\mathrm{c.m.}=526$~MeV with initial orbital angular momenta $\ell=100,200,250$ and 300 (in units of $\hbar$) are shown as functions of time.}
\label{fig:5.2}
\end{figure}

\begin{figure}[!htb]
\includegraphics*[width=0.48\textwidth]{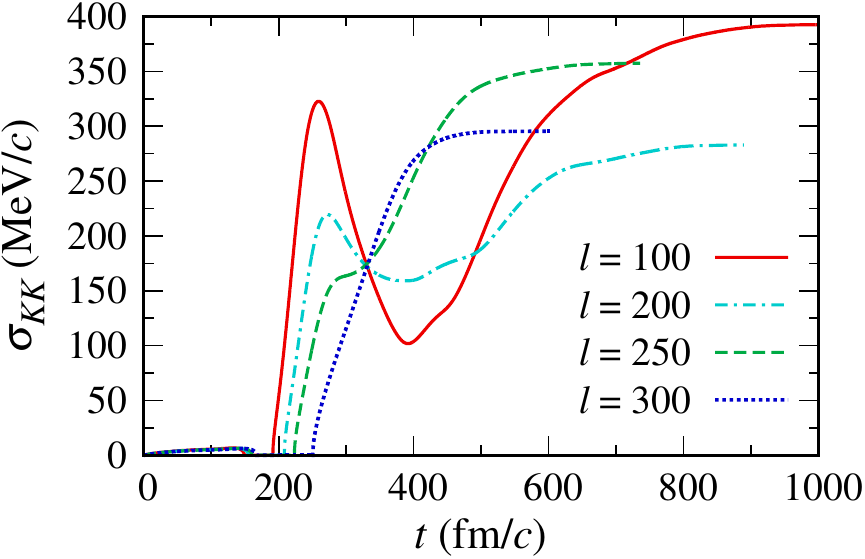}
\caption{Variances of the radial momentum for the ${ }^{136} \mathrm{Xe}+{ }^{208} \mathrm{Pb}$ reaction at $E_\mathrm{c.m.}=526$~MeV with initial orbital angular momenta $\ell=100,200,250$ and 300 (in units of $\hbar$) are shown as functions of time.}
\label{fig:5.3}
\end{figure}

In Figs.~\ref{fig:5.1}--\ref{fig:5.3}, we show examples of computational results for the collisions of the ${ }^{136} \mathrm{Xe}+{ }^{208} \mathrm{Pb}$ reaction at $E_\mathrm{c.m.}=526$~MeV, for four typical angular momenta $\ell$ [($=100$ solid line $)$, 200 (dash-dotted line), 250 (dashed line), and 300 (dotted line) in units of $\hbar$ ], as a function of time. Fig.~\ref{fig:5.1} shows the reduced radial friction coefficient $\gamma_{R}(t,\ell)$ given by Eq.~\eqref{eq89}, which is extracted from TDHF by employing the method explained in detail in Appendix~\ref{sec:appC}. We observe that the radial friction coefficient develops when two nuclei collide at around $t=(200-400) \mathrm{fm} /c$. The magnitude of the friction coefficient increases with decreasing initial angular momentum $\ell$, for which contact times are longer, indicating that larger amount of the relative kinetic energy is converted into internal excitations at smaller orbital angular momenta, as expected. In Fig.~\ref{fig:5.2}, we show the quantal radial momentum diffusion coefficient, $D_{K K}(t)$ given by Eq.~\eqref{eq96}, which is calculated microscopically based on occupied single-particle orbitals within the TDHF approach. The magnitude of the diffusion coefficient increases with decreasing initial angular momentum $\ell$. From these results, we learn that the diffusion coefficient has a relatively long tail when compared to the friction coefficient shown in Fig.~\ref{fig:5.1}. This is related to the fact that the quantal diffusion coefficient is governed by nucleon exchange, which lasts even after the turning point through the neck structure of the di-nuclear system (cf. contact times are shown in Table~\ref{tab:table1}).

Having the radial friction and momentum diffusion coefficients, $\gamma_{R}(t, \ell)$ and $D_{K K}(t)$, we solve the differential equations, and the results are shown in Fig.~\ref{fig:5.3}. From the figure, we see that variance of radial momentum $\sigma_{K K}(t)$ show somewhat of a complicated behavior as a function of time. We notice that the asymptotic value of $\sigma_{K K}(t)$ is largest for $\ell=100 \hbar$, and decreases with increasing $\ell$ values from $100 \hbar$ to $200 \hbar$, then increases $\ell=250 \hbar$ and then decreases again for $\ell=300 \hbar$. This behavior can also be found in Table~\ref{tab:table2}, in which the asymptotic values of the radial momentum and TKE dispersions are presented for a range of initial angular momenta. We consider that in the present analysis the radial momentum dispersion is over predicted for
relatively large initial angular momentum region $(\ell=200-300) \hbar$, which is probably due to the approximate treatment of the radial friction coefficient. Since the primary purpose of the present work is to put the first step towards the microscopic description of the TKE distribution, developing a formalism based on the SMF approach, we leave further improvements of the description as further works.

\begin{table}
    \centering
    \caption{A list of numerical results of the SMF calculations for the ${ }^{136} \mathrm{Xe}+{ }^{208} \mathrm{Pb}$ reaction at $E_\mathrm{c.m.}=526$~MeV for a range of initial orbital angular momenta $\ell$. From left to right columns, it shows the asymptotic values of the radial momentum dispersions, $\sigma_{K K}$ in $\mathrm{MeV} / \mathrm{c}$, the modified radial momentum dispersions, $\quad \tilde{\sigma}_{K K}=\sigma_{K K} / \sqrt{\mu}$ in MeV , and the dispersion of total kinetic energy (TKE), $\sigma_{T K E} \approx 2 \tilde{\sigma}_{K K} \sqrt{E_{\text {kin }}^{\infty}}$ in MeV.}
    \begin{tabular}{c|c|c|c|}
\hline
$l(\hbar)$ & $\sigma_{K K}(\mathrm{MeV} / c)$ & $\widetilde{\sigma}_{K K}\left(\mathrm{MeV}^{1 / 2}\right)$ & $\sigma_{\text {TKE }}(\mathrm{MeV})$ \\
\hline
0 & 553.7 & 1.406 & 52.84 \\
\hline
50 & 497.2 & 1.263 & 47.18 \\
\hline
100 & 392.7 & 0.999 & 37.38 \\
\hline
110 & 367.3 & 0.934 & 35.01 \\
\hline
120 & 342.4 & 0.871 & 32.64 \\
\hline
130 & 319.4 & 0.813 & 30.44 \\
\hline
140 & 298.3 & 0.759 & 28.43 \\
\hline
150 & 277.9 & 0.706 & 26.48 \\
\hline
160 & 268.7 & 0.681 & 25.65 \\
\hline
170 & 266.0 & 0.674 & 25.45 \\
\hline
180 & 261.0 & 0.661 & 25.00 \\
\hline
190 & 265.4 & 0.672 & 25.47 \\
\hline
200 & 282.9 & 0.718 & 27.23 \\
\hline
210 & 302.8 & 0.769 & 29.37 \\
\hline
220 & 317.0 & 0.805 & 31.03 \\
\hline
230 & 329.7 & 0.838 & 32.55 \\
\hline
240 & 344.2 & 0.876 & 34.38 \\
\hline
250 & 357.5 & 0.910 & 36.22 \\
\hline
260 & 366.6 & 0.933 & 37.75 \\
\hline
270 & 365.5 & 0.930 & 38.25 \\
\hline
280 & 352.0 & 0.896 & 37.58 \\
\hline
290 & 330.3 & 0.840 & 36.02 \\
\hline
300 & 295.4 & 0.752 & 32.84 \\
\hline
310 & 233.6 & 0.595 & 26.41 \\
\hline
320 & 152.0 & 0.387 & 17.46 \\
\hline
330 & 85.8 & 0.218 & 9.94 \\
\hline
340 & 34.5 & 0.088 & 4.01 \\
\hline
350 & 36.2 & 0.092 & 4.21 \\
\hline
\end{tabular}      
    \label{tab:table2}
\end{table}

\begin{figure}[!htb]
\includegraphics*[width=0.48\textwidth]{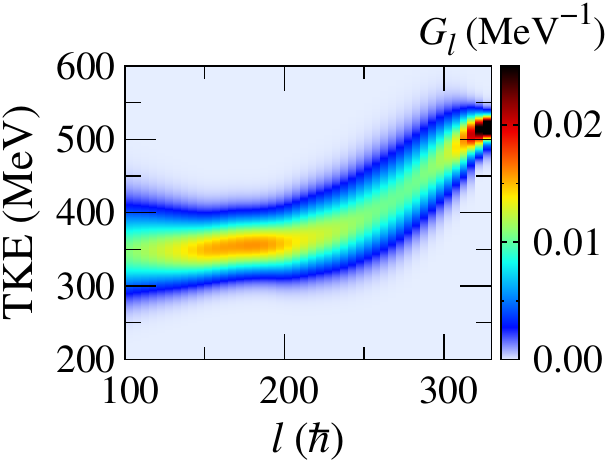}
\caption{The total kinetic energy (TKE) distribution, $G_{\ell}(E)$ defined by Eq.~\eqref{eq103}, is shown in the $\ell$-TKE plane, where $\ell$ represents the initial orbital angular momentum, calculated for ${ }^{136} \mathrm{Xe}+{ }^{208} \mathrm{Pb}$ reaction at $E_{c.m.}=526$~MeV.}
\label{fig:5.4}
\end{figure}

Employing the expression of Eq.~\eqref{eq104}, we can obtain the TKE distribution, and the results are shown in Fig.~\ref{fig:5.4}. Figure~\ref{fig:5.4} illustrates the TKE distribution for the range of initial angular momenta $\ell=(100-300) \hbar$ in the $\ell$-TKE plane. We note that with the TKE distribution, $G_{\ell}$, we can evaluate the mean value of TKE as
\begin{align}
\begin{split}
\overline{T K E(\ell)}=\int d E E G_{\ell}(E) \approx E_{k i n}^{\infty}(\ell)+\tilde{\sigma}_{K K}^{2}(\ell) .\label{eq106}
\end{split}
\end{align}
From Table~\ref{tab:table2}, we see that the asymptotic value of the largest dispersion occurs at $\ell=0$. In this case we find, $\tilde{\sigma}_{K K}^{2}(\ell)=\sigma_{K K}^{2}(\ell) / 2 \mu \approx 1.98 \mathrm{MeV}$, which is much smaller than $E_{\text{kin}}^{\infty}(\ell) \approx 353 \mathrm{MeV}$, confirming the correspondence with the TKE from TDHF. We can also calculate the variance of TKE for each value of angular momentum as,
\begin{align}
\begin{split}
\sigma_{T K E}^{2}(\ell) & =\int d E\left(E-E_{k i n}^{\infty}(\ell)\right)^{2} G_{\ell}(E) \\ & \approx 4 \tilde{\sigma}_{K K}^{2}(\ell) E_{k i n}^{\infty}(\ell) .\label{eq107}
\end{split}
\end{align}
Dispersion of TKE grows linearly with square root of the mean value $\sigma_{T K E}=\tilde{\sigma}_{K K}^{2} \sqrt{E_{\text {kin }}^{\infty}}$. For example for the initial angular momentum, dispersion as large as $\sigma_{T K E}=53$~MeV. This indicates the total excitation energy of the primary fragments have quite large dispersion values. Large values of dispersions of the excitation energy may have an important effect on the de-excitation processes of the primary fragments.

Finally, to make a comparison with experimental data ~\cite{kozulin2012}, we evaluate the yield of the reaction outcomes as a function of the total kinetic energy loss (TKEL, i.e., $E_{\text {c.m.}}-E_{\text {kin}}^{\infty}$ ) by summing up contributions from each initial orbital angular momentum,
\begin{align}
Y\left(E_{\text {c.m. }}-E_{\text {kin }}^{\infty}\right)=Y_{0} \sum_{\ell=100}^{l=350}(2\ell+1) G_{\ell}\left(E_{k i n}^{\infty}\right) \label{eq108}
\end{align}

\begin{figure}[!htb]
\includegraphics*[width=0.48\textwidth]{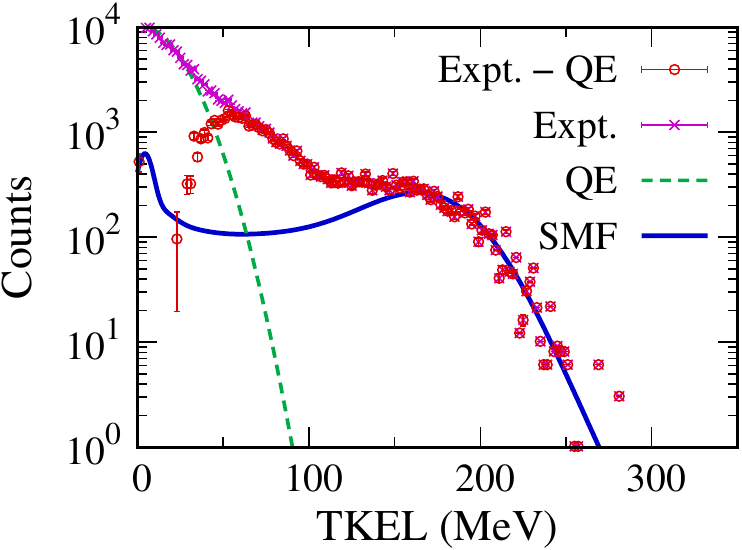}
\caption{The integrated total kinetic energy loss (TKEL) distribution for ${ }^{136} \mathrm{Xe}+{ }^{208} \mathrm{Pb}$ reaction at $E_{c.m.}=526$~MeV : Magenta crosses (open circles) with error bars represent the experimental data with (without) quasi-elastic contribution reported in Ref.~\cite{kozulin2012}. A Gaussian fit of the quasi-elastic contribution is shown by the green dashed line. The result of SMF calculations are shown by the blue solid line, where the normalization constant is set to $Y_{0}=3.5$.}
\label{fig:5.5}
\end{figure}

The normalization constant $Y_{0}$ is adjusted to the data at a suitable point. The experimental setup in the work of Kozulin \textit{et al.}~\cite{kozulin2012} has an energy resolution of 25~MeV. To compare with the data, this experimental uncertainty should be accounted for by, e.g. a folding procedure of the calculated kinetic energy distribution. The folding procedure introduces an approximately uniform shift in the kinetic energy distribution. Therefore, we consider that it does not change the shape of the calculated curve, and the folding effect is absorbed by the normalization constant $Y_{0}$.

Fig.~\ref{fig:5.5} shows a comparison of the calculations with experimental data for the collisions of ${ }^{136} \mathrm{Xe}+{ }^{208} \mathrm{Pb}$ reaction at $E_{c.m.}=526$~MeV. The measured TKE distribution for twp body events (without sequential fission events) is shown by crosses with error bars. Open circles with error bars represent the experimental data from which the quasi-elastic component (a Gaussian fit to the data shown by a dashed line) has been removed~\cite{kozulin2012}. The calculated TKE distribution according to Eq.~\eqref{eq108} is shown by the solid line. From the figure we find that the calculations provide a good description for strongly damped events with large energy losses. TKEL $\geq 150$~MeV. However, it underestimates the count curve over the lower energy-loss segment. This behavior is a result of apparent large dispersions of the TKE distribution over the range $\ell=(200-300) \hbar$, which may be due to the over-prediction of the radial momentum diffusion coefficients and/or the approximate description of the radial friction for the large angular momentum region. Although further improvements of the formalism are necessary, we consider that the quantal diffusion approach based on the SMF theory provides a promising microscopic basis for quantifying kinetic energy dissipation and fluctuations in low-energy heavy-ion reactions.

\section{Summary and Conclusions} \label{sec6}

Phenomenological descriptions of heavy-ion reactions in terms of classical Langevin method have been useful for understanding the energy dissipation and fluctuations, and the multinucleon transfer mechanism in deep inelastic heavy-ion collisions and quasi fission reactions.  In these models, a set of macroscopic variables such as relative distance, relative linear and angular momentum, mass, and charge of the projectile-like and target-like fragments are taken as relevant macroscopic variables. Inertia parameters, potential energy, and transport coefficients are determined by macroscopic means. Because of the presence of adjustable parameters, usually, these phenomenological models cannot make reliable predictions. Time-dependent Hartree-Fock (TDHF) theory provides a microscopic approach for describing the collision dynamics. in TDHF effective mean-field Hamiltonian is specified by an energy density functional, which involves several parameters. These parameters are fixed in terms of ground state properties of a set of nuclei and kept the same values for different nuclear collisions. Although the TDHF theory provides a good description of the most probable path of the collision dynamics, the fluctuations around the mean distribution are severely underestimated. As an extension of the mean-field approach, the time-dependent random phase approximation (TDRPA) has been proposed. In TDRPA, it is in principle possible to calculate dispersions of macroscopic variables around their mean values. However, in the current formulation, this theory is applicable only for symmetric collisions.  

In recent work, the TDHF approach is extended by combining the mean-field description with nuclear transport theory in the microscopic level. This approach is referred to as stochastic mean-filed theory (SMF). In the SMF theory the initial correlations are incorporated into description of dynamics in a stochastic approximation. For a given initial condition, in the standard mean-field approach, a deterministic path is followed. On the other hand, in the SMF theory for a given initial condition, an ensemble of mean-field events is generated by introducing stochasticity only for the initial state. Employing the ensemble of single-particle density matrices, it is possible to determine not only the mean path but also the distribution of events around the mean evolution. 

When a di-nuclear configuration is maintained during the collision process, it is possible to simplify the SMF calculations to a large extent. In this case, we can define macroscopic variables by a geometric projection of the SMF theory with the help of window dynamics. Macroscopic variables, such as the neutron/proton numbers of projectile-like or target-like fragments,  relative coordinate, and relative momenta are defined by integrating local neutron and proton density and local current density on both sides of the window plane in each event. Employing the equation of motion of the SMF theory, it is possible to deduce quantal Langevin equations for the macroscopic variables. For sufficiently small fluctuations, we can linearize these Langevin equations around the mean values of macroscopic variables. In this manner, we can deduce linear coupled Langevin equations for neutron and proton numbers of projectile-like or target-like fragments. In a similar manner, we can derive linear coupled Langevin equations for the radial and tangential components of the relative momentum.  It is well known that Langevin description is equivalent to Fokker-Planck description for the distribution function of the macroscopic variables.  Furthermore, when the drift coefficients are linear functions of the macroscopic variables, the joint distribution functions of neutron numbers/proton numbers, and radial/tangential components of the linear momentum are correlated Gaussian functions specified by their mean values and dispersions of macroscopic variables. As discussed in detail in Section 2 of the review, applications of the SMF theory for interacting di-nuclear complex are greatly simplified.  With the SMF theory, we have achieved remarkable progress in microscopic description of low-energy heavy-ion reactions, as compared to the standard TDHF approach.

\section{Acknowledgments}
The authors acknowledge D. Lacroix, K. Sekizawa, and B. Yilmaz for very stimulating discussions during the studies presented in this work. S. A. gratefully acknowledges Middle East Technical University, Vanderbilt University, and the Nuclear Physics Center at Orsay, for the warm hospitality extended to him during his visits. This work is supported in part by U.S. Department of Energy (DOE) Grant No. DE-SC00155113 (until  July 2022) and by U. S. Department of Energy (DOE) Grant No. DE-SC001347. This work is also supported in part by TUBITAK Grant No. 122F150.

\begin{appendices}
\section{Window Dynamics}\label{sec:appA}
\renewcommand{\theequation}{A\arabic{equation}} 
\setcounter{equation}{0}

We can determine the orientation of symmetry axis of the di-nuclear system with the help of the mass quadruple matrix of the di-nuclear system. In the center-of-mass system, the elements of the quadrupole matrix are given by
\begin{align}
Q_{i j}=3 \sigma_{i j}-\left(\sum_{i} \sigma_{i i}\right) \delta_{i j} ,\label{eqa1}
\end{align}
here $i, j,=1,2,3$ and $x_{1}, x_{2}, x_{3} \rightarrow x, y, z$. In this expression the elements of the sigma matrix are defined in terms of the position co-variances as,
\begin{align}
\begin{split}
\sigma_{i j}(t)= & \sum_{h}<\phi_{h}(t)\left|x_{i} x_{j}\right| \phi_{h}(t)> \\ & - \sum_{h, h^{\prime}}<\phi_{h}(t)\left|x_{i}\right| \phi_{h}(t)><\phi_{h^{\prime}}(t)\left|x_{j}\right| \phi_{h^{\prime}}(t)>  . \label{eqa2}
\end{split}
\end{align}
One can determine the direction of the symmetry axis, with the elements of the quadrupole matrix in the reaction plane which is defined to be at $z=0$. In this case, we simply need to diagonalize the $2 \times 2$ reduced quadrupole matrix on the reaction plane with elements $Q_{x x}, Q_{x y}, Q_{y x}, Q_{y y}$. The eigenvectors $\boldsymbol{E}_{+}$ and $\boldsymbol{E}_{-}$ of the quadrupole matrix specify the principal axes of the di-nuclear mass distribution on the reaction plane. These eigenvectors have the following form,
\begin{align}
\boldsymbol{E}_{ \pm}=\binom{Q_{ \pm}}{1} ,\label{eqa3}
\end{align}
with components given by
\begin{align}
Q_{ \pm}=\frac{Q_{x x}-Q_{y y} \pm \sqrt{\left(Q_{x x}-Q_{y y}\right)^{2}+4 Q_{x y}^{2}}}{2 Q_{x y}} ,\label{eqa4}
\end{align}
The eigenvalues $\Gamma_{ \pm}$corresponding to $Q_{ \pm}$ are given by,
\begin{align}
\Gamma_{ \pm}=\frac{Q_{x x}+Q_{y y} \pm \sqrt{\left(Q_{x x}-Q_{y y}\right)^{2}+4 Q_{x y}^{2}}}{2} .\label{eqa5}
\end{align}

The beam direction is taken to be $x$-direction in the fixed coordinate system $(x, y)$. The eigenvectors of the quadrupole matrix define a rotated orthogonal system in the reaction plane with axes pointing along $\boldsymbol{E}_{+}$ and $\boldsymbol{E}_{-}$ directions. We note that the eigenvalue $\Gamma_{+}$ is larger than the eigenvalue $\Gamma_{-}$. The eigenvector $\boldsymbol{E}_{+}$ associated with the large eigenvalue $\Gamma_{+}$ specifies the direction of the symmetry axis of the di-nuclear system. The angle $\theta$ between the positive direction of the $x$-axis and the direction of $\boldsymbol{E}_{+}=\left(Q_{+}, 1\right)$ is determined by
\begin{align}
\tan \theta=\frac{1}{Q_{+}}=\frac{2 Q_{x y}}{Q_{x x}-Q_{y y}+\sqrt{\left(Q_{x x}-Q_{y y}\right)^{2}+4 Q_{x y}^{2}}} .\label{eqa6}
\end{align}
Using the trigonometric identity,
\begin{align}
\tan \theta=\frac{\tan 2 \theta}{1+\sqrt{1+\tan ^{2} 2 \theta}}, \quad-\frac{\pi}{2}<\theta<+\frac{\pi}{2} ,\label{eqa7}
\end{align}
and
\begin{align}
\tan \theta=\frac{\tan 2 \theta}{1-\sqrt{1+\tan ^{2} 2 \theta}}, \quad+\frac{\pi}{2}<\theta<+\frac{3 \pi}{2} ,\label{eqa8}
\end{align}
we can express the angle $\theta$ of the symmetry axis in terms of the elements of the elements of the quadrupole or the elements of the sigma matrix as,
\begin{align}
\tan 2 \theta=\frac{2 Q_{x y}}{Q_{x x}-Q_{y y}}=\frac{2 \sigma_{x y}}{\sigma_{x x}-\sigma_{y y}}, \label{eqa9}  
\end{align}
which applies to the entire range specified in Eq.~\eqref{eqa7} and Eq.~\eqref{eqa8}.

\section{Analysis of the Closure Relation}\label{sec:apPb}
\renewcommand{\theequation}{B\arabic{equation}} 
\setcounter{equation}{0}
We re-write Eq.~\eqref{eq39} as,
\begin{align}
\begin{split}
\sum_{a \in P, h \in T} A_{a h}^{\alpha}(t) A_{a h}^{* \alpha}\left(t^{\prime}\right)= & \sum_{h \in T}  \int d^{3} R d^{3} r \delta\left(\boldsymbol{r}+\boldsymbol{u}_{h} \tau\right) \\ & \times W_{h}^{\alpha}\left(\boldsymbol{r}_{1}, t\right) W_{h}^{* \alpha}\left(\boldsymbol{r}_{2}, t^{\prime}\right), \label{eqb1}
\end{split}
\end{align}
in which we introduced the coordinate transformation,
\begin{align}
\boldsymbol{R}=\left(\boldsymbol{r}_{1}+\boldsymbol{r}_{2}\right) / 2, \quad \boldsymbol{r}=\boldsymbol{r}_{1}-\boldsymbol{r}_{2},\label{eqb2}
\end{align}
and it's inverse as
\begin{align}
\boldsymbol{r}_{1}=\boldsymbol{R}+\boldsymbol{r} / 2, \quad \boldsymbol{r}_{2}=\boldsymbol{R}-\boldsymbol{r} / 2 .\label{eqb3}
\end{align}
For clarity, we provide quantities $W_{h}^{\alpha}\left(\boldsymbol{r}_{1}, t\right)$ and \\ $W_{h}^{* \alpha}\left(\boldsymbol{r}_{2}, t^{\prime}\right)$ here again,
\begin{align}
\begin{split}
W_{h}^{\alpha}\left(\boldsymbol{r}_{1}, t\right)=& \frac{\hbar}{m} g\left(x_{1}^{\prime}\right) \\ & \times \left(\boldsymbol{\boldsymbol{\hat{e}}} \cdot \nabla_{1} \Phi_{h}^{\alpha}\left(\boldsymbol{r}_{1}, t\right)-\frac{x_{1}^{\prime}}{2 \kappa^{2}} \Phi_{h}^{\alpha}\left(\boldsymbol{r}_{1}, t\right)\right)  ,   
\end{split}
\label{eqb4}
\end{align}
and
\begin{align}
\begin{split}
W_{h}^{* \alpha}\left(\boldsymbol{r}_{2}, t^{\prime}\right)=\frac{\hbar}{m} g\left(x_{2}^{\prime}\right) & \left(\boldsymbol{\boldsymbol{\hat{e}}} \cdot \nabla_{2} \Phi_{h}^{* \alpha}\left(\boldsymbol{r}_{2}, t^{\prime}\right) \right. \\ & \left. -\frac{x_{2}^{\prime}}{2 \kappa^{2}} \Phi_{h}^{* \alpha}\left(\boldsymbol{r}_{2}, t^{\prime}\right)\right)    .
\end{split}
\label{eqb5}
\end{align}

Because of the delta function in the integrand of Eq.~\eqref{eqb1}, we make the substitution $\boldsymbol{r}=-\boldsymbol{u}_{h}^{\alpha}(\boldsymbol{R}, T) \tau$ in the wave functions and introduce the backward diabatic shift to obtain,
\begin{align}
\Phi_{h}^{\alpha}(\boldsymbol{R}+\boldsymbol{r} / 2, t)=\Phi_{h}^{\alpha}\left(\boldsymbol{r}-\boldsymbol{u}_{h}^{\alpha} \tau / 2, t\right) \approx \Phi_{h}^{\alpha}(\boldsymbol{R}, T) ,\label{eqb6}
\end{align}
and
\begin{align}
\Phi_{h}^{\alpha}\left(\boldsymbol{R}-\boldsymbol{r} / 2, t^{\prime}\right)=\Phi_{h}^{\alpha}\left(\boldsymbol{R}+\boldsymbol{u}_{h}^{\alpha} \tau / 2, t^{\prime}\right) \approx \Phi_{h}^{\alpha}(\boldsymbol{R}, T). \label{eqb7}    
\end{align}
The local flow velocity of the wave function $\Phi_{h}^{\alpha}(\boldsymbol{R}, T)$ is calculated in a standard way,
\begin{align}
\begin{split}
\boldsymbol{u}_{h}^{\alpha}(\boldsymbol{R}, T)=\frac{\hbar}{m} & \frac{1}{\left|\Phi_{h}^{\alpha}(\boldsymbol{R}, T)\right|^{2}} \\ & \times \operatorname{Im}\left[\Phi_{h}^{* \alpha}(\boldsymbol{R}, T) \nabla \Phi_{h}^{\alpha}(\boldsymbol{R}, T)\right],
\end{split}
\label{eqb8}
\end{align}
with $T=\left(t+t^{\prime}\right) / 2=t-\tau / 2$. We write the product of Gaussian factors as
\begin{align}
g\left(x_{1}^{\prime}\right) g\left(x_{2}^{\prime}\right)=\tilde{g}\left(X^{\prime}\right) \tilde{G}\left(x^{\prime}\right) ,\label{eqb9}
\end{align}
with
\begin{align}
\tilde{g}\left(X^{\prime}\right)=\frac{1}{\sqrt{\pi}} \frac{1}{\kappa} \exp \left[-\left(\frac{X^{\prime}}{\kappa}\right)^{2}\right] ,\label{eqb10}
\end{align}
and
\begin{align}
\tilde{G}\left(x^{\prime}\right)=\left(\frac{1}{\sqrt{4 \pi}} \frac{1}{\kappa}\right) \exp \left[-\left(\frac{x^{\prime}}{2 \kappa}\right)^{2}\right] .\label{eqb11}
\end{align}
In these Gaussians, following the transformations below Eq.~\eqref{eq25}, the coordinates in the rotating frame are expressed in terms of the coordinates in the fix frame as, $X^{\prime}=\left(X-x_{0}\right) \cos \theta+\left(Y-y_{0}\right) \sin \theta$ and $x^{\prime}=x \cos \theta+y \sin \theta$. Carrying out the product of the factors and making the substitution $\boldsymbol{r}=-\boldsymbol{u}_{h}^{\alpha}(\boldsymbol{R}, T) \tau$, Eq.~\eqref{eqb1} becomes,
\begin{align}
\begin{split}
& \sum_{a \in P, h \in T} A_{a h}^{\alpha}(t) A_{a h}^{* \alpha}\left(t^{\prime}\right)=\left(\frac{\hbar}{m}\right)^{2} \sum_{h \in T} \int d^{3} R \tilde{g}\left(X^{\prime}\right) \\ & \times  \frac{G_{T}^{h}(\tau)}{\left|u_{\perp}^{h}(\boldsymbol{R}, T)\right|}  \left[\left|\boldsymbol{\boldsymbol{\hat{e}}} \cdot \nabla \Phi_{h}^{\alpha}(\boldsymbol{R}, T)\right|^{2}+\frac{X^{\prime 2}-\left(u_{\perp}^{h} \tau / 2\right)^{2}}{4 \kappa^{4}} \right. \\ & \left. \times \left|\Phi_{h}^{\alpha}(\boldsymbol{R}, T)\right|^{2}-\frac{X^{\prime}}{2 \kappa^{2}} \boldsymbol{\boldsymbol{\hat{e}}} \cdot \nabla\left(\left|\Phi_{h}^{\alpha}(\boldsymbol{R}, T)\right|^{2}\right)\right] .\label{eqb12}
\end{split}
\end{align}
Here, $u_{\perp}^{h}(\boldsymbol{R}, T)$ represents the component of the nucleon (proton or neutron) flow velocity perpendicular to the window $u_{\perp}^{h}(\boldsymbol{R}, T)=\boldsymbol{\boldsymbol{\hat{e}}} \cdot \boldsymbol{u}_{T}^{h}(\boldsymbol{R}, T)$ and $G_{h}(\tau)$ indicates the memory kernel
\begin{align}
G_{T}^{h}(\tau)=\frac{1}{\sqrt{4 \pi}} \frac{1}{\tau_{T}^{h}} \exp \left[-\left(\tau / 2 \tau_{T}^{h}\right)^{2}\right] ,\label{eqb13}
\end{align}
with the memory time $\tau_{T}^{h}=\kappa /\left|\boldsymbol{u}_{\perp}^{h}\right|$. In this expression $\tilde{g}\left(X^{\prime}\right)$ is sharp as Gaussian smoothing function centered on the window with a dispersion $\kappa=0.5 \mathrm{fm}$, Due to the fact that $\tilde{g}\left(X^{\prime}\right)$ is centered at $X^{\prime}=0$, the third term in Eq.~\eqref{eqb12} is nearly zero. In the second term, after carrying out an average over the memory, the factor in the middle becomes,
\begin{align}
X^{\prime 2}-\left(u_{x}^{h} \tau / 2\right)^{2} \rightarrow X^{\prime 2}-(\kappa / 2)^{2}. \label{eqb14} 
\end{align}
Since Gaussian $\tilde{g}\left(X^{\prime}\right)$ is sharply peaked around $X^{\prime}=0$, with a variance $(\kappa / 2)^{2}$, the second term in Eq.~\eqref{eqb13} is expected to be very small, as well. Neglecting the second and third terms, Eq.~\eqref{eqb1} becomes,
\begin{align}
\begin{split}
\sum_{a \in P, h \in T} A_{a h}^{\alpha}(t) A_{a h}^{* \alpha}\left(t^{\prime}\right) & =\left(\frac{\hbar}{m}\right)^{2} \sum_{h \in T} \int d^{3} R \tilde{g}\left(X^{\prime}\right) \\ & \times \frac{G_{T}^{h}(\tau)}{\left|u_{\perp}^{h}(\boldsymbol{R}, T)\right|}\left|\boldsymbol{\boldsymbol{\hat{e}}} \cdot \nabla \Phi_{h}^{\alpha}(\boldsymbol{R}, T)\right|^{2} .\label{eqb15}
\end{split}
\end{align}

In the continuation, we express the wave functions in terms of its magnitude and phase as ~\cite{gottfried1966},
\begin{align}
\Phi_{h}^{\alpha}(\boldsymbol{R}, T)=\left|\Phi_{h}^{\alpha}(\boldsymbol{R}, T)\right| \exp \left(i Q_{h}^{\alpha}(\boldsymbol{R}, T)\right) .\label{eqb16}
\end{align}
The phase factor, $Q_{h}^{\alpha}(\boldsymbol{R}, T)$ behaves as the velocity potential of the flow velocity of the wave. Using the definition given by Eq.~\eqref{eqb8}, we observe that the flow velocity is given by $\boldsymbol{u}_{h}^{\alpha}(\boldsymbol{R}, T)=(\hbar / m) \nabla Q_{h}^{\alpha}(\boldsymbol{R}, T)$. Near the window, in perpendicular direction, the phase factors vary faster than the magnitude of the wave functions. Neglecting the variation of the magnitude $\left|\Phi_{h}(\boldsymbol{R}, T)\right|$ in the vicinity of the window, we can express the gradient of the wave function in Eq.~\eqref{eqb13} as,
\begin{align}
\boldsymbol{\boldsymbol{\hat{e}}} \cdot \nabla \Phi_{h}^{\alpha}(\boldsymbol{R}, T) \approx i \Phi_{h}^{\alpha}(\boldsymbol{R}, T)\left(\boldsymbol{\hat{e}} \cdot \nabla Q_{h}^{\alpha}(\boldsymbol{R}, T)\right) ,\label{eqb17}
\end{align}
where the quantity inside the parentheses is the component of the flow velocity of the nucleon perpendicular to the window. As a result, Eq.~\eqref{eqb1} becomes,
\begin{align}
\sum_{a \in P, h \in T} A_{a h}^{\alpha}(t) A_{a h}^{* \alpha}\left(t^{\prime}\right)=\int d^{3} R \tilde{g}\left(X^{\prime}\right) \tilde{J}_{\perp, \alpha}^{T}(\boldsymbol{R}, t-\tau / 2) .\label{eqb18}
\end{align}
Here, $\tilde{J}_{\perp, \alpha}^{T}(\boldsymbol{R}, t-\tau / 2)$ represents the magnitude of the current densities perpendicular to the window due to each wave function originating from the target, and each term multiplied by the memory kernel,
\begin{align}
\tilde{J}_{\perp, \alpha}^{T}(\boldsymbol{R}, T)=\frac{\hbar}{m} \sum_{h \in T} G_{T}^{h}(\tau)\left|\operatorname{Im} \Phi_{h}^{*}(\boldsymbol{R}, T)\left(\boldsymbol{\hat{e}} \cdot \nabla \Phi_{h}(\boldsymbol{R}, T)\right)\right| .\label{eqb19}
\end{align}

In Eq.~\eqref{eqb19} we introduce a further approximation by replacing the individual memory kernels $G_{T}^{h}(\tau)$ by their average values taken over the hole states
\begin{align}
G_{T}(\tau)=\frac{1}{\sqrt{4 \pi}} \frac{1}{\tau_{T}} \exp \left[-\left(\tau / 2 \tau_{T}\right)^{2}\right] ,\label{eqb20}
\end{align}
with the memory time determined by the average speed $u_{T}$ via $\tau_{T}=\kappa /\left|u_{T}(t)\right|$. The average value $u_{T}^{h}(t)$ of the flow speed for each hole state across the window is calculated as $u_{T}^{h}(t)=\int d^{3} R \tilde{g}\left(X^{\prime}\right) \boldsymbol{\hat{e}} \cdot \boldsymbol{j}_{T}^{h}(\boldsymbol{R}, t) / \int d^{3} R \tilde{g}\left(X^{\prime}\right) \rho_{T}^{h}(\boldsymbol{R}, t) \quad$, where $\boldsymbol{j}_{T}^{h}(\boldsymbol{R}, t)$ and $\rho_{T}^{h}(\boldsymbol{R}, t)$ denote the current density and density of hole state originating from target, respectively. The average $\left|u_{T}(t)\right|$ is then calculated by taking the mean value of all the flow speeds $\left|u_{T}^{h}(t)\right|$ of the hole states. It is possible to calculate the average speed from $u_{T}(t)=\int d^{3} R \tilde{g}\left(X^{\prime}\right) \boldsymbol{\hat{e}} \cdot \boldsymbol{j}_{T}(\boldsymbol{R}, t) / \int d^{3} R \tilde{g}\left(X^{\prime}\right) \rho_{T}(\boldsymbol{R}, t)$, where $\rho_{T}(\boldsymbol{R}, t)$ and $\boldsymbol{j}_{T}(\boldsymbol{R}, t)$ are the total density and the total current density of the states originating from the target. We expect both average speeds to have nearly the same magnitude.

\section{Momentum Diffusion Coefficient}\label{sec:appC}
\renewcommand{\theequation}{C\arabic{equation}} 
\setcounter{equation}{0}
In this Appendix, we derive the expression of the momentum diffusion coefficient given by Eq.~\eqref{eq95}. We introduce a partial integration in the expression for $Y_{h a}^{\alpha}(t)$ in Eq.~\eqref{eq81} to get
\begin{align}
\begin{split}
& Y_{h a}^{\alpha}(t)=\frac{\hbar^{2}}{m} \int d^{3} r_{1}\left(g\left(x_{1}^{\prime}\right)\left[\boldsymbol{\hat{e}}_{\alpha} \cdot \boldsymbol{\nabla}_{1} \boldsymbol{\hat{e}}_{R} \cdot \boldsymbol{\nabla}_{1} \phi_{h}^{*}\left(\boldsymbol{r}_{1}, t\right)\right] \right. \\
& \left. +\frac{1}{2}\left[\boldsymbol{\hat{e}}_{\alpha} \cdot \boldsymbol{\nabla}_{1} g\left(x_{1}^{\prime}\right) \boldsymbol{\hat{e}}_{R} \cdot \boldsymbol{\nabla}_{1} \phi_{h}^{*}\left(\boldsymbol{r}_{1}, t\right)\right. \right. \\
& \qquad \qquad \qquad \left. \left. +\boldsymbol{\hat{e}}_{R} \cdot \boldsymbol{\nabla}_{1} g\left(x_{1}^{\prime}\right) \boldsymbol{\hat{e}}_{\alpha} \cdot \boldsymbol{\nabla}_{1} \phi_{h}^{*}\left(\boldsymbol{r}_{1}, t\right)\right] \right. \\
& \left.+\frac{1}{4}\left[\boldsymbol{\hat{e}}_{\alpha} \cdot \boldsymbol{\nabla}_{1} \boldsymbol{\hat{e}}_{R} \cdot \boldsymbol{\nabla}_{1} g\left(x_{1}^{\prime}\right)\right] \phi_{h}^{*}\left(\boldsymbol{r}_{1}, t\right)\right) \phi_{a}\left(\boldsymbol{r}_{1}, t\right) ,\label{eqc1}
\end{split}
\end{align}
and similarly for it's complex conjugate
\begin{align}
\begin{split}
& Y_{h a}^{* \alpha}(t)=\frac{\hbar^{2}}{m} \int d^{3} r_{1}\left(g\left(x_{1}^{\prime}\right)\left[\boldsymbol{\hat{e}}_{\alpha} \cdot \boldsymbol{\nabla}_{1} \boldsymbol{\hat{e}}_{R} \cdot \boldsymbol{\nabla}_{1} \phi_{h}\left(\boldsymbol{r}_{1}, t\right)\right] \right. \\
& +\frac{1}{2}\left[\boldsymbol{\hat{e}}_{\alpha} \cdot \boldsymbol{\nabla}_{1} g\left(x_{1}^{\prime}\right) \boldsymbol{\hat{e}}_{R} \cdot \boldsymbol{\nabla}_{1} \phi_{h}\left(\boldsymbol{r}_{1}, t\right) \right. \\
& \qquad \qquad \qquad \left. +\boldsymbol{\hat{e}}_{R} \cdot \boldsymbol{\nabla}_{1} g\left(x_{1}^{\prime}\right) \boldsymbol{\hat{e}}_{\alpha} \cdot \boldsymbol{\nabla}_{1} \phi_{h}\left(\boldsymbol{r}_{1}, t\right)\right] \\
& \left.+\frac{1}{4}\left[\boldsymbol{\hat{e}}_{\alpha} \cdot \boldsymbol{\nabla}_{1} \boldsymbol{\hat{e}}_{R} \cdot \boldsymbol{\nabla}_{1} g\left(x_{1}^{\prime}\right)\right] \phi_{h}\left(\boldsymbol{r}_{1}, t\right)\right) \phi_{a}^{*}\left(\boldsymbol{r}_{1}, t\right) .\label{eqc2}   
\end{split}
\end{align}
From the diabatic property of the TDHF wave functions, we can shift the wave functions back and forth for short time intervals, $\tau=t-t^{\prime}$, according to
\begin{align}
\phi_{a}\left(\boldsymbol{r}, t^{\prime}\right)=\phi_{a}(\boldsymbol{r}-\boldsymbol{u} \tau, t), \label{eqc3}
\end{align}
where $\boldsymbol{u} \tau$ denotes a small displacement during a short time interval calculated with an appropriate flow velocity $\boldsymbol{u}$. Using the closure relation
\begin{align}
\sum_{a} \phi_{a}^{*}\left(\boldsymbol{r}_{1}, t\right) \phi_{a}\left(\boldsymbol{r}_{2}-\boldsymbol{u} \tau, t\right)=\delta\left(\boldsymbol{r}_{1}-\boldsymbol{r}_{2}+\boldsymbol{u} \tau\right) ,\label{eqc4}
\end{align}
we obtain the following expression,
\begin{align}
\begin{split}
\sum_{h \in T, a \in P} & Y_{h p}^{\alpha}(t) Y_{h p}^{* \beta}\left(t^{\prime}\right)=  \sum_{h \in T} d^{3} r_{1} d^{3} r_{2} \\ & \times\delta\left(\boldsymbol{r}_{1}-\boldsymbol{r}_{2}+\boldsymbol{u} \tau\right) W_{h}^{\alpha}\left(\boldsymbol{r}_{1}, t\right) W_{h}^{*\beta}\left(\boldsymbol{r}_{2}, t^{\prime}\right) .\label{eqc5}
\end{split}
\end{align}
First, we consider of the case for $\alpha=\beta=R$. The radial expression $W_{h}^{R}\left(\boldsymbol{r}_{1}, t\right)$ is given by,
\begin{align}
\begin{split}
& W_{h}^{R}\left(\boldsymbol{r}_{1}, t\right)=\frac{\hbar^{2}}{m}\left(g\left(x_{1}^{\prime}\right)\left[\left(\boldsymbol{\hat{e}}_{R} \cdot \boldsymbol{\nabla}_{1}\right)\left(\boldsymbol{\hat{e}}_{R} \cdot \boldsymbol{\nabla}_{1}\right) \phi_{h}\left(\boldsymbol{r}_{1}, t\right)\right] \right. \\
& +\frac{1}{2}\left[\boldsymbol{\hat{e}}_{R} \cdot \boldsymbol{\nabla}_{1} g\left(x_{1}^{\prime}\right) \boldsymbol{\hat{e}}_{R} \cdot \boldsymbol{\nabla}_{1} \phi_{h}\left(\boldsymbol{r}_{1}, t\right) \right. \\ 
& \left. +\boldsymbol{\hat{e}}_{R} \cdot \boldsymbol{\nabla}_{1} g\left(x_{1}^{\prime}\right) \boldsymbol{\hat{e}}_{R} \cdot \boldsymbol{\nabla}_{1} \phi_{h}\left(\boldsymbol{r}_{1}, t\right)\right] \\
& \left.+\frac{1}{4}\left[\left(\boldsymbol{\hat{e}}_{\alpha} \cdot \boldsymbol{\nabla}_{1}\right)\left(\boldsymbol{\hat{e}}_{R} \cdot \boldsymbol{\nabla}_{1}\right) g\left(x_{1}^{\prime}\right)\right] \phi_{h}\left(\boldsymbol{r}_{1}, t\right)\right) .\label{eqc6}
\end{split}
\end{align}
Using the expression for $g\left(x^{\prime}\right)$ we find,
\begin{align}
\begin{split}
& W_{h}^{K}\left(\boldsymbol{r}_{1}, t\right)=\frac{\hbar^{2}}{m} g\left(x_{1}^{\prime}\right)\left(\left[\left(\boldsymbol{\hat{e}}_{R} \cdot \boldsymbol{\nabla}_{1}\right)^{2} \phi_{h}\left(\boldsymbol{r}_{1}, t\right)\right]\right. \\
& \left.-\frac{x_{1}^{\prime}}{\kappa^{2}}\left[\boldsymbol{\hat{e}}_{R} \cdot \boldsymbol{\nabla}_{1} \phi_{h}\left(\boldsymbol{r}_{1}, t\right)\right] \right. \\ & \left. +\frac{1}{4 \kappa^{4}}\left[x_{1}^{\prime 2}-\kappa^{2}\right] \phi_{h}\left(\boldsymbol{r}_{1}, t\right)\right) ,\label{eqc7}
\end{split}
\end{align}
and the complex conjugate,
\begin{align}
\begin{split}
& W_{h}^{* K}\left(\boldsymbol{r}_{2}, t\right)=\frac{\hbar^{2}}{m} g\left(x_{2}^{\prime}\right)\left(\left[\left(\boldsymbol{\hat{e}}_{R} \cdot \boldsymbol{\nabla}_{2}\right)^{2} \phi_{h}^{*}\left(\boldsymbol{r}_{2}, t\right)\right]\right.  \\
& \left.-\frac{x_{2}^{\prime}}{\kappa^{2}}\left[\boldsymbol{\hat{e}}_{R} \cdot \boldsymbol{\nabla}_{1} \phi_{h}^{*}\left(\boldsymbol{r}_{2}, t\right)\right] \right. \\ & \left. +\frac{1}{4 \kappa^{4}}\left[x_{2}^{\prime 2}-\kappa^{2}\right] \phi_{h}^{*}\left(\boldsymbol{r}_{2}, t\right)\right) .  \label{eqc8}
\end{split}
\end{align}
Let us introduce the following the coordinate transformation,
\begin{align}
\boldsymbol{R}=\left(\boldsymbol{r}_{1}+\boldsymbol{r}_{2}\right) / 2\;,\; \boldsymbol{r}=\boldsymbol{r}_{1}-\boldsymbol{r}_{2} ,\label{eqc9}
\end{align}
and the inverse transformation,
\begin{align}
\boldsymbol{r}_{1}=\boldsymbol{R}+\boldsymbol{r} / 2, \quad \boldsymbol{r}=\boldsymbol{R}-\boldsymbol{r} / 2 .\label{eqc10}
\end{align}
Because of the delta function, we can immediately do the integration over $\boldsymbol{r}$ in Eq.~\eqref{eqc5}, make the substitution for $\boldsymbol{r}=-\boldsymbol{u}_{h} \tau$, and introduce diabatic shifts in the wave functions,
\begin{align}
\begin{split}
\phi_{h}\left(\boldsymbol{r}_{1}, t\right) & =\phi_{h}(\boldsymbol{R}+\boldsymbol{r} / 2, t) \\ & =\phi_{h}\left(\boldsymbol{R}-\boldsymbol{u}_{h} \tau / 2, t\right) \approx \phi_{h}(\boldsymbol{R}, t) ,\label{eqc11}
\end{split}
\end{align}
and
\begin{align}
\begin{split}
\phi_{h}\left(\boldsymbol{r}_{2}, t\right) & =\phi_{h}(\boldsymbol{R}-\boldsymbol{r} / 2, t) \\ & =\phi_{h}\left(\boldsymbol{R}+\boldsymbol{u}_{h} \tau / 2, t\right) \approx \phi_{h}(\boldsymbol{R}, t) ,\label{eqc12}    
\end{split}
\end{align}
with $\bar{t}=\left(t+t^{\prime}\right) / 2$. We can express product of the Gaussian factors as $g\left(x_{1}^{\prime}\right) g\left(x_{2}^{\prime}\right)=\tilde{g}\left(X^{\prime}\right) \tilde{G}\left(x^{\prime}\right), $
where,
\begin{align}
\tilde{g}\left(X^{\prime}\right)=\frac{1}{\kappa \sqrt{\pi}} \exp \left[-\left(\frac{X^{\prime}}{\kappa}\right)^{2}\right] \label{eqc13},    
\end{align}

\begin{align}
\tilde{G}\left(x^{\prime}\right)=\frac{1}{\kappa \sqrt{4 \pi}} \exp \left[-\left(\frac{x^{\prime}}{2 \kappa}\right)^{2}\right] \label{eqc14},
\end{align}
with $x^{\prime}=\boldsymbol{\hat{e}}_{R} \cdot \boldsymbol{r}=-\boldsymbol{\hat{e}}_{R} \cdot \boldsymbol{u}_{h} \tau=u_{R}^{h} \tau$ and $X^{\prime}=\boldsymbol{\hat{e}}_{R} \cdot \boldsymbol{R}$ where $u_{R}^{h}(\boldsymbol{R}, \bar{t})$ denotes the component of the flow velocity of the hole states perpendicular to the window, which in general may depend on the mean position $\boldsymbol{R}=\left(\boldsymbol{r}_{1}+\boldsymbol{r}_{2}\right) / 2$ and mean time $\bar{t}=\left(t+t^{\prime}\right) / 2$. In the product $W_{h}^{\alpha}\left(\boldsymbol{r}_{1}, t\right) W_{h}^{*} \beta\left(\boldsymbol{r}_{2}, t^{\prime}\right)$ there are linear, second, third and fourth order terms in $x_{1}$ and $x_{2}$ in the integrand of Eq.~\eqref{eqc5}. The integrand contains the product of two sharp Gaussians, $\tilde{g}\left(X^{\prime}\right)$ and $\tilde{G}\left(x^{\prime}\right)$ which provides the memory kernel in the integrand. Taking the averages over the memory kernel and the over sharp Gaussian $\tilde{g}\left(X^{\prime}\right)$, all terms in the integrand of Eq.~\eqref{eqc5} which are proportional to the powers of $x_{1}$ and $x_{2}$ vanish. We obtain the similar results for other components of Eq.~\eqref{eqc5} with $\alpha, \beta=R, \theta$, and we find,
\begin{align}
\begin{split}
\sum_{h \in T, a \in P,T} & Y_{h a}^{\alpha}(t) Y_{h a}^{* \beta}\left(t^{\prime}\right)=\left(\frac{\hbar^{2}}{m}\right)^{2} \sum_{h \in T} \int d^{3} R \tilde{g}\left(X^{\prime}\right) \\ & \times \frac{G_{T}^{h}(\tau)}{\left|u_{R}^{h}(\boldsymbol{R}, \bar{t})\right|} 
\left[\left(\boldsymbol{\hat{e}}_{\alpha} \cdot \boldsymbol{\nabla}\right)\left(\boldsymbol{\hat{e}}_{R} \cdot \boldsymbol{\nabla}\right) \phi_{h}(\boldsymbol{R}, t)\right] \\ & \times \left[\left(\boldsymbol{\hat{e}}_{\beta} \cdot \boldsymbol{\nabla}\right)\left(\boldsymbol{\hat{e}}_{R} \cdot \boldsymbol{\nabla}\right) \phi_{h}(\boldsymbol{R}, \bar{t})^{*}\right], \label{eqc15}
\end{split}
\end{align}
where the memory kernel is defined as,
\begin{align}
G_{T}^{h}(\tau)=\frac{1}{\sqrt{4 \pi}} \frac{1}{\tau_{T}^{h}} \exp \left[-\left(\frac{\tau}{2 \tau_{T}^{h}}\right)^{2}\right] ,\label{eqc16}
\end{align}
with the memory time as $\tau_{T}^{h}=\kappa /\left|u_{R}^{h}\right|$. We can write the wave functions as $\phi_{h}(\boldsymbol{r}, t)=\left|\phi_{h}(\boldsymbol{r}, t)\right| \exp \left(i Q_{h}\right)$ ~\cite{gottfried1966}. Since the phase factor behaves like the velocity potential, neglecting derivative of the amplitude of the wave function, we have the approximate result,
\begin{align}
\begin{split}
\left(\boldsymbol{\hat{e}}_{R} \cdot \boldsymbol{\nabla}\right) \phi_{h} & \approx i \phi_{h}\left(\boldsymbol{\hat{e}}_{R} \cdot \boldsymbol{\nabla} Q_{h}\right) \\ & =i \phi_{h}(\boldsymbol{R}, \bar{t}) \frac{m}{\hbar} u_{R}^{h}(\boldsymbol{R}, \bar{t})    .
\end{split}
\label{eqc17}
\end{align}
In a similar manner, we can express the second order derivative of the wave function as,
\begin{align}
\begin{split}
\left(\boldsymbol{\hat{e}}_{\theta} \cdot \boldsymbol{\nabla}\right) & \left(\boldsymbol{\hat{e}}_{R} \cdot \boldsymbol{\nabla}\right) \phi_{h} \approx i\left[\left(\boldsymbol{\hat{e}}_{\theta} \cdot \boldsymbol{\nabla}\right) \phi_{h}(\boldsymbol{R}, \bar{t})\right] u_{R}^{h}(\boldsymbol{R}, \bar{t})  \\ 
& \left. \approx-\phi_{h}(\boldsymbol{R}, \bar{t})\left(\frac{m}{\hbar}\right)^{2} u_{\theta}^{h}(\boldsymbol{R}, \bar{t}) u_{R}^{h}(\boldsymbol{R}, \bar{t})\right] .\label{eqc18}
\end{split}
\end{align}
Using the current density, we can write the radial and the tangential flow velocities in the flowing form,
\begin{align}
\begin{split}
u_{\alpha}^{h}(\boldsymbol{R}, \bar{t})=\boldsymbol{\hat{e}}_{\alpha} & \cdot \boldsymbol{u}_{h}(\boldsymbol{R}, \bar{t})=  \frac{\hbar}{m} \frac{1}{\left|\phi_{h}(\boldsymbol{R}, \bar{t})\right|^{2}} \\ & \times  \operatorname{Im}\left(\phi_{h}^{*}(\boldsymbol{R}, \bar{t}) \boldsymbol{\hat{e}}_{\alpha} \cdot \boldsymbol{\nabla} \phi_{h}(\boldsymbol{R}, \bar{t})\right) .\label{eqc19}
\end{split}
\end{align}
Incorporating this expression, Eq.~\eqref{eqc15} becomes,
\begin{align}
\begin{split}
\sum_{h \in T, a \in P,T} Y_{h a}^{\alpha}(t) Y_{h a}^{* \beta}\left(t^{\prime}\right)=\sum_{h \in T} \int^{3} R \tilde{g}\left(X^{\prime}\right)  \\ \times G_{T}(\tau) J_{\alpha \beta}^{T}(\bar{R}, \bar{t})  .  
\end{split}
\label{eqc20}
\end{align}
Here, $J_{\alpha \beta}^{T}(\bar{R}, \bar{t})$ is given in Eq.~\eqref{eq96}, and $G_{T}(\tau)$ denotes the average memory kernel given by Eq.~\eqref{eqc16}, which is evaluated with the average value of the flow velocity of the hole states originating from the target. The second term on the right side of Eq.~\eqref{eq94} is evaluated in a similar manner, and we obtain the expression given by Eq.~\eqref{eq95} for the momentum diffusion coefficients.

\section{Radial Friction Coefficient}\label{sec:appD}
\renewcommand{\theequation}{D\arabic{equation}} 
\renewcommand{\thefigure}{D\arabic{figure}} 
\setcounter{figure}{0}
\setcounter{equation}{0}
In this Appendix, an analysis of the radial friction coefficient based the mean-field solution of TDHF is explained. The description contains the one-body dissipation of relative kinetic energy and the transfer of the relative angular momentum into the intrinsic degrees of freedom. The dominant mechanism for the one-body dissipation is the multinucleon exchange between the projectile-like and target-like fragments. However, a microscopic derivation of the, so called window formula, for the reduced friction coefficients from the TDHF approach is not trivial. Here, we consider the analogy with the random walk problem to deduce the reduced friction coefficient from the mean-field description of TDHF. By taking the ensemble average in Eq.~\eqref{eq79}, in the limit of small fluctuations, the mean evolution the rate of change of the relative momentum is given by,
\begin{align}
\begin{split}
\frac{\partial}{\partial t} \boldsymbol{P}=+\int d^{3} x g\left(x^{\prime}\right) \dot{x}^{\prime} m \boldsymbol{j}(\boldsymbol{r}, t) & +[\text { pot. terms }] \\ & +\boldsymbol{f}(t),
\end{split}
\label{eqd1}
\end{align}
where $\boldsymbol{j}(\boldsymbol{r}, t)=\frac{\hbar}{m} \sum_{h} \operatorname{Im}\left(\phi_{h}^{*}(\boldsymbol{r}, t) \boldsymbol{\nabla} \phi_{h}(\boldsymbol{r}, t)\right)$. This equation is equivalent to the TDHF description of the relative momentum. The first and the terms on the right are the conservative forces on the relative motion due the motion of the window and the potential terms. In the last term $\boldsymbol{f}(t)$ represents the dynamical force due to nucleon exchange between the projectile-like and target-like fragments with the radial and the tangent components,
\begin{align}
f_{\alpha}(t)=\int d^{3} x g\left(x^{\prime}\right) \sum_{h} \boldsymbol{\hat{e}}_{R} \cdot\left(A_{h h}^{\alpha}-B_{h h}^{\alpha}\right). \label{eqd2}
\end{align}
Using the approximate results of Eq.~\eqref{eqc17} in Appendix~\ref{sec:appC}, we can express the component of the dynamical force due to nucleon exchange as
\begin{align}
\begin{split}
f_{\alpha}(t)=-\frac{\hbar}{m} \int d^{3} x & g\left(x^{\prime}\right)  \sum_{h} m u_{\alpha}^{h}(\boldsymbol{r}, t) \\ & \times \operatorname{Im}\left(\phi_{h}^{*}(\boldsymbol{r}, t) \boldsymbol{\hat{e}}_{R} \cdot \boldsymbol{\nabla} \phi_{h}(\boldsymbol{r}, t)\right) ,\label{eqd3}
\end{split}
\end{align}
where the summation runs over the hole states originating from projectile and target nuclei. The dynamical force involves both the conservative and dissipative forces. To infer the dissipative part of the dynamical force, we use the analogy to the Langevin description of the random walk problem. As seen from Eq.~\eqref{eq95}, the direct momentum terms of the diffusion coefficients are determined by the sum of nucleon fluxes that carry the product of momentum components from projectile to target and from target to projectile. In analogy to the description of the random walk, the components of the dissipative force are determined by the net momentum flux across the window as follows,
\begin{align}
\begin{split}
& f_{\alpha}^{d i s s}(t)=-\frac{\hbar}{m} \int d^{3} x g\left(x^{\prime}\right) \sum_{h \in P} m u_{\alpha}^{h}(\boldsymbol{r}, t) \\ & \times \left|\operatorname{Im}\left(\phi_{h}^{*}(\boldsymbol{r}, t) \boldsymbol{\hat{e}}_{R} \cdot \boldsymbol{\nabla} \phi_{h}(\boldsymbol{r}, t)\right)\right| +\frac{\hbar}{m} \int d^{3} x  \\ 
& \times g\left(x^{\prime}\right) \sum_{h \in T} m u_{\alpha}^{h}(\boldsymbol{r}, t)\left|\operatorname{Im}\left(\phi_{h}^{*}(\boldsymbol{r}, t) \boldsymbol{\hat{e}}_{R} \cdot \boldsymbol{\nabla} \phi_{h}(\boldsymbol{r}, t)\right)\right|. \label{eqd4}
\end{split}
\end{align}
Here, we express the net momentum flux as the difference of the momentum flux carried by the hole orbitals originating from the projectile and the momentum flux carried by the hole orbitals originating from the target. This approximate description may overestimate the net momentum flux, in particular in collisions at large impact parameters, and may require improvements. The quantity $\left|\operatorname{Im}\left(\phi_{h}^{*}(\boldsymbol{r}, t) \boldsymbol{\hat{e}}_{R} \cdot \boldsymbol{\nabla} \phi_{h}(\boldsymbol{r}, t)\right)\right|$ represents the magnitude of the nucleon flux from one fragment to the other.

\begin{figure}[!htb]
\includegraphics*[width=0.48\textwidth]{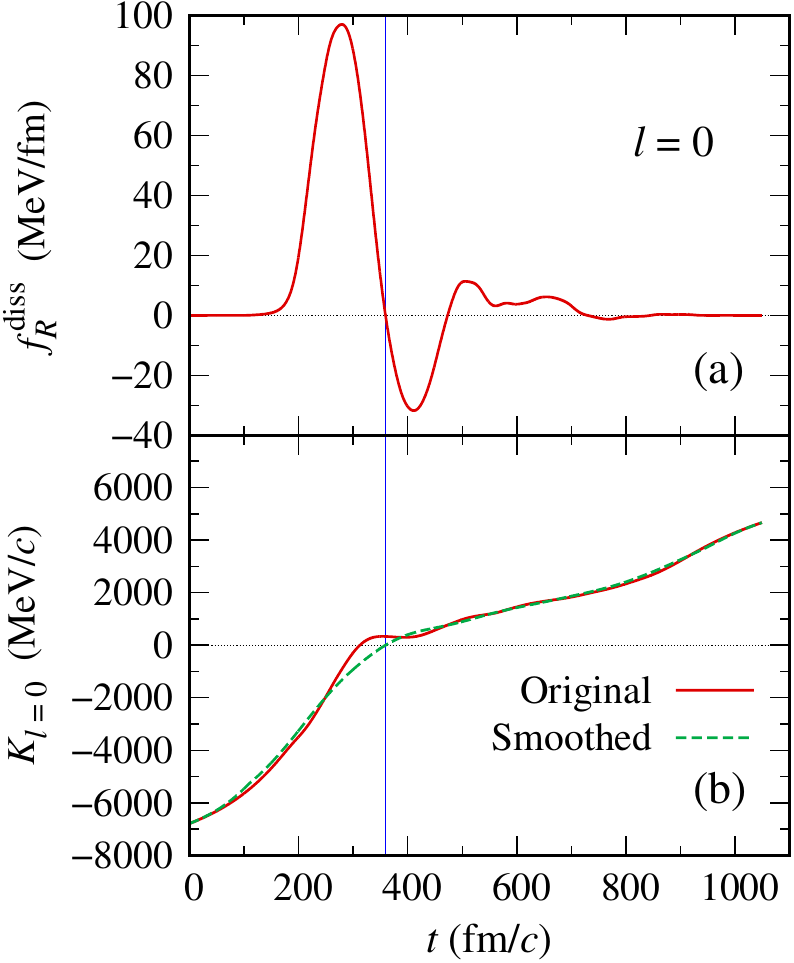}
\caption{Radial friction force $F_{R}^{\text {diss}}(t, \ell=0)$ (a) and radial momentum $K(t, \ell=0)$ (b) are shown as a function of time for central collision ( $\ell=0$ ). In (b), the original value of the radial momentum obtained from TDHF is shown by solid line, while smoothed curve is represented by dashed line. Vertical line indicates the time at which $F_{R}^{\text {diss}}(t, \ell=0)$ and the smoothed $K(t, \ell=0)$ vanish.}
\label{figd1}
\end{figure}

In this work, we consider the radial friction force and the reduced radial friction coefficient. To derive the expressions for the radial friction coefficient $\gamma_{R}(t, \ell)$ in the collision with the initial angular momentum $\ell$, we set the radial dissipative force to the phenomenological expression
\begin{align}
f_{R}^{\text {diss }}(t, \ell)=-\gamma_{R}(t, \ell) K(t, \ell) ,\label{eqd5}
\end{align}
where $K_{\ell}=\boldsymbol{\hat{e}}_{R} \cdot \boldsymbol{P}_{\ell}$ is the radial component of the relative linear momentum. From this relation, in principle, it should be possible to deduce the radial friction coefficient for each value of the initial angular momenta $\ell$. Fig.~\ref{figd1} shows the radial friction force in (a) and the radial momentum in (b) for the central collisions with initial angular momentum $\ell=0$ as a function of time. The radial momentum vanishes at the turning point, which occurs at about $t=360 \;\mathrm{fm} / \mathrm{c}$ [see Fig.~\ref{figd1}(b)]. We expect that the radial friction force vanishes at the turning points. However, we note that the friction force vanishes at a slightly later time, The time shift may originate from the approximate expression of the friction force in Eq.~\eqref{eqd4}, which overestimates the net momentum flux across the window, and the shifts become larger for increasing angular momentum.

\begin{figure}[!htb]
\includegraphics*[width=0.48\textwidth]{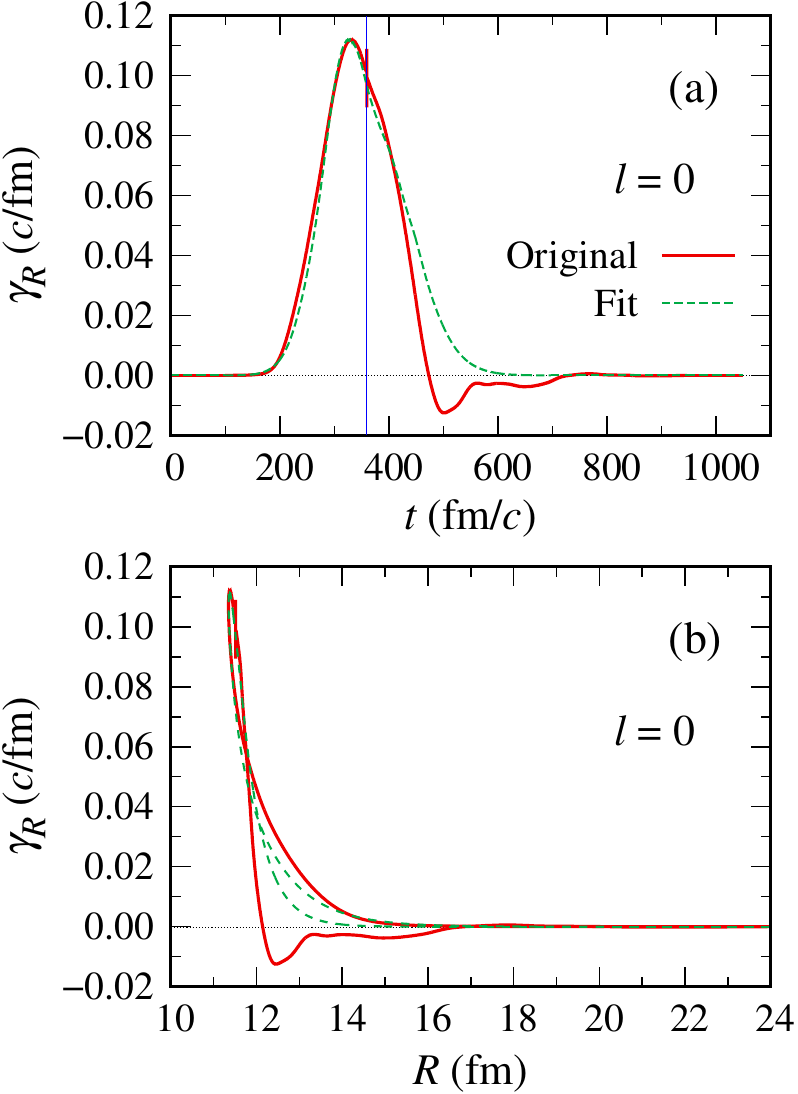}
\caption{The extracted reduced radial friction coefficient $\gamma_{R}(t, \ell=0)=-f_{\ell=0}^{d i s s}(t) / \bar{K}_{\ell=0}(t)$ is shown by the solid line for the central collision $(\ell=0)$ where $\bar{K}_{\ell=0}(t)$ is the smoothed radial momentum shown in Fig.~\ref{figd1} by dashed line. In (a)it is shown as a function of time, while the same quantity is shown as a function of relative distance in (b). The dashed line represents the parameterized function given by Eq.~\eqref{eqd7} and Eq.~\eqref{eqd8}.}
\label{figd2}
\end{figure}

\begin{figure}[!htb]
\includegraphics*[width=0.48\textwidth]{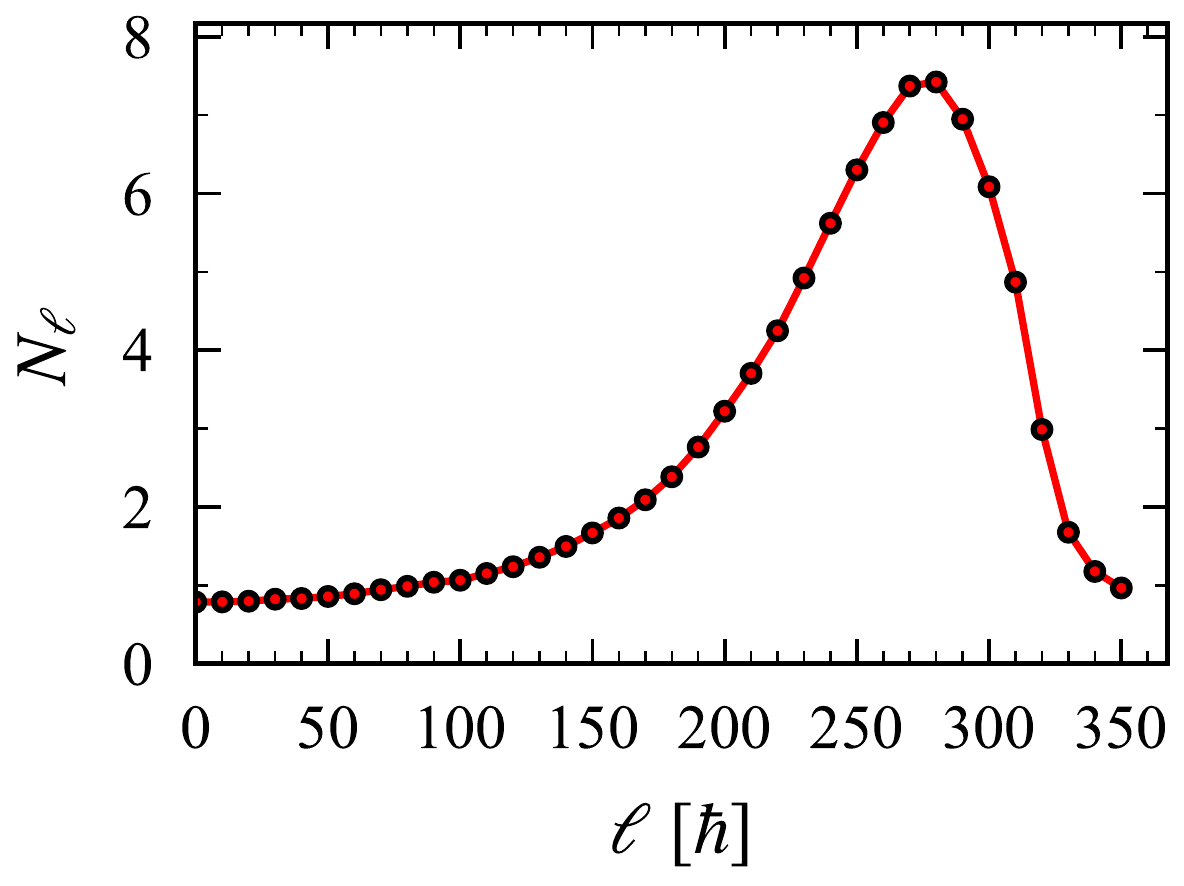}
\caption{The normalization constant $N_{\ell}$ in Eq.~\ref{eqd9} as a function of the initial orbital angular momentum, $\ell$}
\label{figd3}
\end{figure}

To extract the radial friction coefficients for all values of the initial angular momenta $\ell$, we employ an approximate method as described below. In Fig.~\ref{figd1}(b), we introduce a smoothing of the radial momentum so that the friction force and the radial momentum vanish at the same instant. The smoothed (averaged over a time interval of $220 \;\mathrm{fm} / \mathrm{c}$ ) radial momentum is shown by the dashed line in Fig.~\ref{figd1}(b), and the vertical line indicates the instant at which both the friction force and the radial momentum become vanishingly small. We can then determine the reduced friction coefficient for $\ell=0$ from the ratio,
\begin{align}
\gamma_{R}(t, \ell=0)=-\frac{1}{\bar{K}_{\ell=0}(t)} f_{R}^{d i s s}(t, \ell=0). \label{eqd6}
\end{align}
Note that the phenomenological relation in Eq.~\eqref{eqd5} is used. Figure~\ref{figd2} shows the friction coefficient in the central collision as a function of time. The turning point is reached around time $t=360~\mathrm{fm} / \mathrm{c}$. Dissipation occurs mainly during the incoming phase until the turning point. Only a small fraction of the dissipation occurs during the outgoing phase after, the turning point until the separation of the fragments. As a result, we can ignore the unphysical negative tail after $t=460 \;\mathrm{fm} / \mathrm{c}$. In Fig.~\ref{figd3}, we plot the friction coefficient $\gamma_{R}[R(t, \ell=0), \ell=0]$ for $\ell=0$ in terms of the radial distance $R(t, \ell=0)$. It is possible to parameterize the friction coefficient during the incoming phase until it reaches the maximum value in terms of an exponential form as
\begin{align}
\gamma_{R}^{\text {inc }}[R(t)]=c_{1} \exp \left[-c_{2}\left(R(t)-c_{3}\right)\right] ,\label{eqd7}
\end{align}
where $c_{1}=9.97 \;\mathrm{c} / \mathrm{fm}, c_{2}=1.04\; \mathrm{fm}$ and $c_{3}=6.86 \;\mathrm{fm}$. The friction coefficient reaches the maximum value at the minimum distance $R_{\min }$. In the outgoing segment from the minimum distance $R_{\min }$ until the fragment are separated, we parameterize the friction coefficient with multiplying by another exponential form as
\begin{align}
\gamma_{R}^{\text {out }}[R(t)]=\gamma_{R}^{\text {inc }}[R(t)] c_{4} \exp \left[-c_{5}\left(R(t)-R_{\min }\right)\right], \label{eqd8}
\end{align}
where, $c_{4}=1.96$ and $c_{5}=0.95$. Note that $c_{4}>1$ is used, because the minimum distance is reached at $t=313~\mathrm{fm} / \mathrm{c}$, while the obtained friction coefficient has a peak value at slightly later time. We joined the two expressions for the incoming and outgoing phases smoothly around the turning point. Assuming that the friction coefficients scale with the relative distance for all initial angular momentum in a similar manner as for the central collision, we express the reduced radial friction coefficient for all values $\ell$ as,
\begin{align}
\gamma_{R}(t, \ell)=N_{\ell} \gamma_{R}\left[R_{\ell}(t)\right]. \label{eqd9}
\end{align}
The factor $\gamma_{R}\left[R_{\ell}(t)\right]$ is the friction coefficient given by Eq.~\eqref{eqd7} and Eq.~\eqref{eqd8} is determined by the same form as given in Eq.~\eqref{eqb7} and Eq.~\eqref{eqb8} extracted from $\ell=0$ case as a function relative distance. The normalization factor $N_{\ell}$ is determined by matching the dissipated energy with the total energy loss calculated from the TDHF dynamics for each initial angular momentum,
\begin{align}
E_{\ell}^{d i s s}=\int d t \gamma_{R}(t, \ell) \frac{K_{\ell}^{2}(t)}{\mu}=E_{k i n}^{\infty}(\ell) .\label{eqd10}
\end{align}

\begin{figure}[!htb]
\includegraphics*[width=0.48\textwidth]{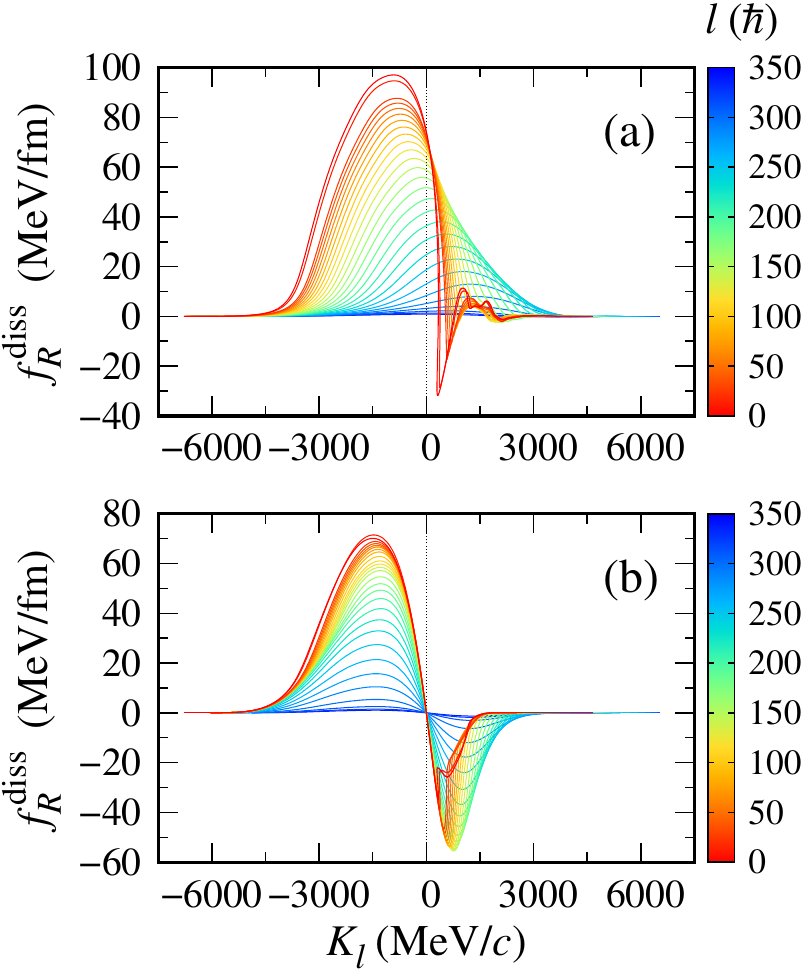}
\caption{The friction force $f_{\ell}^{diss}$ is shown as a function radial momentum $K_{\ell}$ for the ${ }^{136} \mathrm{Xe}+{ }^{208}\mathrm{Pb}$ reaction at $E_\mathrm{c.m.}=526$~MeV for a range of $\ell$ values. Colors represent the values of the initial angular momenta. (a) The friction force obtained with Eq.~\ref{eqd4} based on the single-particle orbitals in TDHF. (b) The reconstructed radial friction force according to phenomenological expression, $f_{l}^{d i s s}=-\gamma_{\ell} K_{\ell}$, where $\gamma_{\ell}$ is expressed as fitted functions given by Eq.~\eqref{eqd7} and Eq.~\eqref{eqd8}.
}
\label{figd4}
\end{figure}

Figure~\ref{figd3} shows the magnitude of the normalization constant as a function of $\ell$. Figure~\ref{fig:5.2} in the main text presents the reduced radial friction coefficients determined in the manner obtained above as a function of time for typical angular momenta. In Fig.~\ref{figd4}, we compare the original radial friction force as predicted by Eq.~\eqref{eqd4} and the reconstructed radial friction force using the approximate treatment of Eq.~\eqref{eqd9} in panel (b) as a function of the radial momentum for the range of initial angular momenta. It is visible that the original radial friction forces shown in Fig.~\ref{figd4}(a) do not vanish at the turning point at which the radial momentum changes its sign. On the other hand, the reconstructed friction force as a function of the radial momentum shown in Fig.~\ref{figd4}(b) gives rise the expected behavior and provides a support for the reduced friction coefficients that we obtain using the approximate procedure.

\section{Einstein Relations}\label{sec:appE}
\renewcommand{\theequation}{E\arabic{equation}} 
\setcounter{equation}{0}

We consider a macroscopic variable $Q$ and its conjugate momentum $P$. In the semiclassical approximation, the time evolution of the macroscopic variables are determined by a Langevin equation,

\begin{equation}
\frac{d P}{d t}=-\frac{d U}{d Q}-\beta \frac{d Q}{d t}+\xi_{P}(t) \label{eq_e1} .
\end{equation}

Here, the first term is the macroscopic driving force, the second and third terms denote the dissipative force and the stochastic force. Momentum diffusion coefficient is determined by the auto-correlation function of the stochastic force as

\begin{equation}
D_{p}(t)=\int_{0}^{t} d t^{\prime} \overline{\xi_{P}(t) \xi_{P}(t^{\prime})} \label{eq_e2},
\end{equation}
where bar indicates the average value taker over the ensemble generated by the Langevin Eq.~\eqref{eq_e1}. In the Markovian limit, this average value is expressed in terms of the friction coefficient and the effective temperature $T^{*}$ as
\begin{align}
D_{P}=\beta T^{*}  .  \label{eq_e3}
\end{align}

The effects of shell structure and Pauli blocking can be incorporated into $\beta$ and the effective temperature $T^{*}$. In the overdamped limit, the rate of change of momentum is small and can be neglected in Eq.~\eqref{eq_e1}. Hence, in the overdamped limit the macroscopic variable is determined by a simple form of the Langevin equation

\begin{equation}
\frac{d Q}{d t}=-\frac{1}{\beta} \frac{d U}{d Q}+\xi_Q(t) .\label{eq_e4}
\end{equation}
Here, the stochastic force for the variable $Q$ is $\xi_{Q}(t)=\frac{1}{\beta} \xi_{P}(t)$. The diffusion coefficient $D_{Q}$ for the macroscopic variable is given by the autocorrelation function of $\xi_{Q}(t)$,
\begin{equation}
D_{Q}=\int_{0}^{t} d t^{\prime} \overline{\xi_{Q}(t) \xi_{Q}\left(t^{\prime}\right)}. \label{eq_e5}
\end{equation}
Employing the result in Eq.~\eqref{eq_e3}, the diffusion coefficient $D_{Q}$ is expressed as
\begin{equation}
D_{Q}=\frac{T^{*}}{\beta} .\label{eq_e6}
\end{equation}
Taking the mean value of Eq.~\eqref{eq_e4} and using Eq.~\eqref{eq_e6}, we find

\begin{align}
\frac{d Q}{d t}=-\frac{D_Q}{T^{*}} \frac{d U}{d Q}  .  \label{eq_e7}
\end{align}
In the Einstein relations of Eq.~\eqref{eq50} and Eq.~\eqref{eq51}, the macroscopic variables are the neutron and proton numbers of projectile-like or target-like fragments, $N(t)$ and $Z(t)$. The rate of change of the macro variables $v_{n}(t)=\frac{d N}{d t}$ and $v_{p}(t)=\frac{d Z}{d t}$ denote neutron and proton drift velocities.

\end{appendices}
\bibliographystyle{ieeetr}
\bibliography{VU_bibtex_master.bib}

\end{document}